\documentclass[a4paper,fleqn,usenatbib,useAMS]{mnras}

\usepackage{graphicx}	% Including figure files
\usepackage{amsmath}	% Advanced maths commands
\usepackage{amssymb}	% Extra maths symbols
\usepackage{multicol}        % Multi-column entries in tables
\usepackage{bm}		% Bold maths symbols, including upright Greek
\usepackage{pdflscape}	% Landscape pages
\usepackage{float}
\usepackage{subcaption}  % For the perturbation analysis figures

\usepackage[T1]{fontenc}
\usepackage{ae,aecompl}

\usepackage{newtxtext,newtxmath}
\title[Pitch Angle Selection Effects]{\textit{SpArcFiRe: morphological selection effects due to reduced visibility of tightly winding arms in distant spiral galaxies}}

\author[Peng, English, Silva, Davis, and Hayes]
{
Tianrui (Rae) Peng$^{1,2}$,
John Edward English$^1$,
Pedro Silva$^{1}$,
Darren R. Davis$^{1}$,
Wayne B. Hayes$^{1}$
\\
$^{1}$Department of Computer Science, UC Irvine\\
$^{2}$Current address: Department of Computer Science, Columbia University
}

\date{Last updated 2015 May 22; in original form 2013 September 5}

\pubyear{2017}

\begin{document}
\label{firstpage}
\pagerange{\pageref{firstpage}--\pageref{lastpage}}
\maketitle

\begin{abstract}
The Galaxy Zoo project has provided a plethora of valuable morphological data on a large number of galaxies from various surveys. Several biases have been identified in the Galaxy Zoo data, which users of the data must be aware.
Here we report on a newly discovered selection effect. In particular, astronomers interested in studying spiral galaxies may select a set of spiral galaxies based upon a threshold in {\it spirality}, which we define as the fraction of Galaxy Zoo humans who have reported seeing spiral structure.
One tool that can be used to analyze spiral galaxies is SpArcFiRe, an automated tool that decomposes a spiral galaxy into its constituent spiral arms, providing objective, quantitative data on their structure. One of SpArcFiRe's measures is the pitch angle of spiral arms. We have observed that, when selecting a set of spiral galaxies based on a threshold on Galaxy Zoo spirality, the spiral arms appear to have a mean pitch angle that very clearly increases linearly with redshift for $0.05 \le z \le 0.085$ even after accounting for the Malmquist bias.
We hypothesize that this is a selection effect, based on the fact that tightly-wound spiral arms become less visible as spatial resolution and noise degrade the image with increasing redshift, leading to fewer such galaxies being included in the sample at higher redshifts.  We corroborate this hypothesis by artificially degrading images of nearby galaxies, then using a machine learning algorithm trained on Galaxy Zoo data to provide a spirality for each artificially degraded image. It correctly predicts that the detected spirality of a fixed galaxy decreases as image quality degrades. We then use these spiralities to corroborate the hypothesis that the mean pitch angle of those galaxies remaining above a fixed spirality threshold is higher than those eliminated by the selection effect.
This demonstrates that users who select samples of galaxies using a threshold of Galaxy Zoo votes must carefully consider the possibility of selection effects on morphological measures, even if the measure itself is believed to be objective and unbiased. Finally, we also perform an empirical sensitivity analysis to demonstrate that SpArcFiRe's output changes in a smooth and predictable fashion to changes in its internal algorithmic parameters.
\end{abstract}


%%%%%%%%%%%%%%%%%%%%%%%%%%%%%%%%%%%%%%%%%%%%%%%%%%

%%%%%%%%%%%%%%%%% BODY OF PAPER %%%%%%%%%%%%%%%%%%

% The MNRAS class isn't designed to include a table of contents, but for this document one is useful.
% I therefore have to do some kludging to make it work without masses of blank space.
%\begingroup
%\let\clearpage\relax
%\tableofcontents
%\endgroup
%\newpage

\section{Introduction}

The arms of spiral galaxies are still not fully understood \citep{BinneyTremaine2011}, in part because there is no widely accepted method of quantifying their visible structure. While the light profiles of elliptical galaxies are fairly easy to model \citep{Peng2010}, no such easy quantification exists for spirals. The human-based classification scheme {\it Galaxy Zoo} provides an initial idea of the structure of galaxies \citep{Lintott2008,Lintott2010,WillettEtAlGZ2}. However, human classifications provide only a very rough quantification, and even more troubling, humans have known biases \citep{Lintott2008,Land2008,hayes2016nature}, and possibly unknown ones, that can skew our view of galaxy structure.

\begin{figure*}
    \centering
    \includegraphics{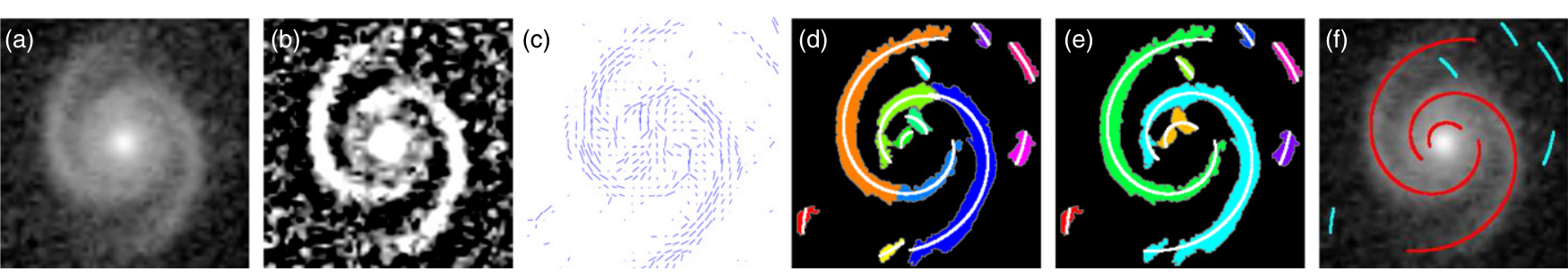}
    \caption{SpArcFiRe's steps in parsing a spiral galaxy image.
{\bf a)} Center and de-projected the image.
{\bf b)} Contrast-enhancement.
{\bf c)} Orientation field (at reduced resolution for display purposes).
{\bf d)} Initial arm segments found via Hierarchical Agglomerative Clustering (HAC) of nearby pixels with similar orientations and consistent logarithmic (log-spiral) spiral shape, overlaid with the associated log-spiral arcs fitted to these clusters.
{\bf e)} Final pixel clusters (and associated arcs) found by merging adjacent compatible arcs.
{\bf f)} Final arcs superimposed on image (a).  Red arcs wind S-wise, cyan arcs wind Z-wise.
}
\label{fig:SpArcFiRe} 
\end{figure*}

SpArcFiRe \citep{DavisHayesCVPR2012,DavisHayes2014,DarrenDavisThesis2014} is the latest and most comprehensive among recent attempts \citep{Odewahn2001,Au2006,Au2006a,Shamir2011,BenDavis2012} to quantify spiral structure in an automated fashion. It seems to agree with humans as well as can be expected for all the measures given by the Galaxy Zoo 1 catalog, and has played a hand in analyzing exactly how the Galaxy Zoo 1 human classifications have a selection bias (but not a disagreement bias) when selecting which direction the arms wind in a spiral galaxy \citep{hayes2016nature}.  SpArcFiRe's analysis of a spiral galaxy includes parameters that describe each individual spiral arm segment found in an image: these parameters include the pixels corresponding to each arm segment, its average length, width, location, pitch angle, and bounds (error estimates) in each of these.  (See Figure \ref{fig:SpArcFiRe}; \cite{DavisHayes2014,DarrenDavisThesis2014} provide more detail.)

\subsection{The Observation}
\begin{figure}
\centering
 \includegraphics[width=\linewidth]{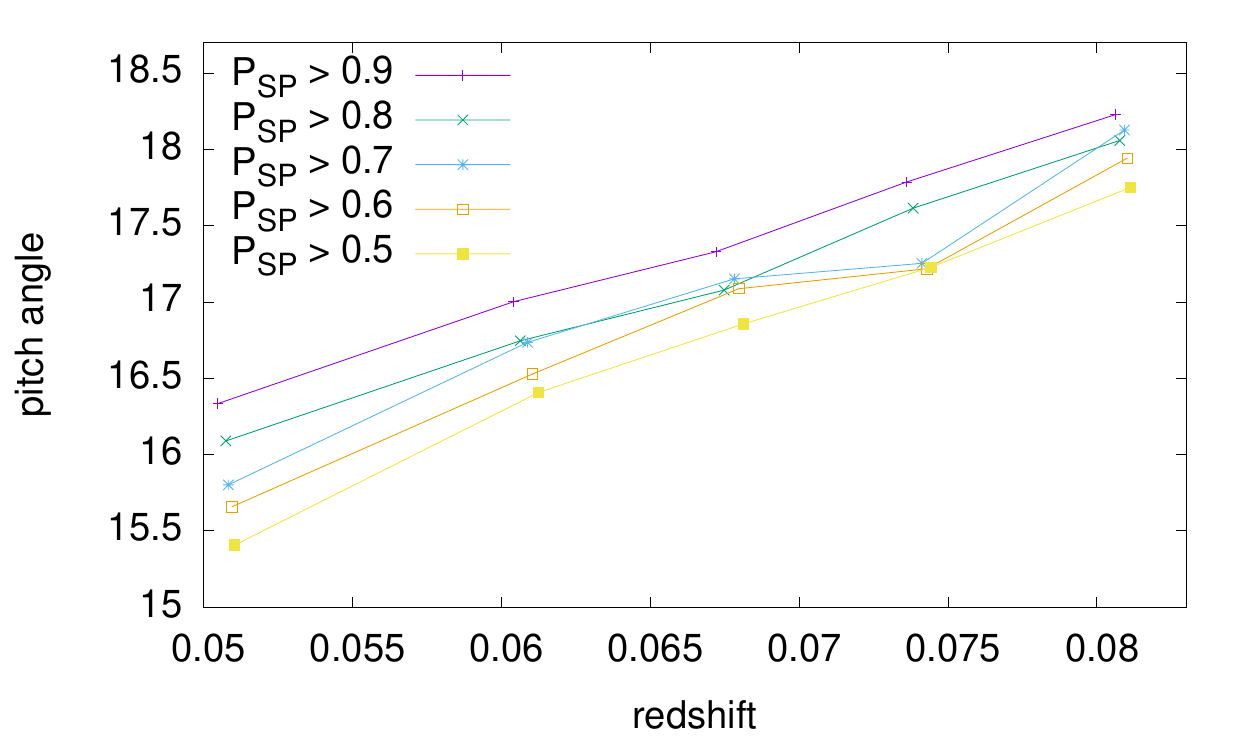}
 \caption{{\bf The observation that spurred this study.} Within a volume-limited sample accounting for the Malmquist bias ($z<0.085$, absolute magnitude brighter than $-22.25$ in the $r$ band) of SDSS spiral galaxies (GZ1 spirality $P_\mathrm{SP}\ge x$ for the values of $x$ displayed), SpArcFiRe observes their mean pitch angle to increase with redshift. Linear extrapolation predicts that this value would reach 90 degrees at about $z=1.2$, suggesting the observed increase is either spurious, or must become sublinear with redshift. The lines, for $P_\mathrm{SP}\ge\{0.9, 0.8, 0.7, 0.6, 0.5\}$ have a total of 2896, 3639, 4106, 4473, and 4843 galaxies, respectively. (Galaxies accumulate as $P_\mathrm{SP}$ decreases.) For each line, the set of galaxies were divided equally into 5 bins sorted by redshift. The mark on the curve then represents the point (mean redshift, mean pitch angle) across the set of galaxies in that bin. Comparing the 5 curves, note that the mean pitch angle increases slightly with increasing spirality, suggesting a possible selection effect: loosely winding arms (corresponding to larger pitch angles) are more visible than tightly-wound arms, leading to more such galaxies being included in the sample. The same effect (loosely winding arms being more visible) may cause a selection bias that increases with redshift due to image degradation. The purpose of this paper is to test that hypothesis.}
 \label{fig:pa-vs-z}
\end{figure}

We have run SpArcFiRe on the same set of SDSS \citep{SDSS,SDSS3} galaxies as did the Galaxy Zoo 1 survey, and intend to publish our output data soon. However, in our initial excitement, we looked for as many correlations as we could think of between our new analysis of SDSS galaxies, and existing data.  One of the most interesting we found was an apparent positive correlation between average pitch angle of spiral galaxy arms and redshift, even after carefully selecting a uniform volume-limited sample of galaxies to account for the Malmquist bias (Figure \ref{fig:pa-vs-z}). In our initial examination of the results, we realized that it might be a selection effect: tightly winding arms (ie., arms with a low pitch angle) may become less distinct and eventually disappear as a galaxy image degrades with distance.  Thus, at large distances, galaxies with loosely winding (ie., high pitch angle) arms may be selected more easily than tightly-winding (low pitch angle) galaxies, all other properties being equal.  At this point we realized that it would be wise to perform a detailed analysis of how SpArcFiRe responds to images of decreasing resolution and increasing noise.  That is the primary purpose of this paper.
%A second purpose is to show how SpArcFiRe's output depends upon its most important internal algorithmic parameters. These parameters determine how SpArcFiRe responds to noise in the image, and large-scale differences in spiral structure, such as grand design vs. flocculent galaxies. It is likely that the best parameters to use when analysing a grand design spiral would be different than the best parameters to use when analysing a flocculent galaxy. While, for now, we have analyzed the entire SDSS survey using one uniform set of internal SpArcFiRe parameters, future work may wish to focus on how to best adjust SpArcFiRe's internal parameters on a per-galaxy basis; the parameter analysis herein may help guide such work.

\section{Image degradation}
\subsection{``Clear'' set of spiral galaxy images}
We define the {\it spirality} of a galaxy image as the fraction of humans from GZ1 who voted that they saw either S-wise or Z-wise arms.\footnote{We emphasize that spirality is technically associated with an {\em image}, not with an object.  The spirality of an image of a spiral galaxy can decrease towards zero as the image becomes more degraded---which is precisely the effect we're trying to quantify in this paper. Spirality can also become zero if a disk galaxy it tilted so far as to be edge-on, because then there is no {\em visible} spiral structure.} More precisely, if {\tt P\_S} and {\tt P\_Z} constitute the fraction of people who voted for each\footnote{Called {\tt P\_CW} and {\tt P\_ACW} in the dataset but since ``clockwise'' is ambiguous, the terms S-wise and Z-wise have since been adopted.}, the {\it spirality} {\tt P\_SP=P\_S+P\_Z}.

We picked 7536 clear spiral galaxy images from SDSS for our experiment. The precise selection came from the union of the following two sets: (a) a complete volume-limited sample of galaxies out to $z=0.085$ with magnitude in the SDSS $r$-band brighter than $-22.25$ and spirality greater than 0.7 and a Petrosian 90\% radius greater than 6 arc-seconds (15 pixels), which we subjectively determined to be the smallest galaxy in which arms could be seen; this set comprises 4106 spirals, and is exactly the same set depicted on the $P_\mathrm{SP}>0.7$ curve in Figure \ref{fig:pa-vs-z}.  The second set is (b) any galaxy with a GZ1 spirality greater than 0.9, which added another 3435 galaxies to our set.\footnote{In hindsight perhaps we should have chosen 0.7 for the lower spirality cutoff for both sets. However, considering our blurred set contains almost a million images, we think the sample size is big enough.}
The goal now is to degrade these images artificially and observe the effect.

\subsection{Image degradation using Sunpy}
Sunpy ~\citep{Sunpy} is a tool primarily used to generate artificial images from the Illustris simulation \citep{nelson2015illustris}; since images from a simulation can be generated with almost arbitrary clarity, Sunpy was used to produce degraded images of simulated galaxies by mimicking the degradation that occurs to real images of galaxies. We add a point spread function (full width half maximum or FWHM(PSF)), redefine the pixel size, and add noise to an image, in that order. In the latter case, SunPy assumes an input image with ostensibly zero noise. Then, to add noise to arrive at a S/N of $K$, it adds up all the signal $S$ across the entire image, and then adds Gaussian pixel noise of total value $KS$.  Here we have used Sunpy to artificially degrade a set of high quality SDSS images of real galaxies; these do not have zero noise to start with, so our degraded images will have S/N slightly lower than specified.

Starting with our clear set of 7536 SDSS galaxies, we degrade them with the following parameters: we vary the FWHM(PSF) from 4" (clear) to 128" (very blurry) in geometric steps of $2^{1/4}$ arc seconds; and we vary the S/N from 256 (clear) down to 8 (very noisy) in geometric steps of 2. We set the pixels-per-arcsecond ratio to a constant of 1/3.2 of the FWHM(PSF), corresponding roughly to SDSS that has an average FWHM(PSF) of about 1.3 arc seconds and a pixel size of 0.4 arc seconds. Throughout the paper the term ``PSF'' refers to FWHM(PSF).

\subsection{Blurring Pipeline}
We ran our blurring program on all 7536 galaxy images across the above set of FWHM(PSF) and S/N values, generating 949,536 blurred images. Running SpArcFiRe on these images produced 622,585 galaxies with usable output from SpArcFiRe; the remaining 326,951 images were so badly degraded that SpArcFiRe failed to find anything in the image, either because it could not isolate the galaxy at all, or because it could not find any spiral arcs.

\begin{figure}
\centering
 \includegraphics[width=\linewidth]{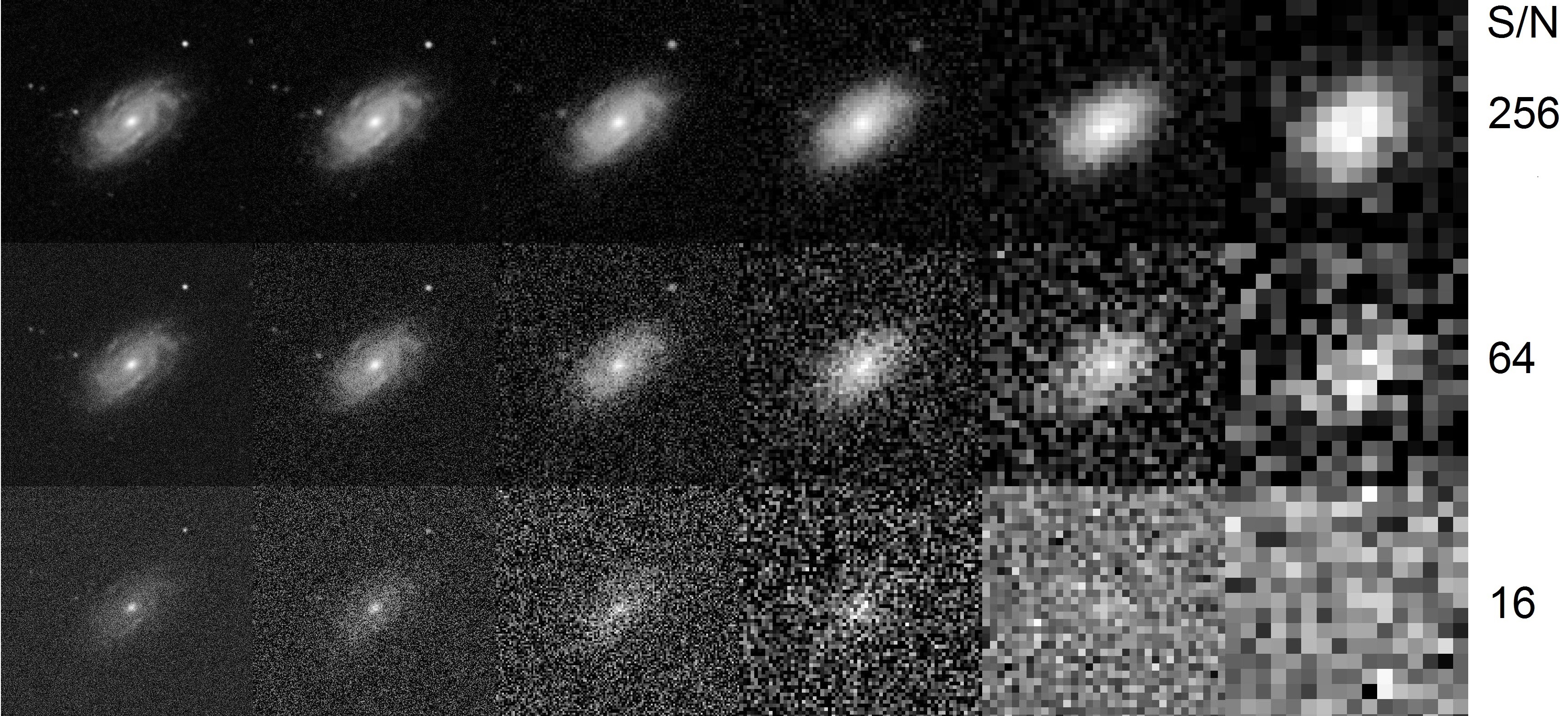}
 \includegraphics[width=\linewidth]{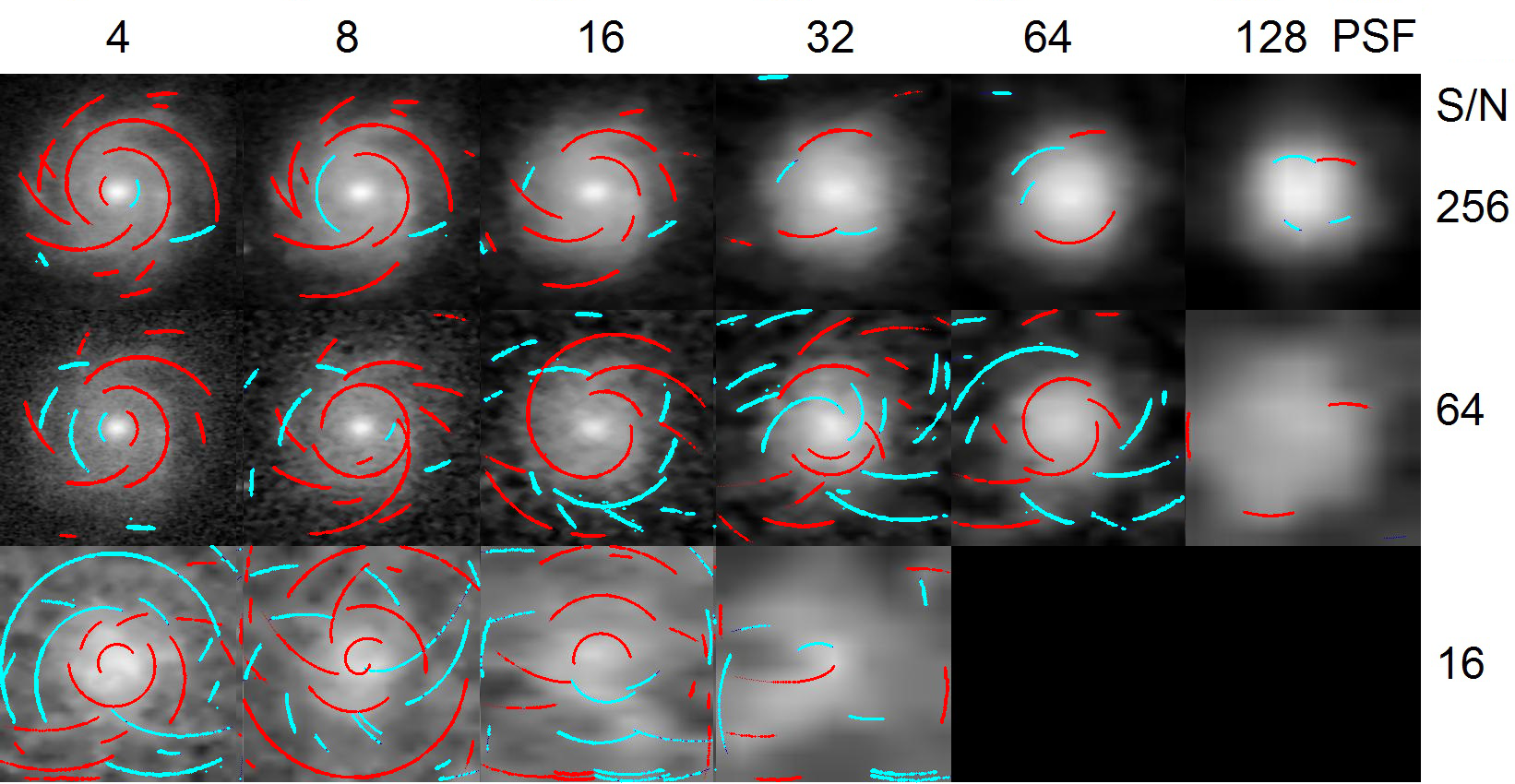}
 \caption{{\bf Top}: FWHM(PSF) blurring and noise added to SDSS galaxy 1237648702972625038 using Sunpy. {\bf Bottom}: The corresponding output images generated by SpArcFiRe, which have been cropped and de-projected so the disk appears face-on.  The two black squares indicate that SpArcFiRe failed to find an object in the image. Observe that at least the global chirality is correct up to FWHM(PSF) 16 for S/N as low as 64, but by S/N 16 the output arcs are mostly noise---unsurprising because the arms seem to be invisible to the human eye in the input images as well.}
 \label{fig:sp_result}
\end{figure}

Figure ~\ref{fig:sp_result} depicts an example galaxy (SDSS DR12 19-digit ID 1237648702972625038) and its degraded images. The top half of Figure \ref{fig:sp_result} depicts the original unblurred image at the far top-left, along with several samples of the degraded image: moving to the right increases the FWHM(PSF) (note the pixelation observable to the right since we set the pixels-per-FWHM(PSF) to the constant 3.2), and moving down decreases (degrades) the S/N ratio. 
The bottom half of Figure ~\ref{fig:sp_result} is the corresponding chart of the spiral arcs found by SpArcFiRe. The bottom right has 2 dark squares because the images are so badly degraded that SpArcFiRe failed to produce any output (not surprising, considering the input images they came from).

\section{Results}
\subsection{Pitch angle {\it vs.} image degradation: raw data}
In this section we study the ``raw'' effect of image degradation on mean pitch angle.  We first take the unblurred images of our 7536 ``clear'' spiral galaxies, and compute a galaxy-level pitch angle.  As explained in \cite{DavisHayes2014}, these galaxy-level pitch angles are computed from an arc-length weighted mean pitch angle across the spiral arcs discovered by SpArcFiRe. We define two different types of means: one of them includes {\em all} arcs found, including sign, even though some of the arcs are almost certainly noise; we simply call this one the mean pitch angle. The second type includes only those arcs that agree with the {\it dominant chirality}, which is defined as the winding direction with the longest total length of all arcs of one sign. This dominant chirality has been shown to agree very well with the GZ1 human majority vote on which direction the whole galaxy ``winds'', to the point that SpArcFiRe is actually a more reliable indicator of winding direction than the average human.\footnote{Galaxy Zoo ranks human voters based upon how often they agree with the majority; SpArcFiRe ranks in the upper quartile of this quality measure, making it more reliable than at least 75\% of humans for determining winding direction.} Pitch angles computed using only arcs that agree with the dominant chirality are labelled {\bf DCO} (Dominant Chirality Only), and we believe it is a more reliable indicator of global pitch angle, since arcs whose sign disagree with the dominant chirality  are much more likely to be noise arcs than real arms \citep{DavisHayes2014}. Obviously, since the non-DCO mean pitch angle includes arcs of opposite sign, their mean will be closer to zero (ie., smaller in magnitude) than the DCO mean pitch angles.  So for example in Figure \ref{fig:sp_result}, the red (S-wise) arcs are clearly the dominant chirality in the most clear images, and most of the cyan arcs are noise in those images; the absolute value of the mean pitch angle including sign will obviously be less than the absolute value of the DCO mean pitch angle, since in the former case there will be cancellation.

\begin{figure}
 \includegraphics[width=\columnwidth]{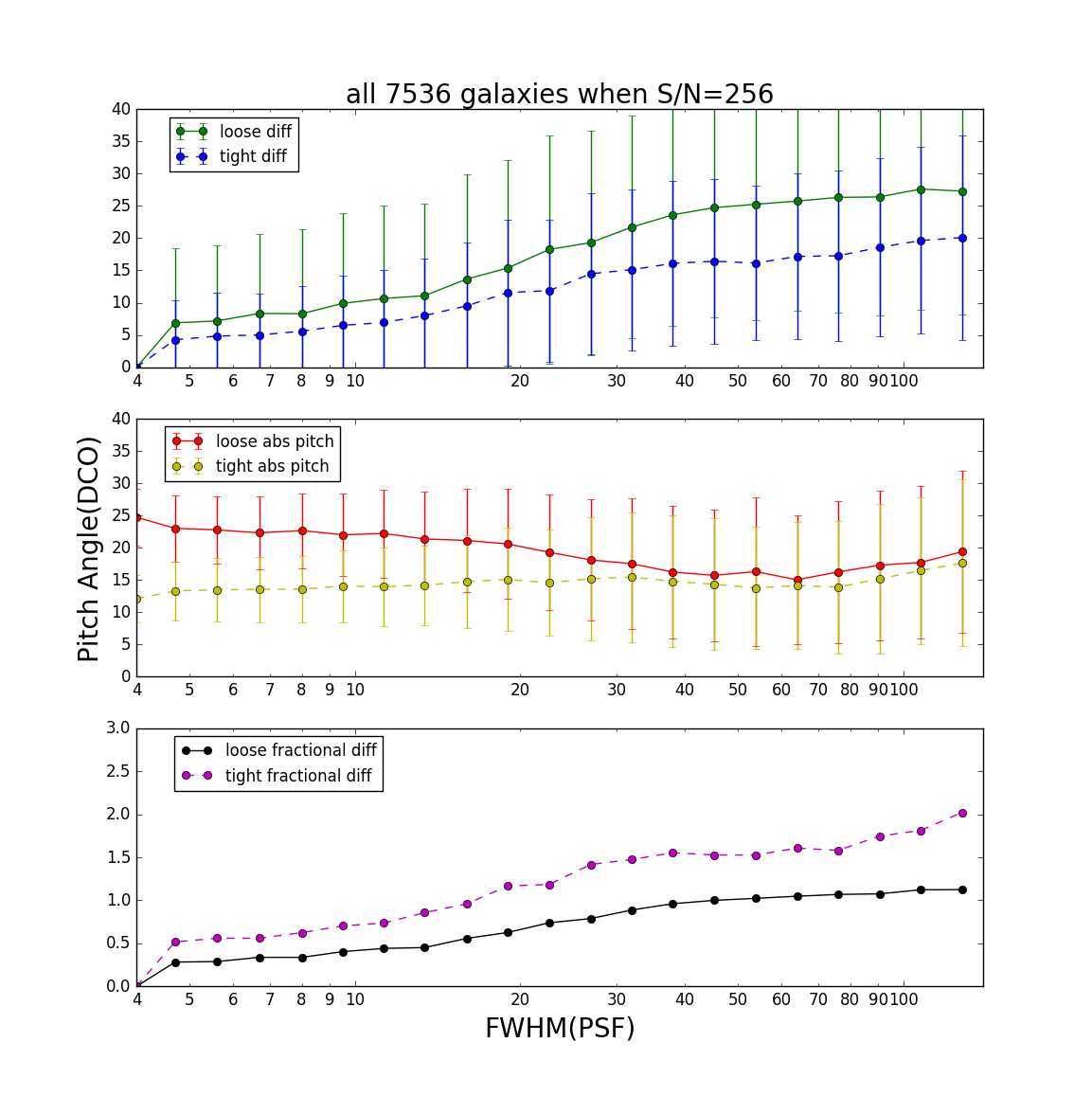}
 \caption{Various types of changes in the pitch angle as a function of FWHM(PSF), comparing galaxies in the upper (``loose'') and lower (``tight'') quartiles of pitch angle measured on the unblurred images. In the {\bf re-order FIRST?} vertical plot, the solid red (loose) and dashed yellow (tight) curves show the mean pitch angles of the two groups as a function of FWHM(PSF); clearly the red (loose) arms have a higher pitch angle at the low end of FWHM(PSF); interestingly, the two averages meet at around FWHM(PSF) 40, meaning that effectively the difference between the two has become invisible. In the {\bf re-order SECOND?} vertical plot, the solid green (loose) and dashed blue (tight) curves demonstrate how the error in absolute value of pitch angle increases with FWHM(PSF), as expected, and that the error in the loosely winding arms increases more rapidly.  However, in the third vertical plot, the solid black and dashed purple lines demonstrate that {\em as a fraction}, the error of the tightly wound arms increases more rapidly. All of these measures are averages that discard arcs that wind in the non-dominant direction. (``DCO'' on the vertical axis means ``dominant chirality only''.) The S/N ratio is held constant at 256 (the clearest S/N) throughout. Error bars in all cases are 1 sigma.}
 \label{fig:PSFChiralityOnly}
  \includegraphics[width=\columnwidth]{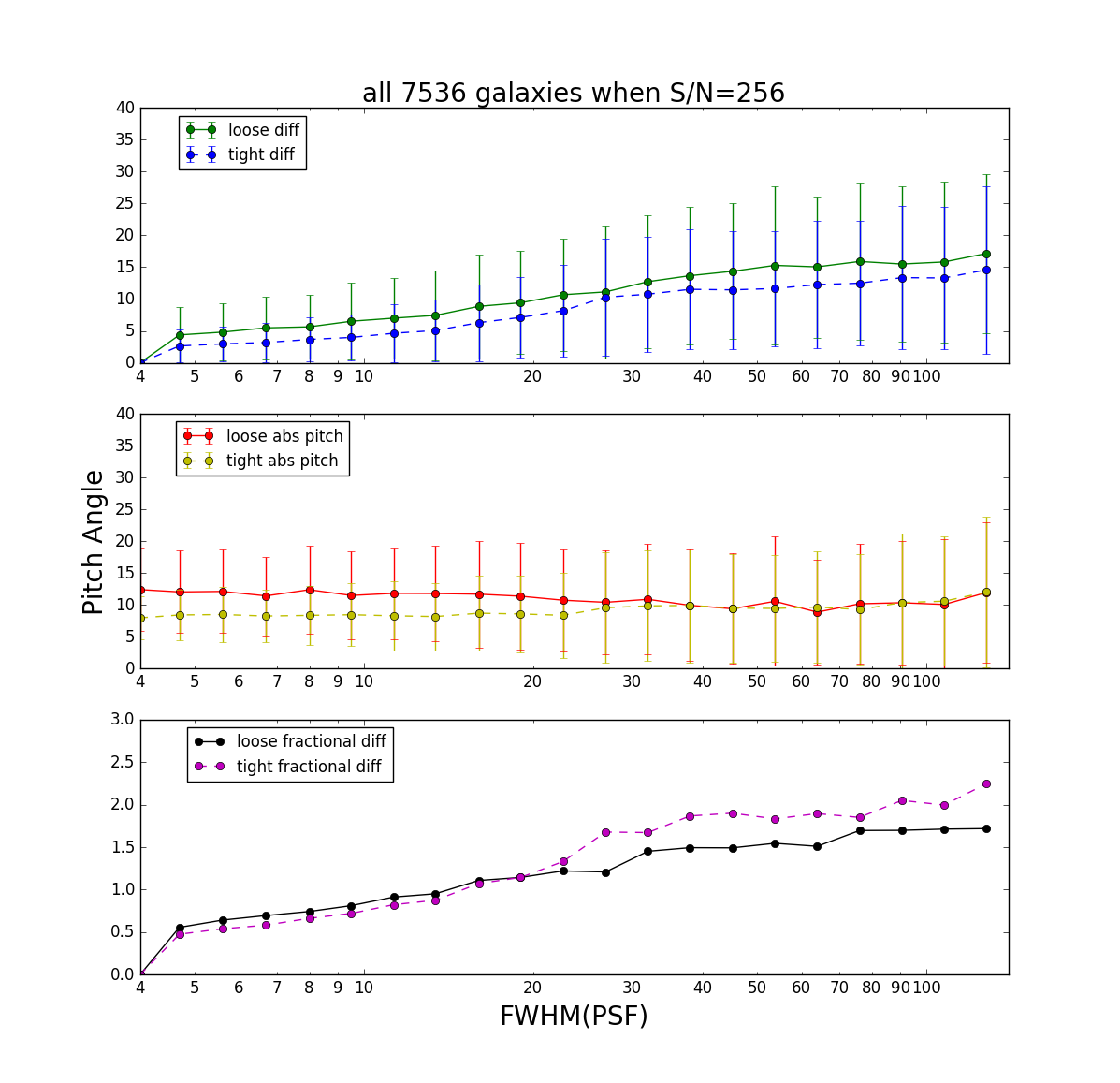}
 \caption{Exactly the same description as above, except the means now include arcs that wind in the non-dominant direction (which are often but not always noise). Most of the above observations still hold qualitatively.}
 \label{fig:PSF}
\end{figure}

\begin{figure}
 \includegraphics[width=\columnwidth]{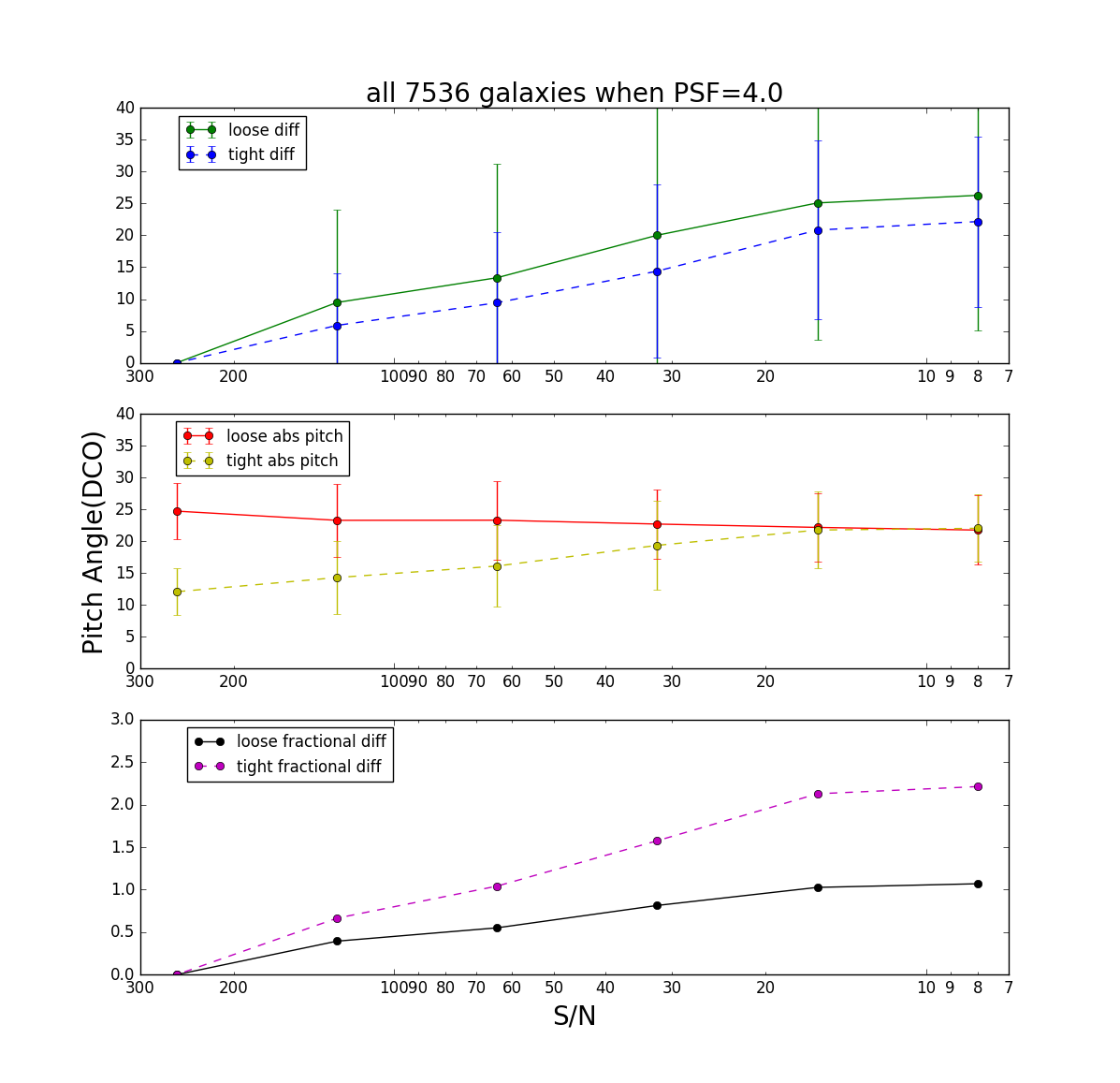}
 \caption{Similar curve descriptions as Figure \ref{fig:PSFChiralityOnly} except now the FWHM(PSF) is held constant and the S/N is changed; highest S/N are on the left, with images degrading towards the right.  The tight and loose pitch angles become indistinguishable at about S/N=16.  Note the change in tight pitch angles is quite a bit more strongly affected by noise than it was in the blurring case; we are not sure why. As with Figure \ref{fig:PSFChiralityOnly}, arcs of the wrong winding direction are excluded.}
 \label{fig:SNChiralityOnly}
  \includegraphics[width=\columnwidth]{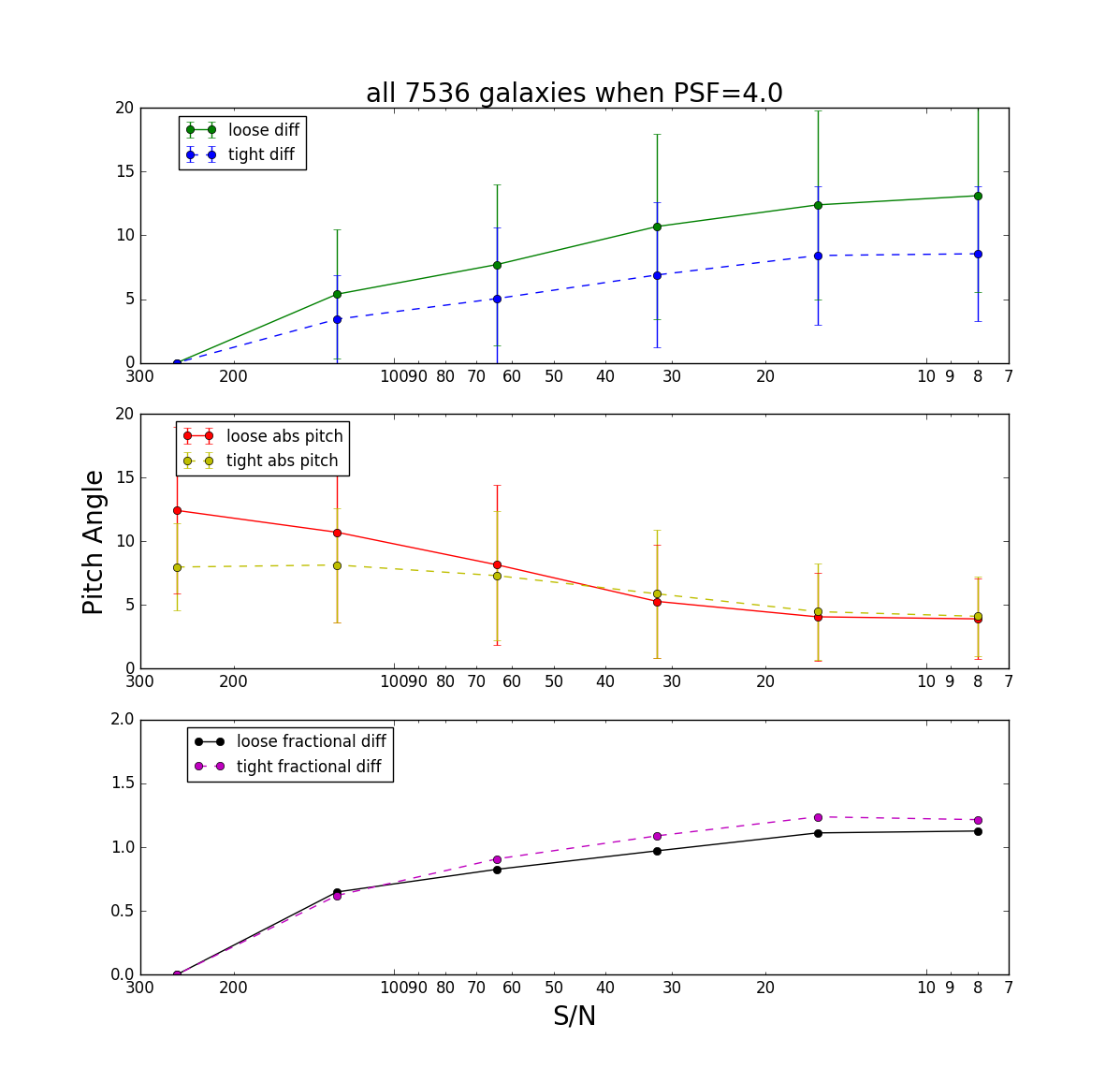}
 \caption{Similar to Figure \ref{fig:SNChiralityOnly} but including arcs of the wrong winding direction.}
 \label{fig:SN}
\end{figure}

\begin{figure}
 \includegraphics[width=\columnwidth]{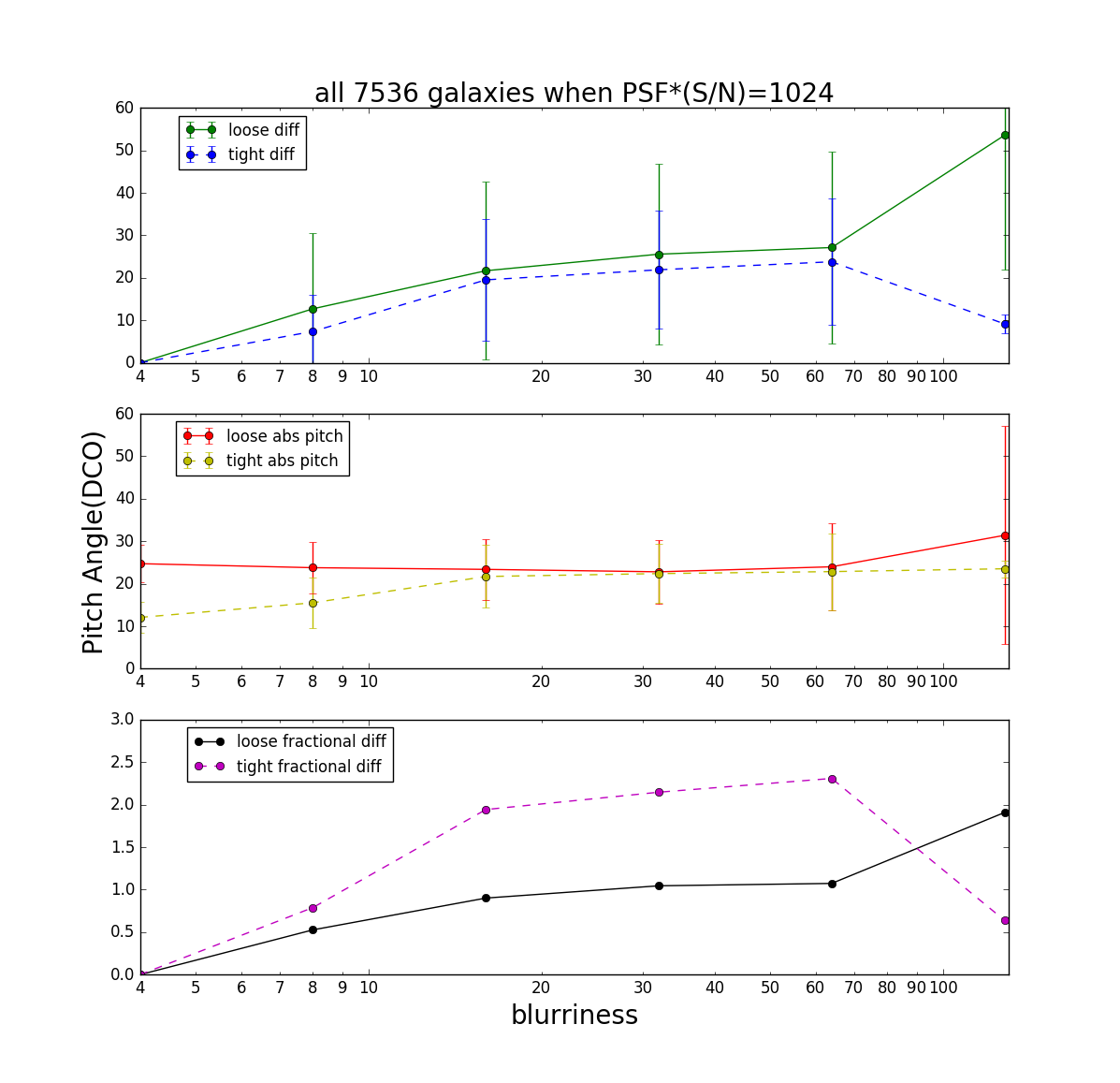}
 \caption{An attempt to account for degrading both S/N and blurring in one plot: on the left we have FWHM(PSF)=4 and S/N=256. As we move to the right, we increase FWHM(PSF) and decrease S/N simultaneously to maintain their product at 1024. As may be expected, the loose (solid red) and tight (dashed yellow) meet each other earlier, at a FWHM(PSF) of 16 (S/N=64), which is earlier in both measures than occurred in either individually. Excludes arcs of the wrong chirality.}
 \label{fig:BRChiralityOnly}
  \includegraphics[width=\columnwidth]{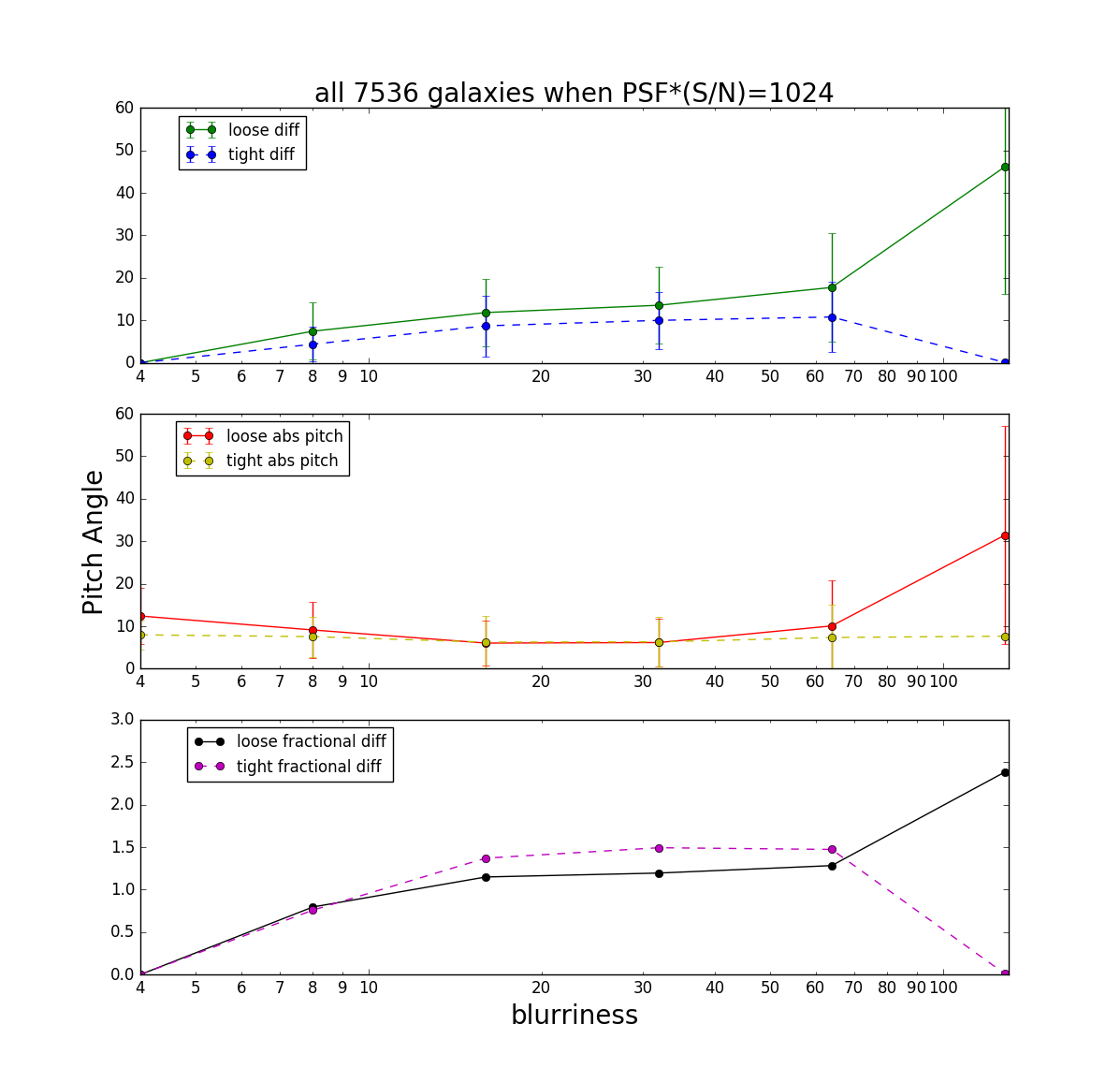}
 \caption{As above, but now including arcs of the wrong chirality.  Now the loose and tight arcs become indistinguishable even earlier, around FWHM(PSF) 8 or perhaps slightly higher (S/N $\approx$ 128).}
 \label{fig:BR}
\end{figure}

In order to determine the effect of image degradation on pitch angle, we first sort the clear (unblurred) images of our 7536 galaxies according to absolute value of their galaxy-level pitch angle. We refer to the upper quartile of this list as the ``loosely winding'' (high pitch angle) set, and the lower quartile as the ``tightly winding'' (low pitch angle) set.
%Figures \ref{fig:PSFChiralityOnly}--\ref{fig:BR} then show how these galaxies (which remain a fixed set throughout all the analysis, based on their unblurred pitch angles) respond to image degradation, by plotting 6 curves as a function of various degradation parameters: (i,ii) the mean pitch angle of these upper and lower quartile galaxies; (iii,iv) the {\em difference} between the mean pitch angle and the unblurred pitch angle; and (v,vi) the percentage difference.
For all plots that we will show in Figures \ref{fig:PSFChiralityOnly}--\ref{fig:BR}, we plotted three different values for both tightly and loosely winding galaxies. The first value is the absolute value of the mean pitch angle across the group. The second value is the absolute value of the difference between a galaxy's unblurred pitch angle and the blurred pitch angle. The third value is the {\em fractional} difference between the unblurred and blurred pitch angle. In each case we provide results both for the DCO and non-DCO cases.

% fig:Spirality_all_noise}
%  \includegraphics[width=\columnwidth]{histogram_all_noise}
%  \caption{The vertical axis of this plot is  number of galaxies that have bigger than a 0.7 spirality. And the horizontal axis of this plot is PSF. we also used different colors to represent S/N values.}
%  \label{fig:histogram_all_noise}
%  \includegraphics[width=\columnwidth]{Spirality_bigger_07_all_noise}
%  \caption{The vertical axis of this plot is the mean pitch angle that have bigger than a 0.7 spirality. And the horizontal axis of this plot is PSF. we also used different colors to represent S/N values.}
%  \label{fig:Spirality_bigger_07_all_noise}

Figures ~\ref{fig:PSFChiralityOnly} and ~\ref{fig:PSF} show the above values as a function of FWHM(PSF), for DCO and non-DCO pitch angles, respectively. Figure ~\ref{fig:PSFChiralityOnly} shows how the DCO pitch angle changes as we increase the FWHM(PSF) with a constant S/N of 256 (the best S/N value). The red and yellow lines depict absolute pitch angles of loosely wound and tightly wound galaxies, respectively; the pitch angles of these two groups of galaxies become indistinguishable at around FWHM(PSF)=40. The green and the blue lines depict the (absolute value of) difference between the blurred and unblurred pitch angles for loose and tight, respectively; the curve depicting loosely wound galaxies is always above that for tightly wound, meaning, presumably in proportion to the fact that loosely wound pitch angles are larger than tightly wound ones. Finally, we also plot the {\em fractional} differences which are the black and the purple dashed lines, as the purple (tightly wound) and black (loosely wound) lines. We see that, as a fraction, the tightly wound galaxies are more strongly affected by blurring than the loosely wound ones.

Figure  ~\ref{fig:PSF} is similar to Figure  ~\ref{fig:PSFChiralityOnly}, but including all arcs (not just DCO ones). The qualtitative observations are similar to those of Figure \ref{fig:PSFChiralityOnly}, thought not as stark since there is more noise to start with in the non-DCO analysis of pitch angle.

Figures ~\ref{fig:SNChiralityOnly} and ~\ref{fig:SN} show the average pitch angles (DCO and non-DCO, respectively) as a function of signal to noise ratio, with the FWHM(PSF) held constant at 4 (the best FWHM(PSF)). Again, most of the qualitative observations of ~\ref{fig:PSFChiralityOnly} and ~\ref{fig:PSF} hold, and the difference in pitch angle between the tightly (red) and loosely (yellow) wound arms disappears at about a S/N of 16, just as it disappears at FWHM(PSF) 40 in the previous plots; and again, the tightly wound (purple) arms suffer a larger percentage perturbation than the loosely wound (black) arms. Also, the non-DCO pitch angles (Figure \ref{fig:SN}) are more adversely affected than the DCO (Figure \ref{fig:SNChiralityOnly}) ones.

In Figures ~\ref{fig:BRChiralityOnly} and ~\ref{fig:BR}, we combine the effects of FWHM(PSF) and S/N together by setting the product of the two to 1024. Thus, as the FWHM(PSF) increases, the S/N decreases in tandem to keep the product at 1024. We can see that when both types of degradation are applied together, the resulting pitch angle measures degrade far more quickly, with the difference between the two sets vanishing at about the point where the FWHM(PSF) is about 16 (S/N of 64) for the DCO arcs, and even earlier for non-DCO arcs---about FWHM(PSF) 8, S/N 128.

\subsection{The spirality selection effect}
As we have seen, a galaxy's measured pitch angle will change as the image degrades, and the mean measured pitch angle in a group of selected galaxies will change correspondingly.  Note, however, that it only makes sense to measure the pitch angle of a spiral galaxy; elliptical galaxies don't have arms. Recall that to make the observation depicted in Figure \ref{fig:pa-vs-z}, we selected galaxies for which some fraction of Galaxy Zoo Humans said that they saw spiral structure; this fraction determines the {\it spirality} of the galaxy's image, and putting a threshold on the spirality gives us a selection of galaxies for which we have some level of certainty of them having visible spiral structure.
In order to duplicate the observation of Figure \ref{fig:pa-vs-z} on a set of artificially blurred galaxies, we need a way to estimate how humans would have voted on their spiralities. For this we turn to our machine learning expertise.  We have previously described a machine learning algorithm that demonstrated how we could eliminate a human-created bias in a winding direction survey \citep{Hayes:2017hb}.  Here we extend that algorithm and make it more accurate and robust.
The main difference here is which attributes our random forest\footnote{A random forest is a set of decision trees; each tree is an ``expert'' at classification using a random set of attributes. Together these ``experts'' form a forest which has better performance than one decision tree using all attributes. See \cite{SilvaCaoHayes2017} for more details.} is allowed to use as inputs.  Since we ran SpArcFiRe on all the blurred images, we are allowing the machine to use most of SpArcFiRe's outputs.  We also allow it to use colors and magnitudes from SDSS, because we assume those would not change when the image becomes blurred or noisy.  However, some parameters that come from SDSS may vary with the noise added by blurring the objects.  A more sophisticated analysis would be required to decide if, for example, Petrosian radii vary with noise, but that is outside of the scope of this paper. In order to avoid bias introduced by this potential issue we decided not to use such attributes. We ended up with 105 features per object, and we only used the following classes of variables (the ones that we believe are unaffected by blurriness) from SDSS: absolute colors, de-reddened magnitudes in the $g$, $i$ and $r$ bands, and $k$-correction for $z=0$ in all bands. 

We used a random forest with 35 trees, with a 95-5 split for training and testing. 
As a measure of how good our spirality prediction is, we compute the root mean squared error (RMSE) between the predicted and real spirality for the 5\% of the set called the test set, after training on 95\% of the initial set.
Our final RMSE was 0.14, slightly higher than the ones reported in \cite{Hayes:2017hb}, which is expected since we are using fewer features.

In \cite{Hayes:2017hb}, the RMSE in predicting spirality was about 0.137 across all spiralities, but more detailed analysis has shown that the RMSE in predicting spiralities was heavily skewed: near spirality zero (ie., for elliptical galaxies), our machine was extremely accurate, having RMSE in the 0.02 range; but at the high end of spirality (above about 0.7), the RMSE was much larger, in the 0.30 range.  This is precisely the opposite of what we want, because we want to be able to precisely pinpoint which galaxy images indeed show spiral structure, not which ones do not. Further study into the issue revealed that the problem was that there are somewhere between 2x and 6x more elliptical galaxies than face-on spiral ones in the Galaxy Zoo sample\footnote{There are 2x as many ellipticals as spirals if you insist on 90\% certainty in both; if you only insist on 50\% majority then the number is closer to 6x.}, which simply made our machine train more intensively to reduce its error on that most voluminous sample of galaxies (the ellipticals), at the expense of making larger errors on the much smaller sample of spirals.\footnote{Note that the issue is {\em not} that there are not enough spirals to train on. Tens of thousands of spirals is plenty. The problem was just that the machine ``tried harder'' to reduce the error of prediction for the largest sub-sample it could find, which was the ellipticals.}  The solution is simple: we reduced the number of ellipticals that we trained on by a factor of 16, thereby reversing the trend: there are now 2-6 times as many spirals (a few tens of thousands) and only a few thousand ellipticals to train on, so now the machine is very good at recognizing galaxies with high spirality (say, above 0.7) and less good at predicting the spirality of galaxies at the low-spirality end.  This is fine, because we really don't care how bad the machine is at predicting spirality of any galaxy with spirality less than 0.5, so long as it doesn't say the galaxy {\em is} a spiral galaxy. This we have achieved: our RMSE is now about 0.15 across the entire spectrum of spiralities, which has reduced our RMSE at the high end by a factor of 2.\footnote{As we have mentioned in \cite{Hayes:2017hb}, this RMSE is worse than the Kaggle winner's 0.07, but the Kaggle winner \citep{GZ2Kaggle2015} made no attempt to detect or reduce human-induced biases.}

With this new machine, we can now estimate spiralities for our set of 622,585 degraded galaxy images.  The results we report below are of an analysis made on the blurred objects only, which were not part of the training nor test set.

\begin{figure}
 \includegraphics[width=\columnwidth]{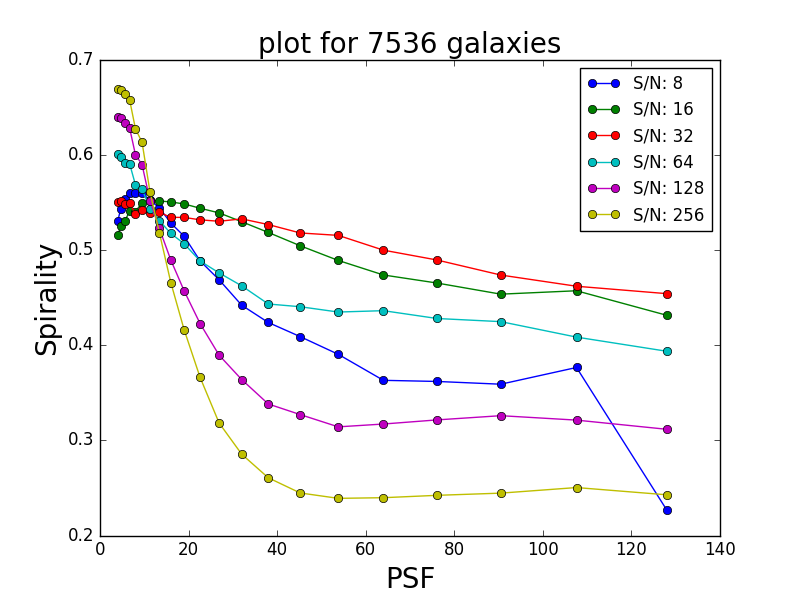}
 \caption{Mean spirality decreases with FWHM(PSF), although the interaction with S/N seems more complex.}
 \label{fig:Spirality_all_noise}
 \includegraphics[width=\columnwidth]{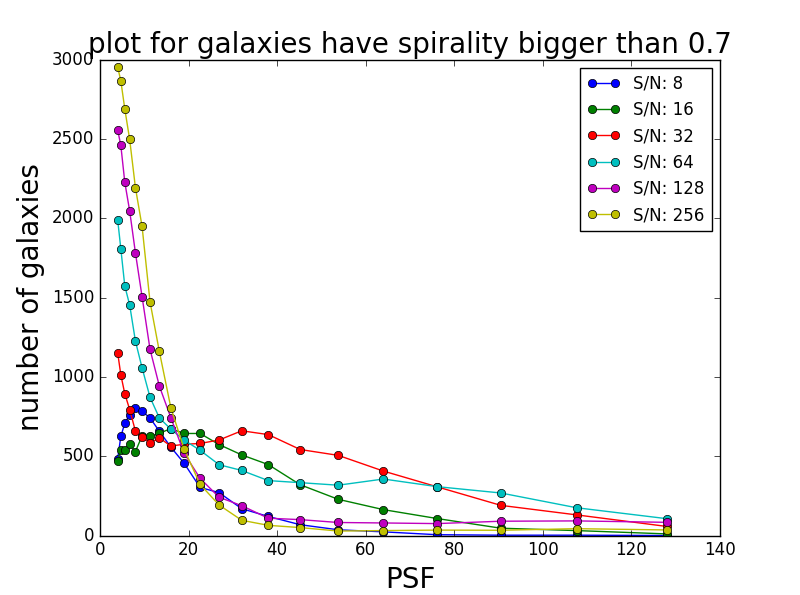}
 \caption{Heading towards explaining the selection effect in SDSS, we impose a cutoff of 0.7 in spirality, and count how many galaxies in our sample have spirality above 0.7 after being degraded in both FWHM(PSF) and S/N.  We see the number of galaxies above the 0.7 spirality cutoff decreases rapidly with increasing blurring and decreased S/N.}
 \label{fig:histogram_all_noise}
 \includegraphics[width=\columnwidth]{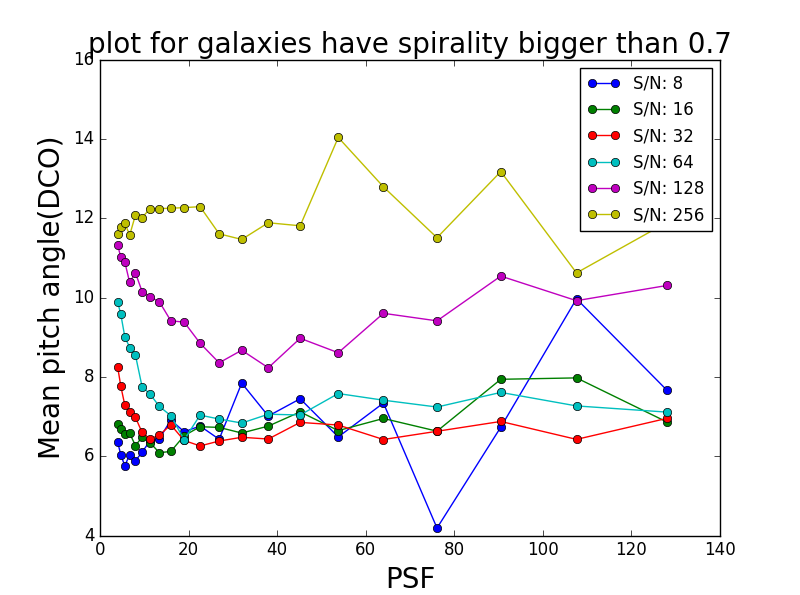}
 \caption{The vertical axis of this plot is the mean pitch angle that have bigger than a 0.7 spirality. And the horizontal axis of this plot is FWHM(PSF). we also used different colors to represent S/N values.}
 \label{fig:Spirality_bigger_07_all_noise}
\end{figure}

Figure ~\ref{fig:Spirality_all_noise} shows spirality as a function of FWHM(PSF) and S/N. We can see that the spiralities of galaxies almost uniformly become smaller as the FWHM(PSF) increases, although the interaction with S/N is more complex. Note that in the high S/N case (256), the spirality drops very quickly with increasing FWHM(PSF): this is because, in the absence of also adding noise, the images become smooth blobs which effectively look elliptical, so the machine correctly labels these images as having low {\em observable} spiral structure. As the S/N decreases, the added noise is sometimes interpreted as arcs, and our machine (which looks for arcs) mistakes these for spiral structure; this is why, as we decrease the S/N, the spirality decreases more slowly with increasing FWHM(PSF). This effect is greatest around S/N 16 or 32, but once the S/N drops to 8, the spirality again drops sharply with increasing FWHM(PSF) as the machine starts to recognize the images as nothing but blurry noise.

Figure ~\ref{fig:histogram_all_noise} depicts the number of artificially degraded galaxy images that have a spirality greater than a threshold of 0.7, as a function of FWHM(PSF) and S/N. It shows that as the blurriness gets bigger, there are fewer images that meet the threshold of 0.7 in spirality. 

Figure ~\ref{fig:Spirality_bigger_07_all_noise} depicts the mean pitch angle (DCO) of galaxies meeting the 0.7 threshold in spirality, as a function of FWHM(PSF) and S/N.  It is fairly clear that the mean pitch angle decreases with decreasing S/N, but the relationship between pitch angle and FWHM(PSF) is less clear in this group; except for the highest S/N case, there is a sharp drop in mean pitch angle as the FWHM(PSF) grows at the small end, but then the pitch angle slowly increases again once the FWHM(PSF) is above about 20. We are not sure why this is; however, as the next section shows, all of this data comes together nicely when we set the same selection criteria for real and artificially blurred galaxies.

\subsection{Comparing real {\it vs.} blurred images}

\begin{figure}
 \includegraphics[width=\columnwidth]{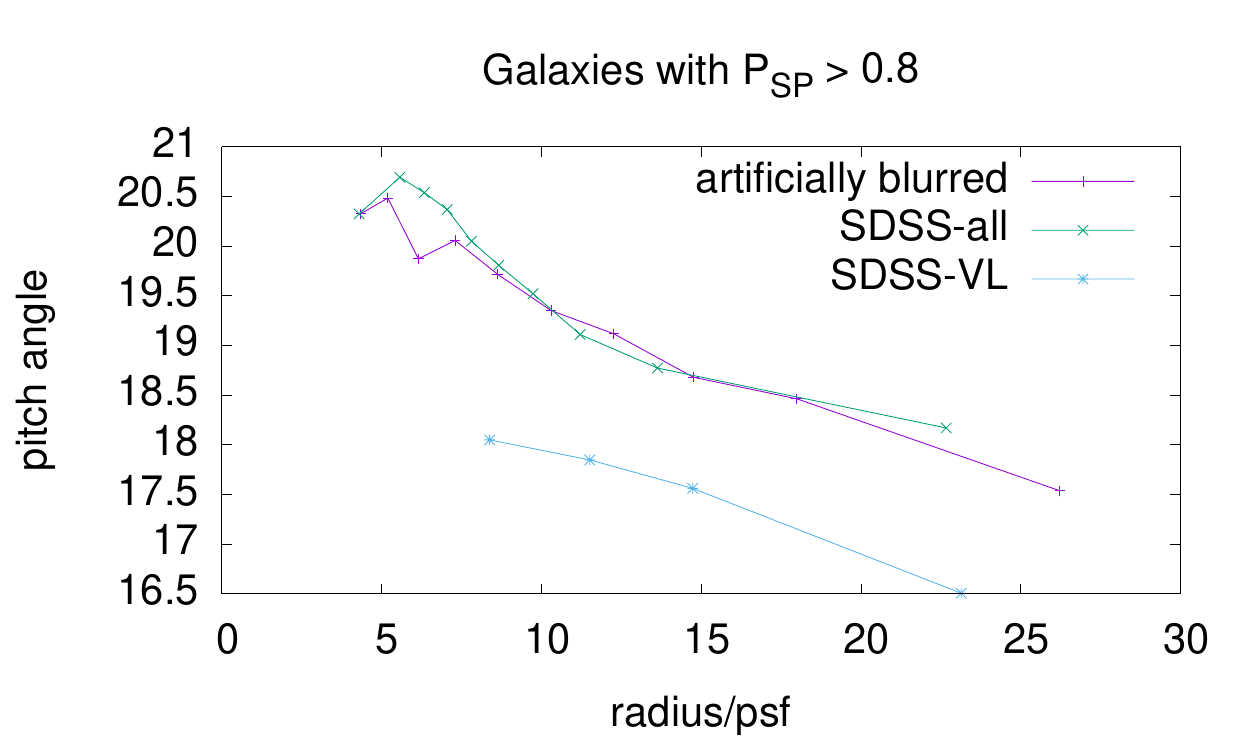}
 \caption{The line SDSS-VL is the exact same set of (3622 volume limited) galaxies with $P_\mathrm{SP}>0.8$ as Figure \ref{fig:pa-vs-z}, but plotted against radius/FWHM(PSF); the SDSS-all curve is all (36,384) SDSS galaxies for which PS>0.8, which will include some dimmer, closer galaxies than the volume limited sample; and the green curve are (24,347) images from our 7536 nearby galaxies that have been artificially blurred, chosen using similar criteria to the ``SDSS-all'' sample.  As can be seen, the artificially blurred sample does a very good job of matching the ``SDSS-all'' sample, corroborating our selection effect hypothesis.}
 \label{fig:real-fake-0.8}
 \end{figure}

\begin{figure}
 \includegraphics[width=\columnwidth]{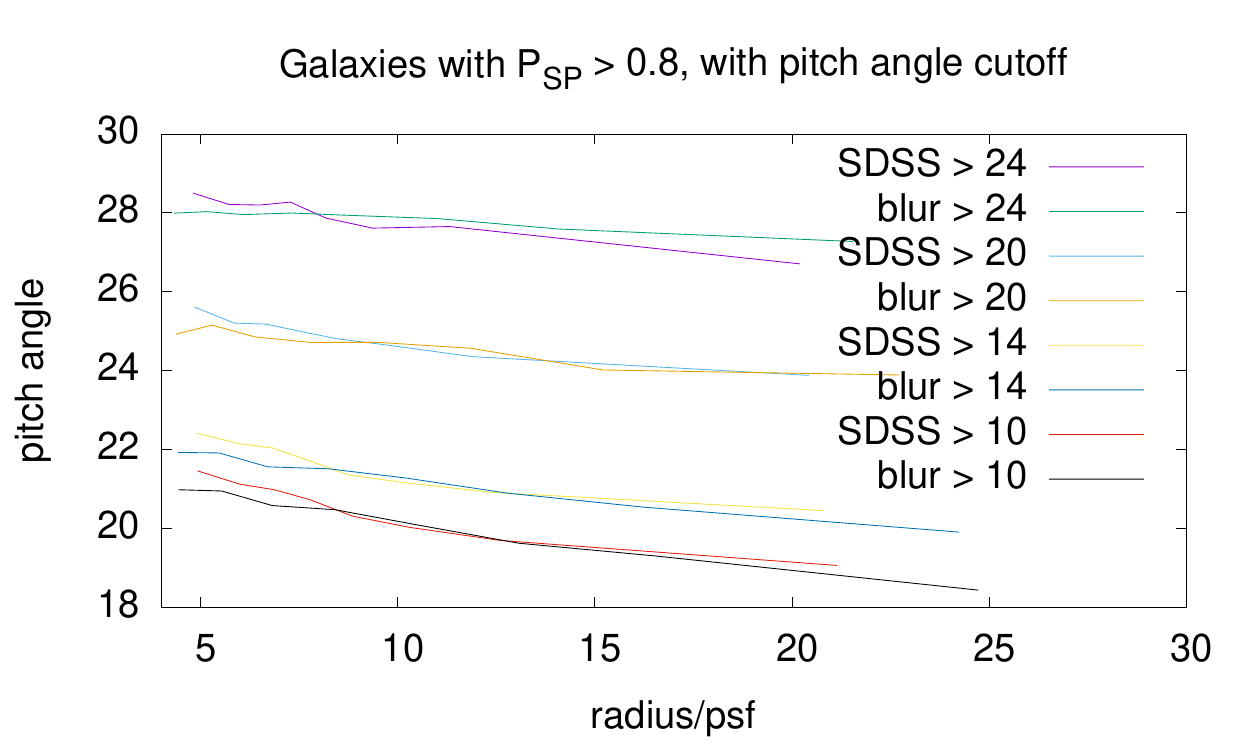}
 \caption{Another comparison of measured pitch angles between real SDSS galaxies vs. artificially blurred ones. Here, all galaxies have a spirality $P_\mathrm{SP}>0.8$, and we have attempted to account for the selection effect by imposing a lower limit on pitch angle, to remove galaxies with too-low pitch angles from the sample. As we can see, with equal selection limits, the real and artificially blurred galaxies have a statistically identical mean pitch angles for the 4 values of pitch angle thresholds 10, 14, 20, and 24 degrees.  This again suggests we are correctly modelling how observed pitch angle changes with image degradation.  However, the slope of all the lines are still negative, suggesting that none of the thresholds we have chosen are strong enough, although the slopes become less negative with increasing threshold. Above a threshold of 24 degrees, the sample sizes become too small to be meaningful.}
 \label{fig:pa-with-cutoff}
 \end{figure}

%\begin{figure}
%    \centering
%    \includegraphics[width=\linewidth]{plot_radius.pdf}
%    \caption{
%Plot of pitch angle {\it vs.} diameter of the galaxy in units of the PSF FWHM (SDSS r band), for galaxies in SDSS that have GZ1 measured spirality > $x=0.5, 0.6, 0.7, 0.8, 0.9$ (respectively 74680, 60750, 47880, 36260, and 22300 galaxies), split into 10 bins with an equal number of galaxies (exactly 1/10th of each of the numbers above) per bin.  Note the horizontal axis is reversed compared to previous plots (see text for why). Towards the right are highly resolved galaxies, which seem to have a mean pitch angle of about 9-11 degrees; towards the left are less resolved galaxies. As we can see, as the galaxy becomes less resolved, the mean pitch angle rises sharply.  We believe this is a selection effect resulting from the arms of low pitch angle galaxies becoming invisible as resolution decreases, meaning that only galaxies with loosely winding arms meet the spirality threshold of 0.7. We are not sure why there is a dip in the middle. It is also interesting that as the human confidence in spirality increases, the entire curve is lifted towards higher pitch angle; again we are not sure why, although one could hypothesize that even humans find it easier to ``notice'' loosely winding arms than tightly wound ones.}
%    \label{fig:pa-vs-diameter-sdss}
%\end{figure}

Despite the complex interplay between pitch angle, FWHM(PSF), S/N, and spirality threshold described above, we now show in Figures \ref{fig:real-fake-0.8} and \ref{fig:pa-with-cutoff} that our blurred images successfully mimic how pitch angle is affected by image degradation in the real sky.

We hypothesize that the real issue is not with any one of galaxy angular size, FWHM(PSF), S/N, or spirality.  Instead, we observe that on any given night, there could be different FWHM(PSF) values at the observing site, differing levels of noise, etc.  Thus, we choose a relatively stringent threshold of 0.8 in spirality, and look at the pitch angle of galaxies as a function of {\it radius in units of the FWHM(PSF)}, and allowing any S/N that gives a galaxy above the spirality threshold.  Figure \ref{fig:real-fake-0.8} demonstrates that, using these criteria, the pitch angle as a function of (radius/FWHM(PSF)) for real SDSS images (``SDSS-all''), vs. images from our artificially blurred set, give mean pitch angle curves that are virtually indistinguishable from each other.  Similar plots occurs for other spirality thresholds.

At this point, it appears we have shown that there is a selection effect such that tightly wound spiral arms are harder to see than loosely wound ones.  It stands to reason, then, that it should be possible to mimic the procedure used to produce complete volume-limited samples. In particular, if our hypothesis is correct, we should be able to create a lower threshold in pitch angle, and eliminate all galaxies below that threshold. With a high enough threshold, we should be able to produce a sample of galaxies that avoid the selection effect because their pitch angles are all big enough to avoid the selection effect. We test this procedure in Figure \ref{fig:pa-with-cutoff}. While we do see that the selection effect is reduced as the pitch angle threshold is increased, there is no obvious cutoff that seems capable of completely eliminating the effect of ``lower pitch angle with increasing radius in units of the FWHM(PSF)''.  Clearly, there is more here than we have been able to discern;
the details of how to account for or correct for this pitch angle selection effect will presumably depend on the details of the data being used and the measurements and analysis being done with the data. The results presented here clearly demonstrate that a selection bias exists in terms of detecting spirality; future programs to quantify spiral structure (or its evolution) in survey data will need to take this effect into account, but this will probably require simulations tailored to the details of the data and methodology being used.

\section{Conclusion}
We have demonstrated that there exists a morphological selection effect when attempting to isolate ``spiral'' galaxies based upon a threshold in the human Galaxy Zoo 1 votes.  In particular, it seems that tightly wound spiral arms, being less visible with increasing image degradation, are less prone to be included in a selected set of galaxies based on said spirality threshold.  We were able to reproduce the effect to high precision by creating a set of artificially degraded images in tandem with a machine learning algorithm to predict the spiralities of said blurred images.  However, we were not fully successful in attempting to account for the selection effect by adding a threshold to pitch angle in an effort to explicitly eliminate tightly-wound spiral galaxies. Further work will be required to determine if the pitch angle variations we observe are some deeper selection effect, or a physical effect in the real universe.

% End of mnras_guide.tex

\section*{Appendix: Sensitivity Analysis of SpArcFiRe to Algorithm Parameter Changes}

Another way to describe the main body of this paper is to say that we have performed a {\it sensitivity analysis} of how SpArcFiRe's results depend upon the image quality. Such studies are crucial in understanding how any analysis may be affected by image quality.
A related issue is how SpArcFiRe's results change as a function of its internal algorithmic parameters. We have performed a detailed analysis along this vein, to the SpArcFiRe algorithm described in \cite{DavisHayesCVPR2012} and \cite{DavisHayes2014}. This appendix describes these sensitivity experiments.

\newcommand{\sensitivitycaption}[1]{
\caption[Changes in several measures computed from our output when varying #1.]{
Changes in several measures computed from our output when varying #1 (x axis).
The median change is given as the red line, with the red error bars giving the upper and lower quartiles.
The green error bars give the 10\textsuperscript{th} and 90\textsuperscript{th} percentiles of the changes to our output, and the blue error bars give the 5\textsuperscript{th} and 95\textsuperscript{th} percentiles.}
}

The algorithmic parameters discussed in \cite{DarrenDavisThesis2014} manage tradeoffs encountered during spiral galaxy structure extraction.
Changing these parameters will, by their nature, at least slightly alter the behavior of our method.
Here, we characterize the effects on our output when varying six algorithmic parameters at the core of our procedure.
For each such parameter, we run our method on our Galaxy Zoo comparison set (used in \cite{DarrenDavisThesis2014}) with five alternate parameter values.\footnote{Due to CPU time constraints, we present results from a slightly old version of our code.
The main change relevant here is that the default unsharp mask was 10 instead of 6, but other than the difference in this default, we do not expect the minor code changes to substantially affect our results, especially since we saw no substantial difference when performing these tests on a version of the code earlier than the one used here.}
If output could not be produced for at least one of these parameter values, the galaxy was dropped from the analysis for that parameter.\footnote{The number of affected galaxies was too small to substantially affect our plots, except in the case of the unsharp mask amount.
We investigate this potential change when examining the effect of this parameter.}
We then compare, on a per-galaxy basis, changes to several aspects of our output.
Specifically, we examine effects on the fragmentation and volume of arcs detected by our method, as measured by the average arc length, total arc length, and total number of arcs.
We also measure effects on our detected winding direction and pitch angle.
For the pitch angle we again use the arc-length-weighted average of all arcs agreeing with the arc-length-weighted winding direction, referred to here as the ``signed pitch angle'' (the sign determines winding direction).
To isolate the effect on the measured arm tightness, we also measure changes in the absolute value of the pitch angle (higher absolute values indicate looser arms).
Lastly, we also consider changes to winding direction.
Winding direction is expressed as $1$ for S-wise, $-1$ for Z-wise, and $0$ for no winding direction (a galaxy-level pitch angle of exactly zero).
Consequently, a S-to-Z winding direction change has a winding direction difference of $-2$ and a Z-to-S winding direction change has a winding direction difference of $2$.
Similarly, changes to or from a non-winding-direction value of $0$ are expressed as differences of $-1$ or $1$, but this type of change is rare because almost all galaxies have a measured winding direction.

\begin{figure}
	\centering
	\begin{subfigure}[b]{0.32\linewidth}
		\centering
		\includegraphics[width=\linewidth]{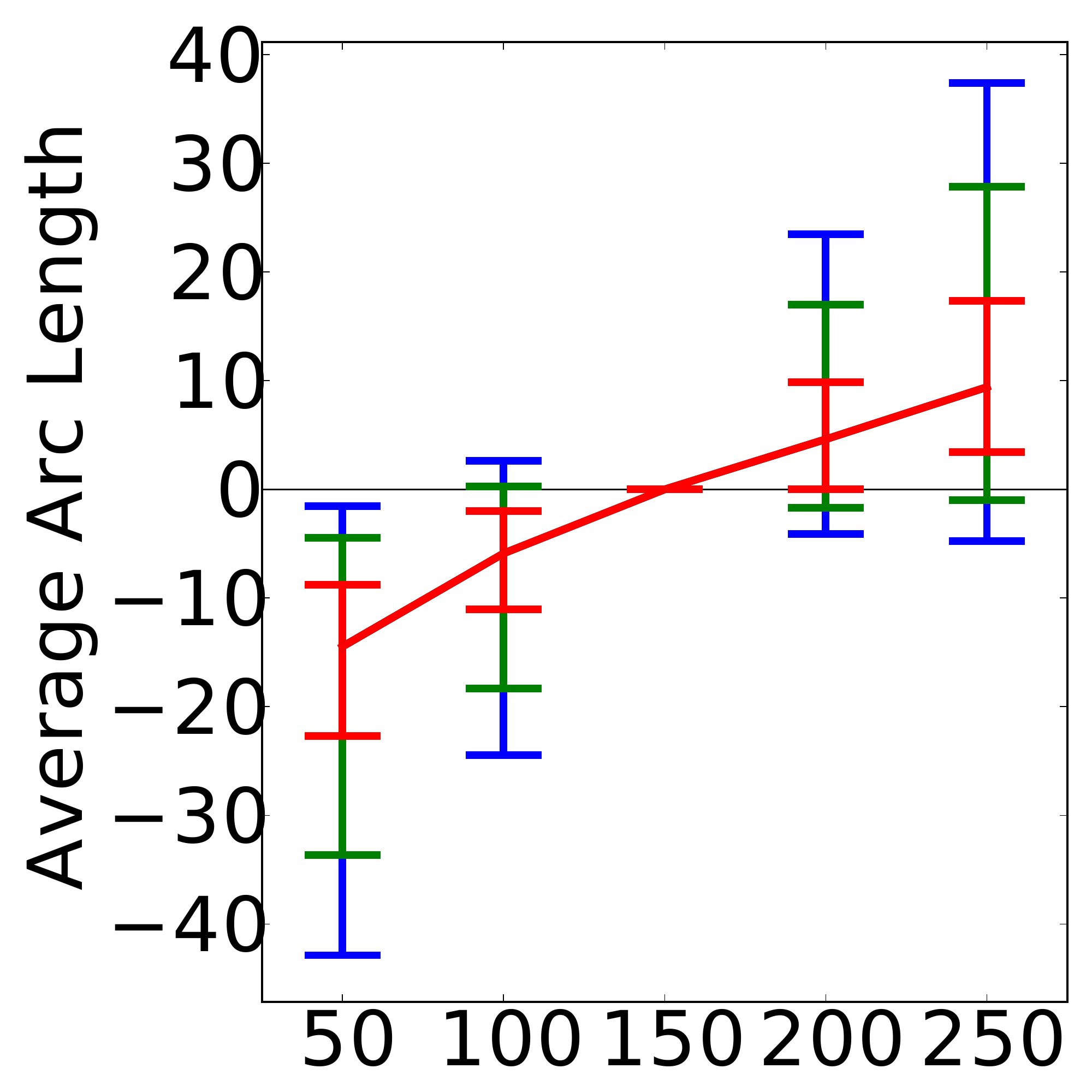}
	\end{subfigure}
	\begin{subfigure}[b]{0.32\linewidth}
		\centering
		\includegraphics[width=\linewidth]{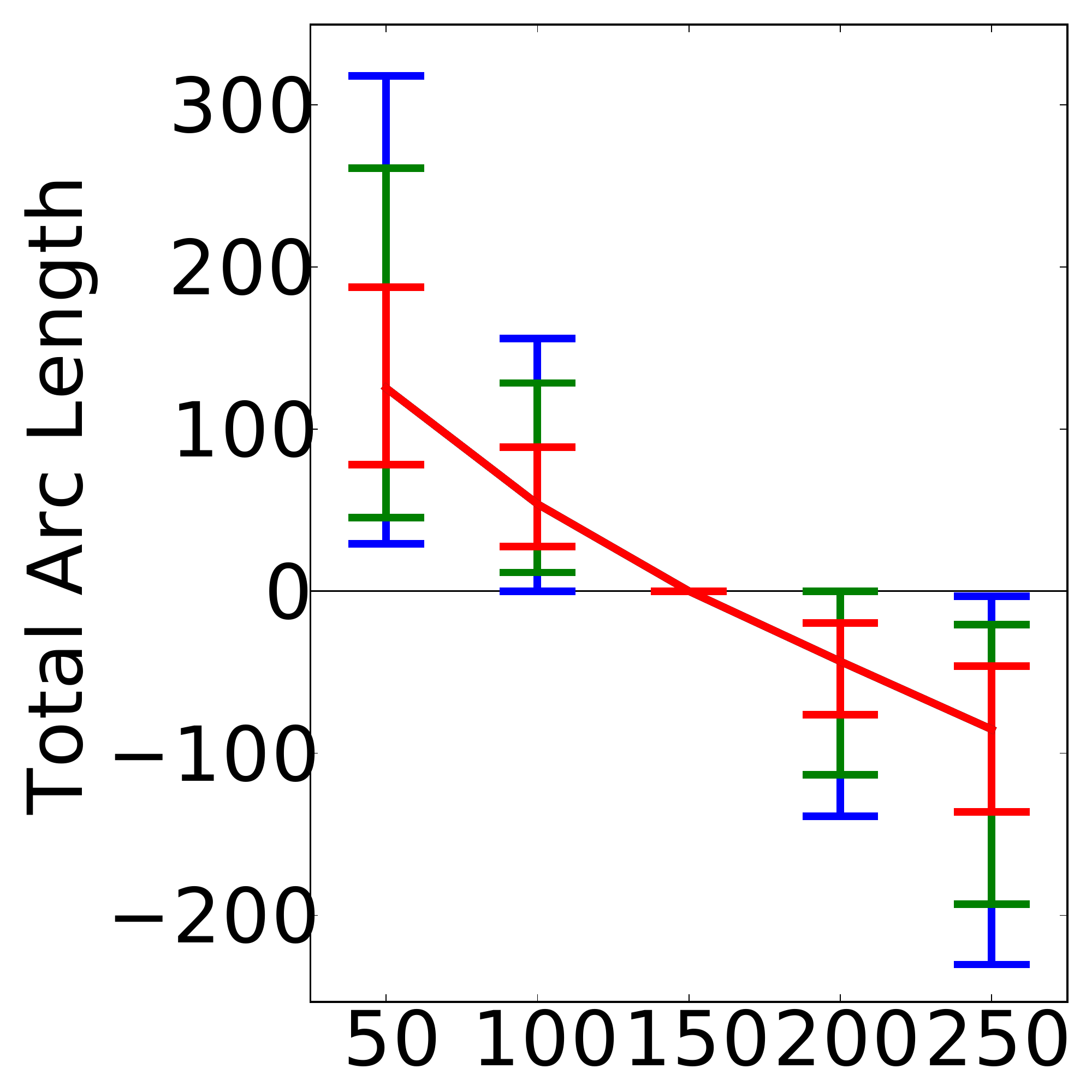}
	\end{subfigure}
	\begin{subfigure}[b]{0.32\linewidth}
		\centering
		\includegraphics[width=\linewidth]{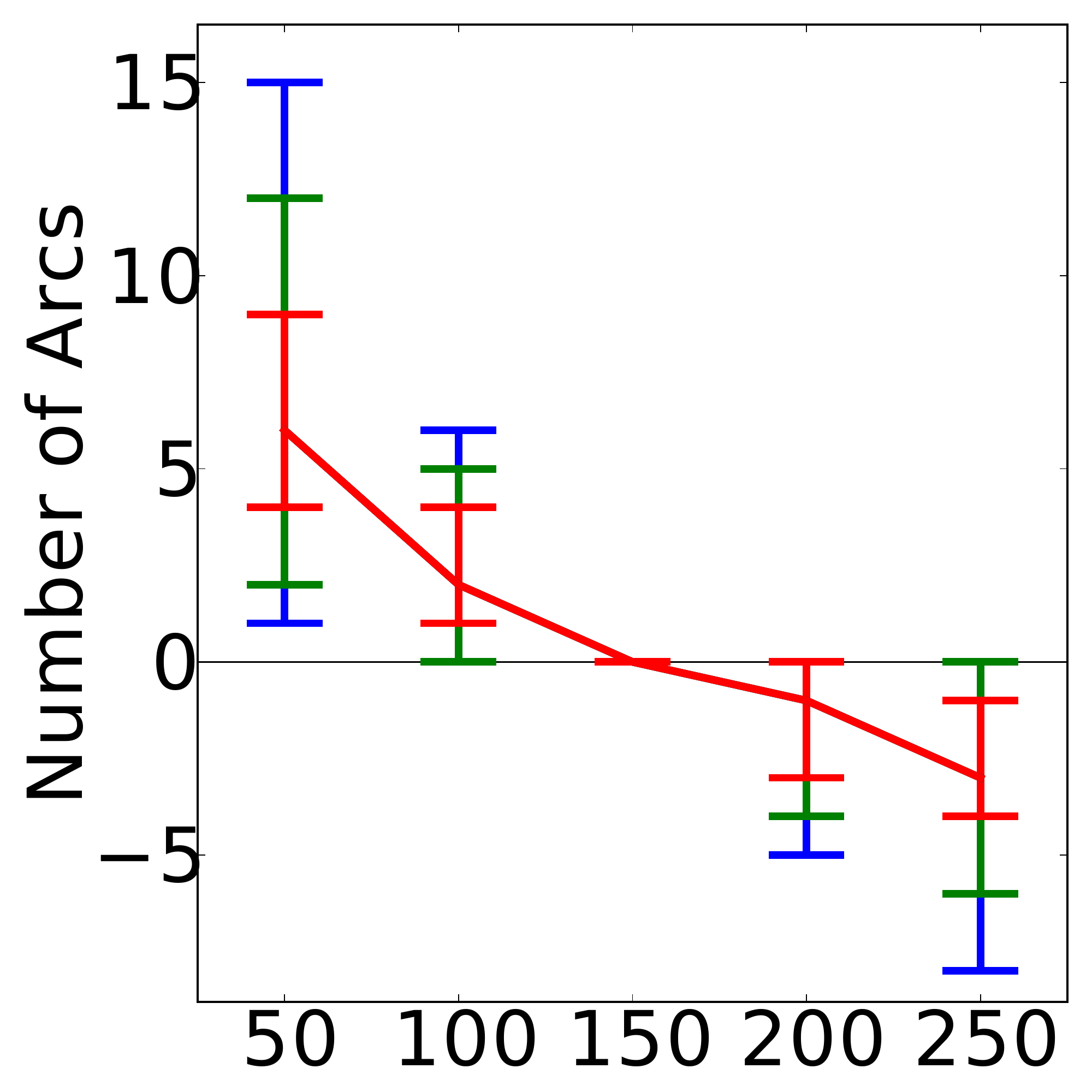}
	\end{subfigure}
	\\
	\begin{subfigure}[b]{0.32\linewidth}
		\centering
		\includegraphics[width=\linewidth]{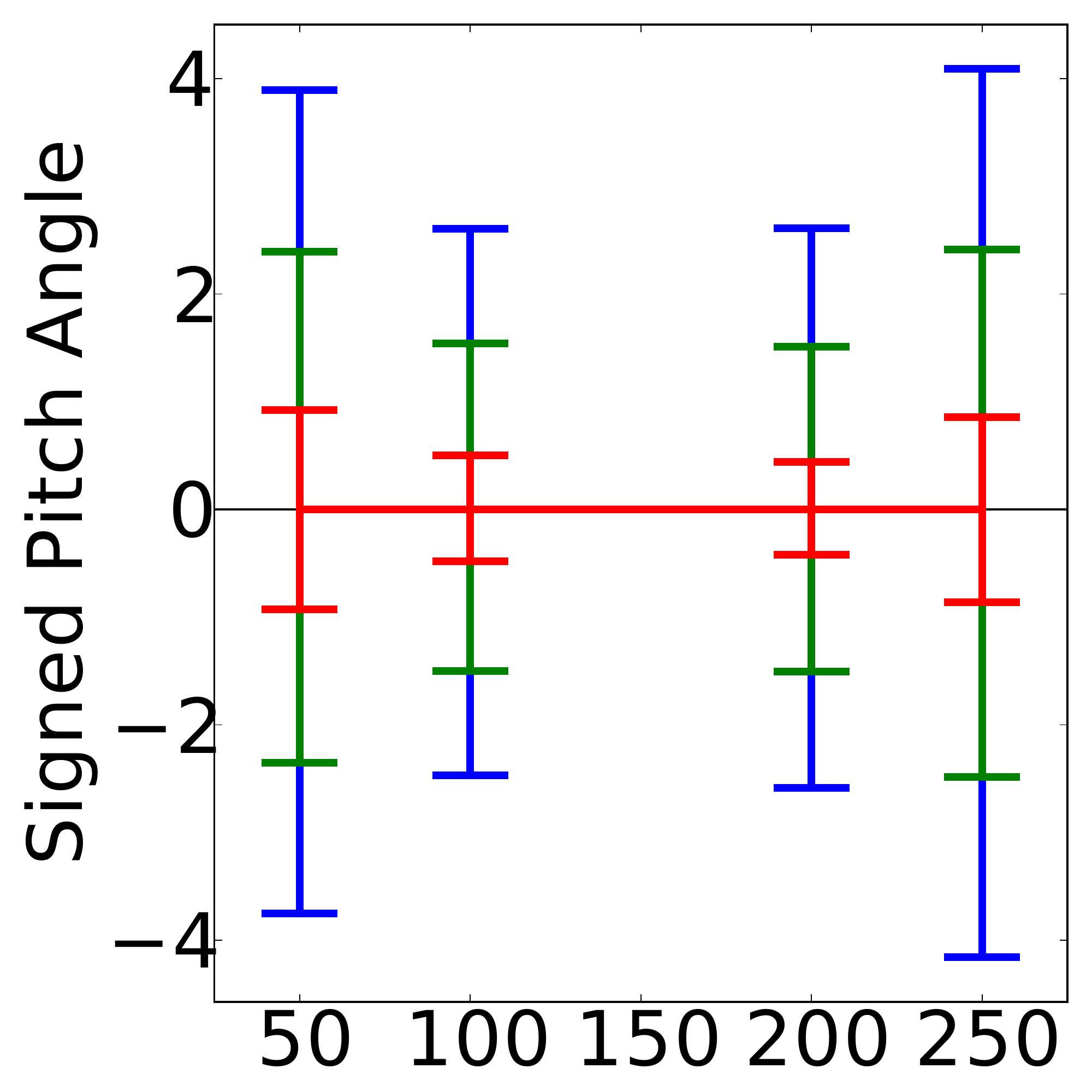}
	\end{subfigure}
	\begin{subfigure}[b]{0.32\linewidth}
		\centering
		\includegraphics[width=\linewidth]{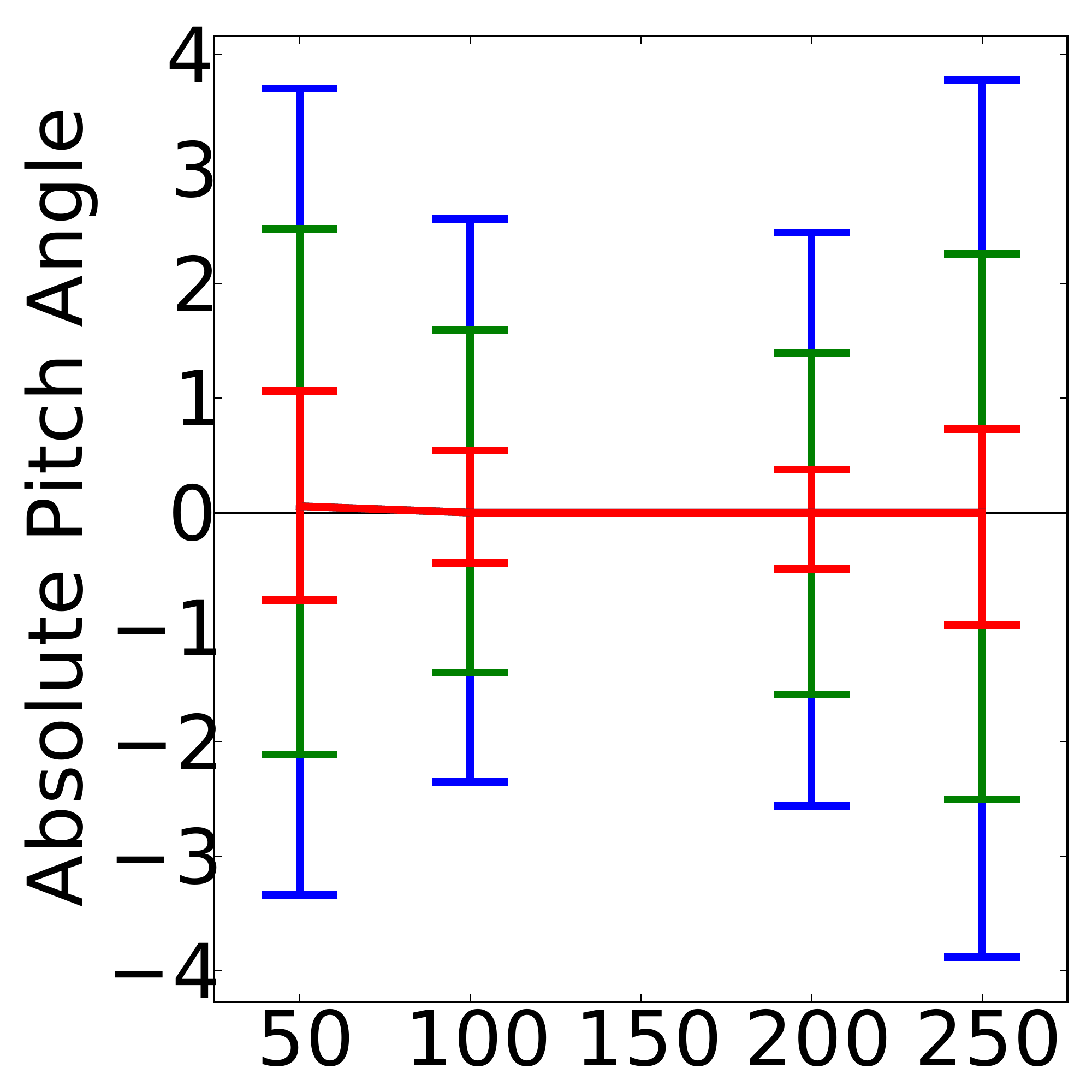}
	\end{subfigure}
	\begin{subfigure}[b]{0.32\linewidth}
		\centering
		\includegraphics[width=\linewidth]{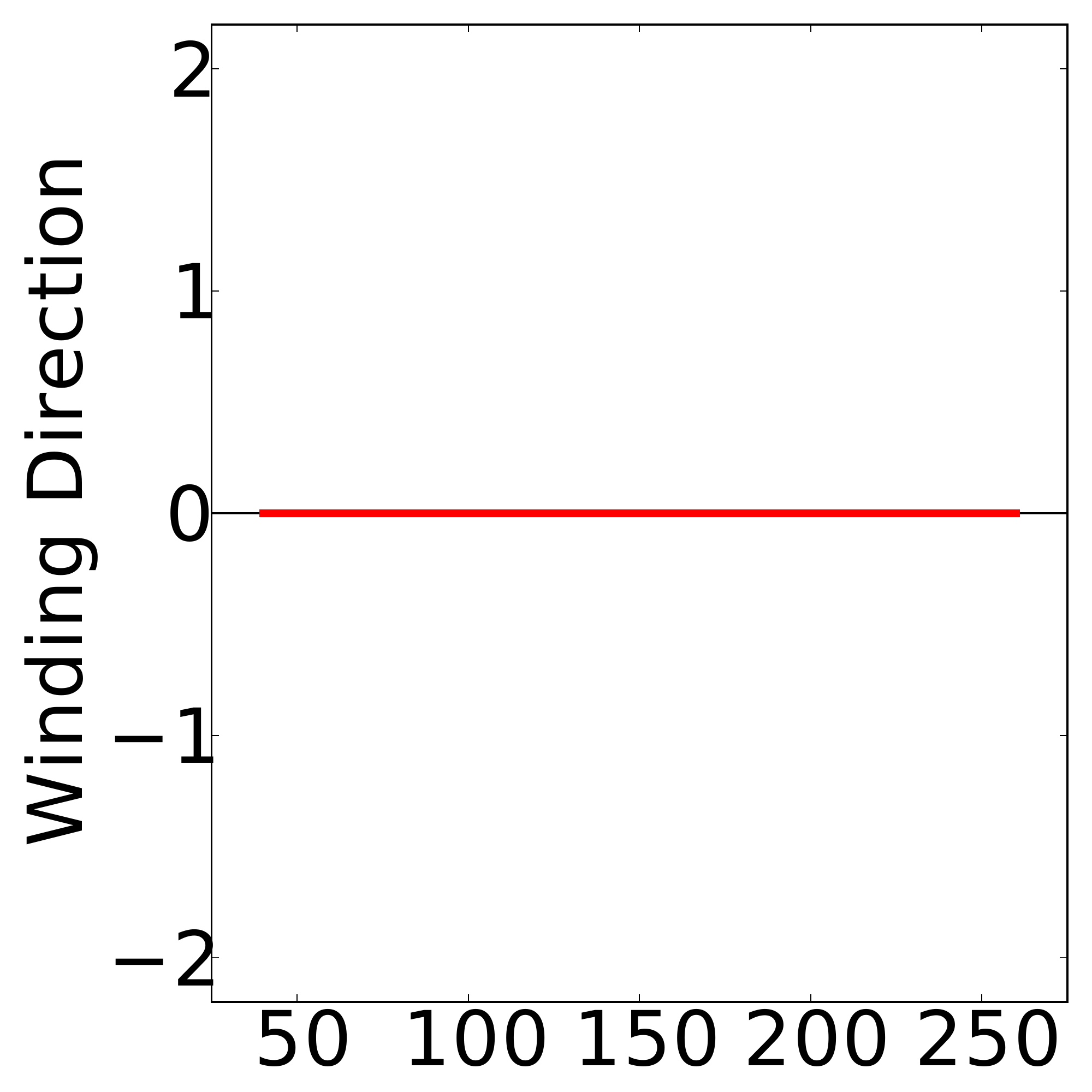}
	\end{subfigure}
	\sensitivitycaption{the minimum size needed for a cluster to be included in our output}
	\label{fig:sensitivity_clusSizeCutoff}
\end{figure}

Figure \ref{fig:sensitivity_clusSizeCutoff} illustrates the effects of varying the minimum cluster size (in pixels) needed for a cluster (and its associated arc) to be included in our output.
%This parameter was discussed in Section \ref{sec:pixel_clustering}.
The red lines give the median changes (in the measured aspects of our output) resulting from each setting of this cluster size threshold (relative to the baseline value of 150 used elsewhere in this work).
The red error bars give the upper and lower quartiles of this difference.
Similarly, the green error bars give the 10\textsuperscript{th} and 90\textsuperscript{th} percentiles of this difference, and the blue error bars give the 5\textsuperscript{th} and 95\textsuperscript{th} percentiles.
Unsurprisingly, the average arc length increases to the extent that we require larger cluster sizes.
The typical change in average arc length is about 10, but this difference can increase to about 40 in some cases.
Allowing fewer small clusters into the image also decreases the total arc length, as expected.
It is also obvious that a less permissive size threshold will reduce the number of clusters; the degrees of change and variation indicate how many clusters tend to be discarded in this manner.
Finally, we note that changes to the measured pitch angle are minor, with no (or negligible) change to the typical pitch angle measured, and fairly low scatter.
The use of arc-length weighting likely increases the stability of the measured pitch angle, but all clusters (that agree with the dominant winding direction) nevertheless contribute to the pitch angle measurements.
Winding direction also tends to be unaffected; from the figure (in particular, the absence of visible blue bars) we note that at least 90\% of the images (and probably more) are unaffected in winding direction.

\begin{figure}
	\centering
	\begin{subfigure}[b]{0.32\linewidth}
		\centering
		\includegraphics[width=\linewidth]{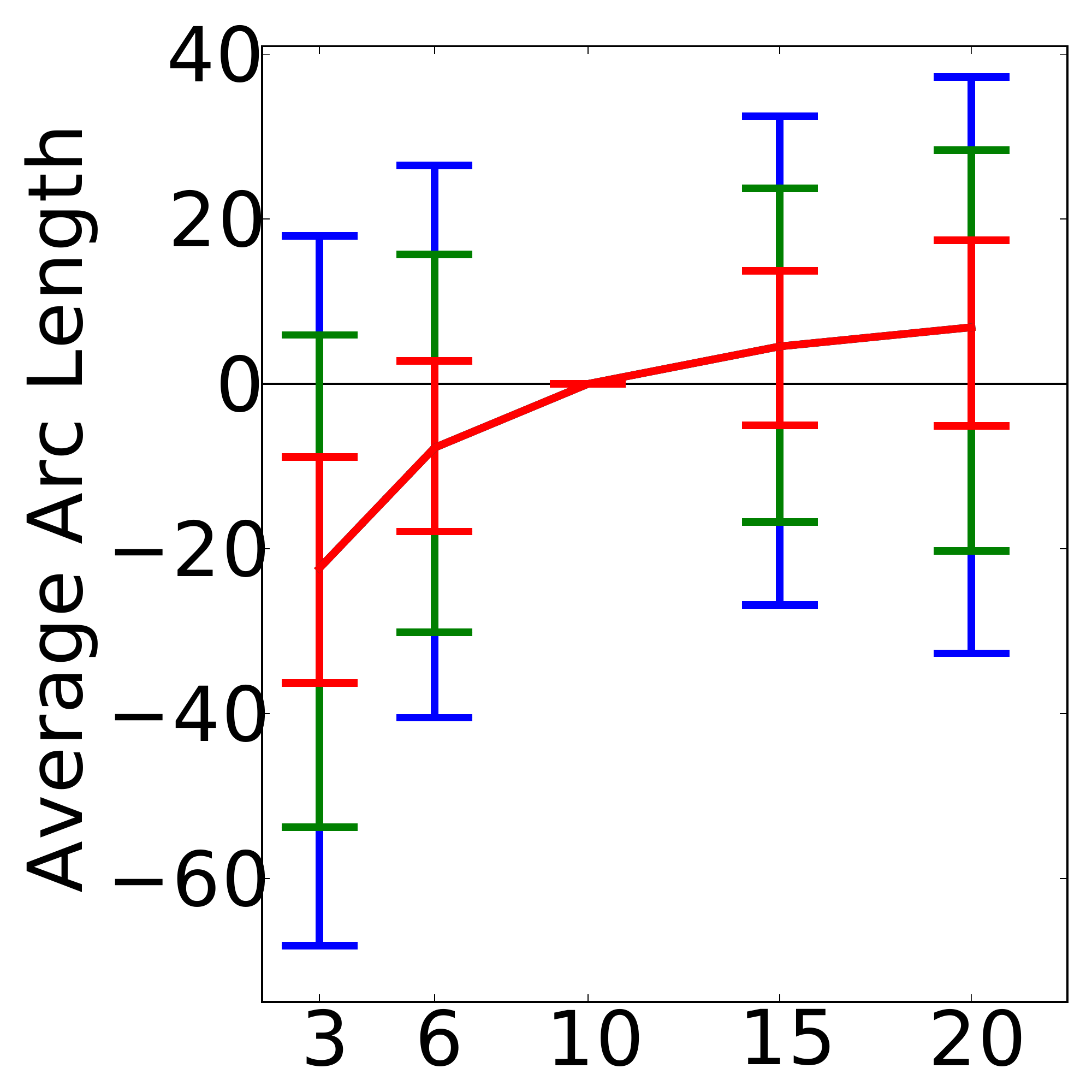}
	\end{subfigure}
	\begin{subfigure}[b]{0.32\linewidth}
		\centering
		\includegraphics[width=\linewidth]{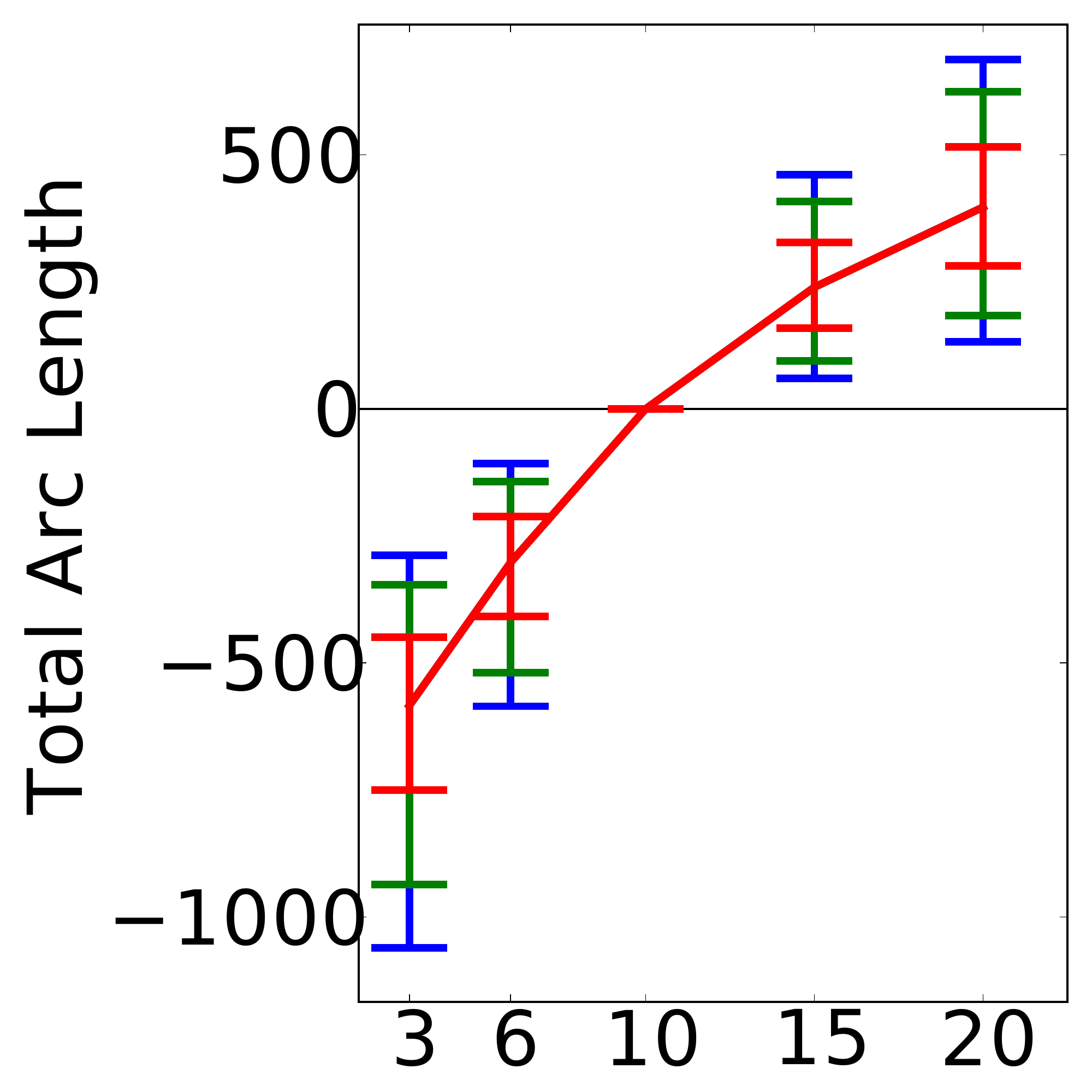}
	\end{subfigure}
	\begin{subfigure}[b]{0.32\linewidth}
		\centering
		\includegraphics[width=\linewidth]{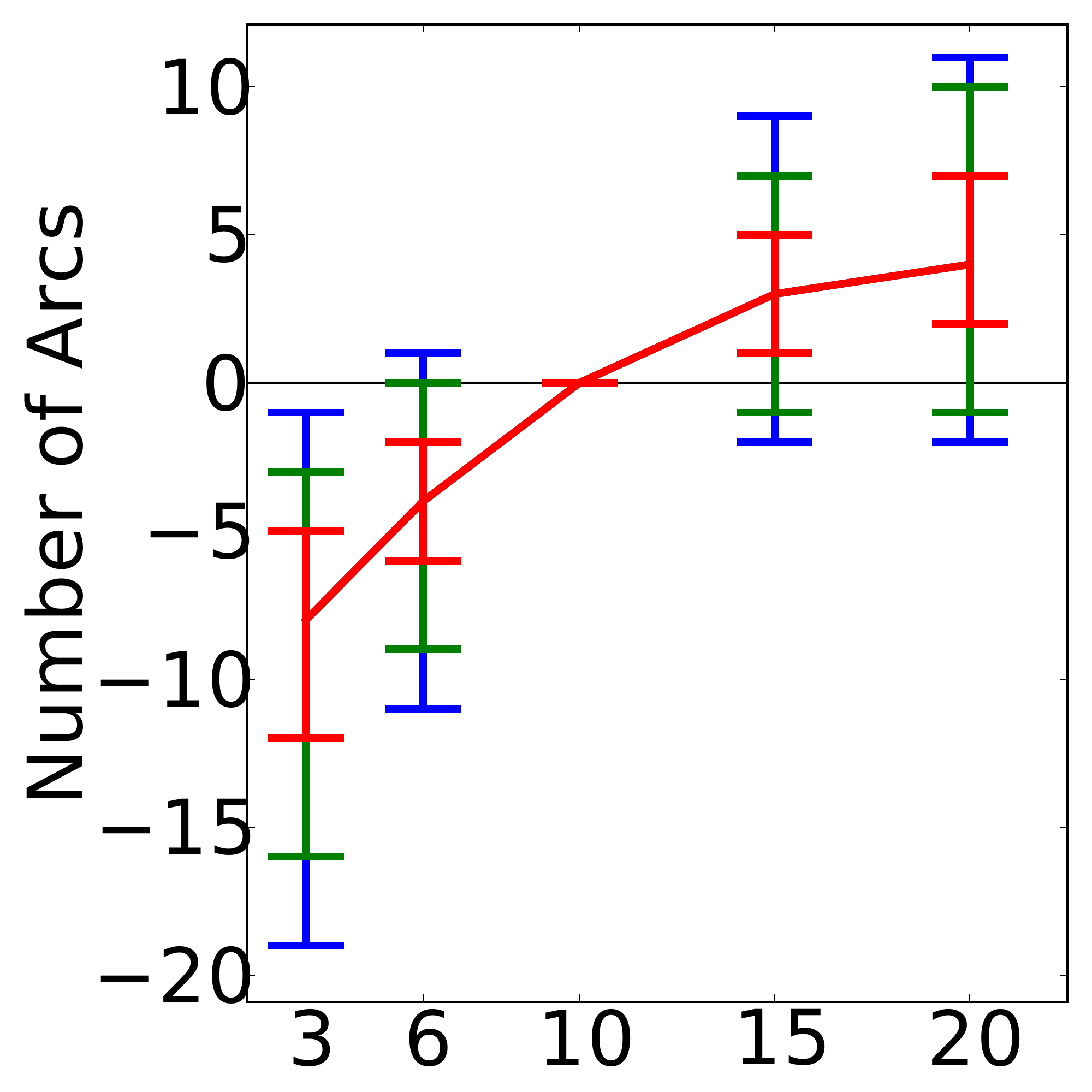}
	\end{subfigure}
	\\
	\begin{subfigure}[b]{0.32\linewidth}
		\centering
		\includegraphics[width=\linewidth]{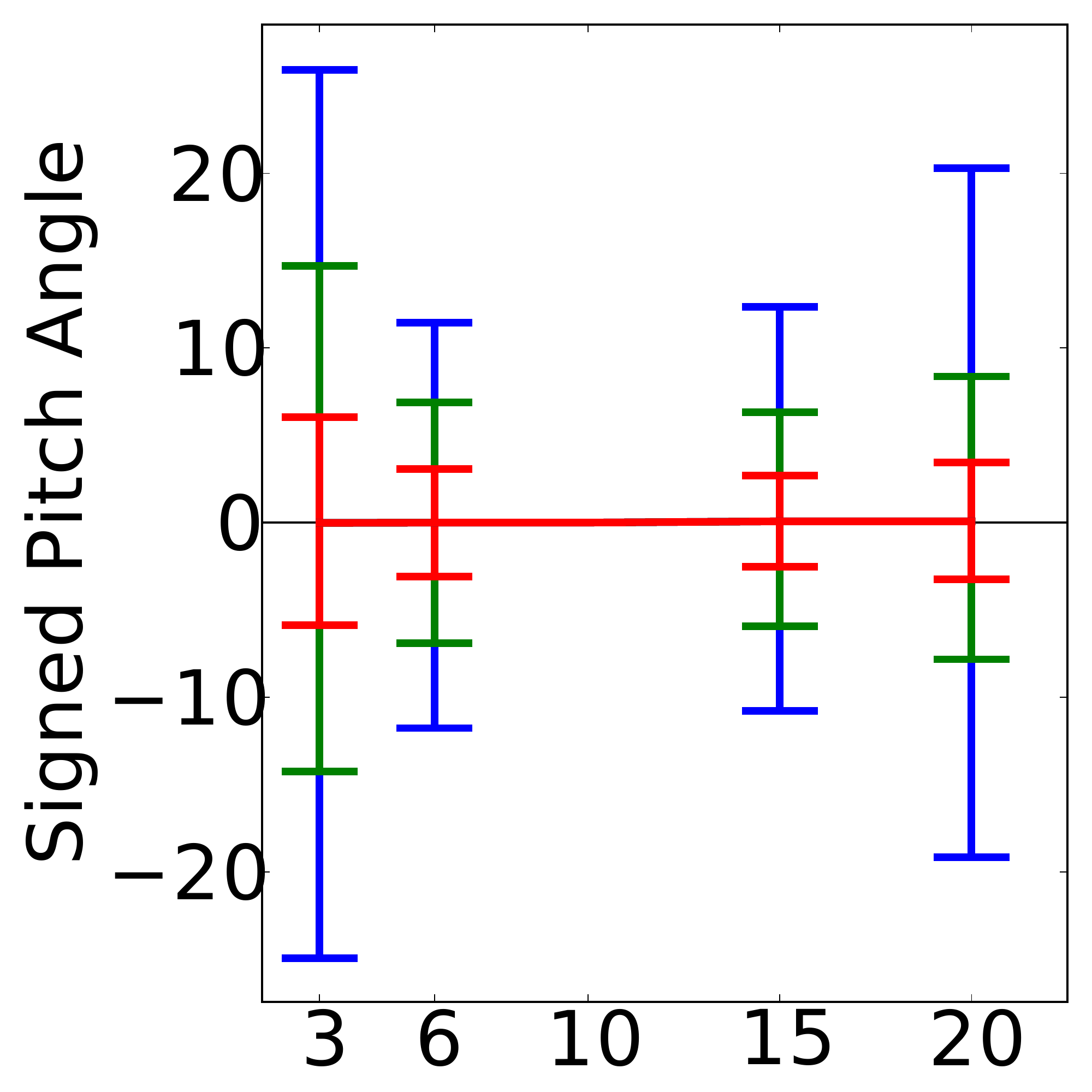}
	\end{subfigure}
	\begin{subfigure}[b]{0.32\linewidth}
		\centering
		\includegraphics[width=\linewidth]{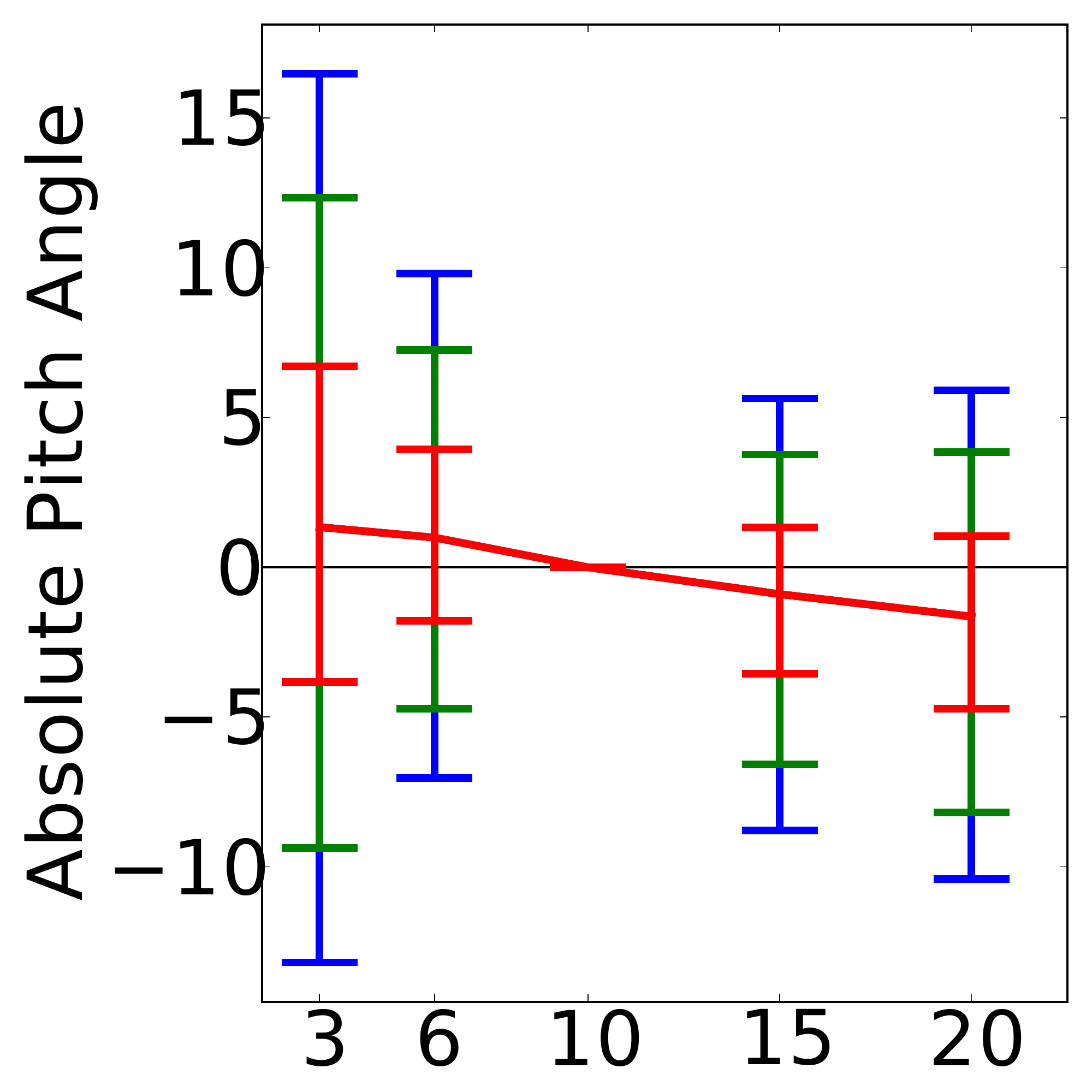}
	\end{subfigure}
	\begin{subfigure}[b]{0.32\linewidth}
		\centering
		\includegraphics[width=\linewidth]{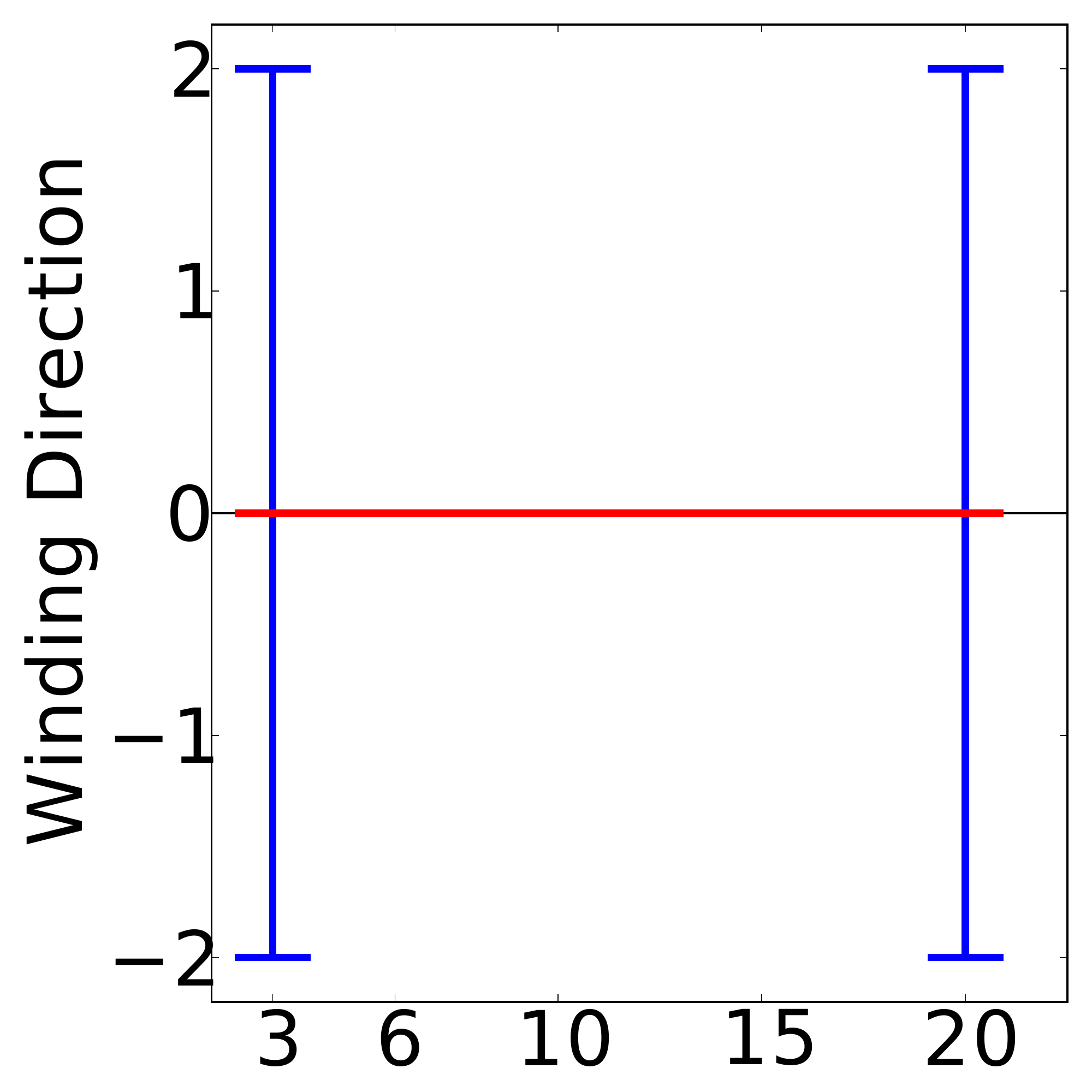}
	\end{subfigure}
	\sensitivitycaption{the unsharp mask amount}
	\label{fig:sensitivity_unsharpMaskAmt}
\end{figure}

Figure \ref{fig:sensitivity_unsharpMaskAmt} is set up in the same way as Figure \ref{fig:sensitivity_clusSizeCutoff}, but instead examines the effects of varying the unsharp mask amount.\footnote{As discussed earlier, the baseline unsharp mask amount is 10 instead of 6 because we use a slightly old version of our code.}
Increasing the unsharp mask amount increases the contrast between bright and dark regions, increasing the magnitudes of arm-aligned orientation vectors and thus increasing the pixel similarity scores within spiral arms (since the similarity scores are computed by a dot product.
This, in turn, increases the number of pixel and cluster merges, which means that clusters tend to be larger (more pixels merge into the clusters) and have a higher tendency to merge (since clusters are more likely to be adjacent to each other when they grow).
Since larger clusters are more likely to reach the minimum size threshold, and since clusters are more likely to be picked up by the orientation field when the unsharp mask is stronger (up to a point), we see an increase in the total number of arcs.
This trend overpowers the reduction in cluster count caused by an increased number of merges, but the cluster-count increase slows for larger unsharp mask amounts, suggesting that large-cluster merges accelerate for larger unsharp mask amounts, or that incremental increases in the unsharp mask amount have a stronger impact on orientation field sensitivity when increasing from a lower value of the unsharp mask (both of these potential effects may play a role).
The higher propensity for cluster merges increases the average arc length, in spite of the decrease in this length induced by the introduction of smaller clusters that meet the size threshold under a stronger unsharp mask.
The slowdown in the increase in average arc length could indicate that larger numbers of smaller clusters are being introduced for higher unsharp mask amounts, or that cluster growth becomes more width-wise for larger unsharp mask amounts.
The total arc length increases strongly and steadily for all increases in the unsharp mask, due to both the introduction of new clusters and the growth of existing clusters.

Changes in the unsharp mask amount do not change the typical signed pitch angle measured by our method, although the variance in pitch angle changes is relatively high.
The measured arm tightness does increase slightly (the absolute pitch angle decreases slightly) with a stronger unsharp mask.
The median change in typical arm tightness is relatively small (about 0.99 and 1.34 degrees with unsharp mask amount decreases, and about 0.90 and 1.64 degrees with unsharp amount increases), but since it varies smoothly with unsharp mask amount, the change in tightness might be a real effect.
We are uncertain as to the cause, but possibilities include: slightly looser arms (higher absolute pitch angles) at the brightest parts of the spiral arms (which do not need a stronger unsharp mask to be picked up by the orientation field); the possibility that ``real'' spiral arms are, on average, slightly looser than noisy arcs (that are more prevalent with a stronger unsharp mask); a (slightly) elevated risk of detecting the edge of the galaxy disk as a low-pitch-angle spiral arm (or increased sensitivity to legitimate ring patterns at the edge of the galaxy disk) when the unsharp mask amount is increased; a slightly increased tendency to merge very-low-pitch-angle arcs into a zero-pitch-angle ring when the unsharp mask amount is increased; a slight tendency for pitch angle to be reduced when clusters are merged (but our later analysis of cluster-merge leniency does not find this to be the case); and a possible tendency for brightness gaps to have a sharper brightness decrease (to the point where the gap is exaggerated instead of reduced by the unsharp mask) for arms with a higher pitch angle (e.g., arms with a high (or at least not-extremely-low) pitch angle may have a higher chance of containing a severely-interfering dust lane, in which case a stronger unsharp mask is more likely to fill gaps in low-pitch-angle arcs).
Winding direction is mostly unaffected, although there are some flips in both directions for the severe unsharp mask values, since the unsharp mask can strongly affect the prominence and inclusion of clusters (and thus the presence and strength of winding direction votes).

At the lowest tested unsharp mask amount, the number of images where output could not be produced (due to a lack of sufficiently-sized clusters) was much higher than normal, requiring us to discard many more galaxies than normal (7094 out of 29250).
If we remove the lowest unsharp mask amount value from the analysis (so that the number of unavailable galaxies is reduced to 755, which is still relatively high, but now only about 2.6\% of the galaxies), the results are almost the same, except that winding direction flips are visible in the ``blue'' error bars (5\% and 95\% difference percentiles) for an unsharp mask amount of 15.

\begin{figure}
	\centering
	\begin{subfigure}[b]{0.32\linewidth}
		\centering
		\includegraphics[width=\linewidth]{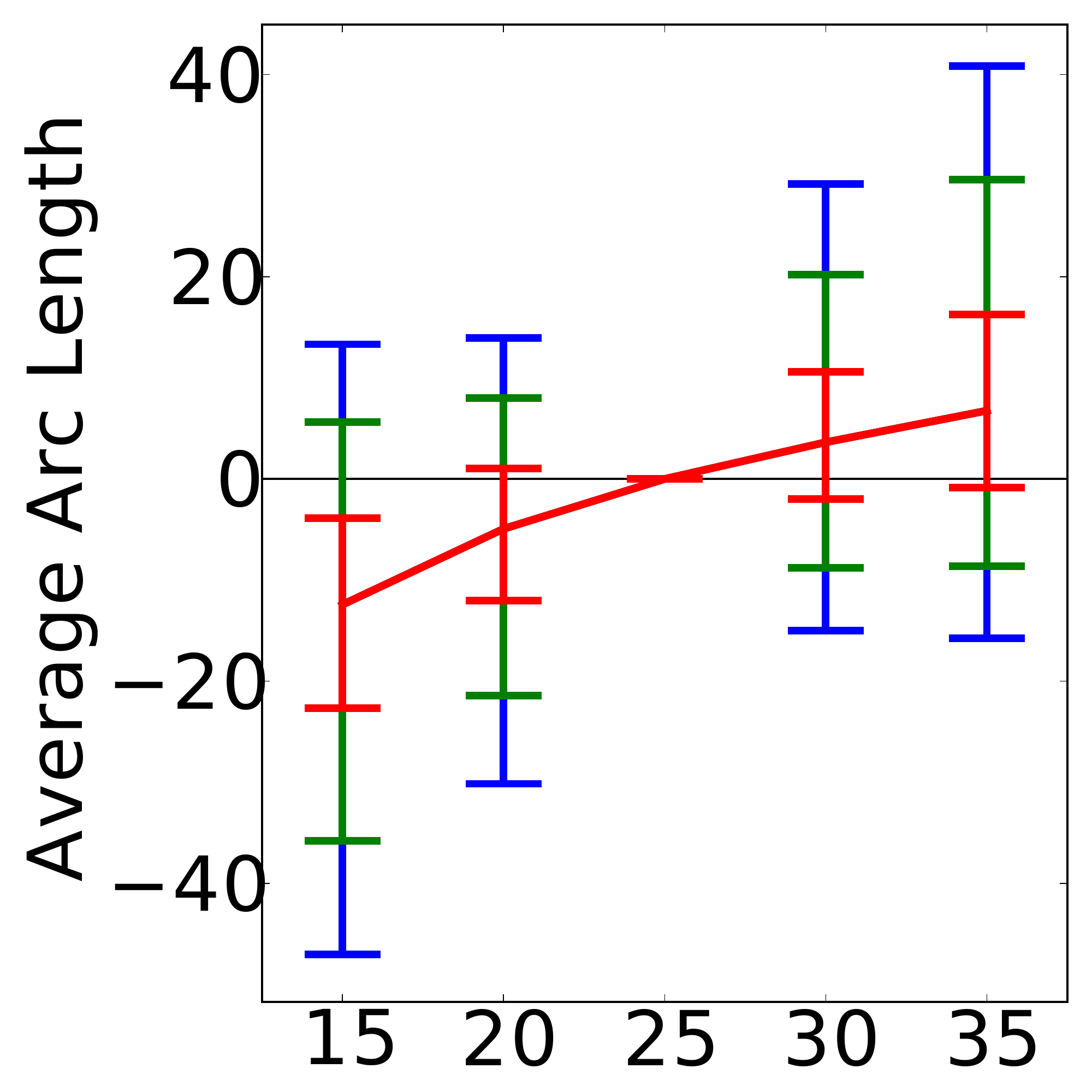}
	\end{subfigure}
	\begin{subfigure}[b]{0.32\linewidth}
		\centering
		\includegraphics[width=\linewidth]{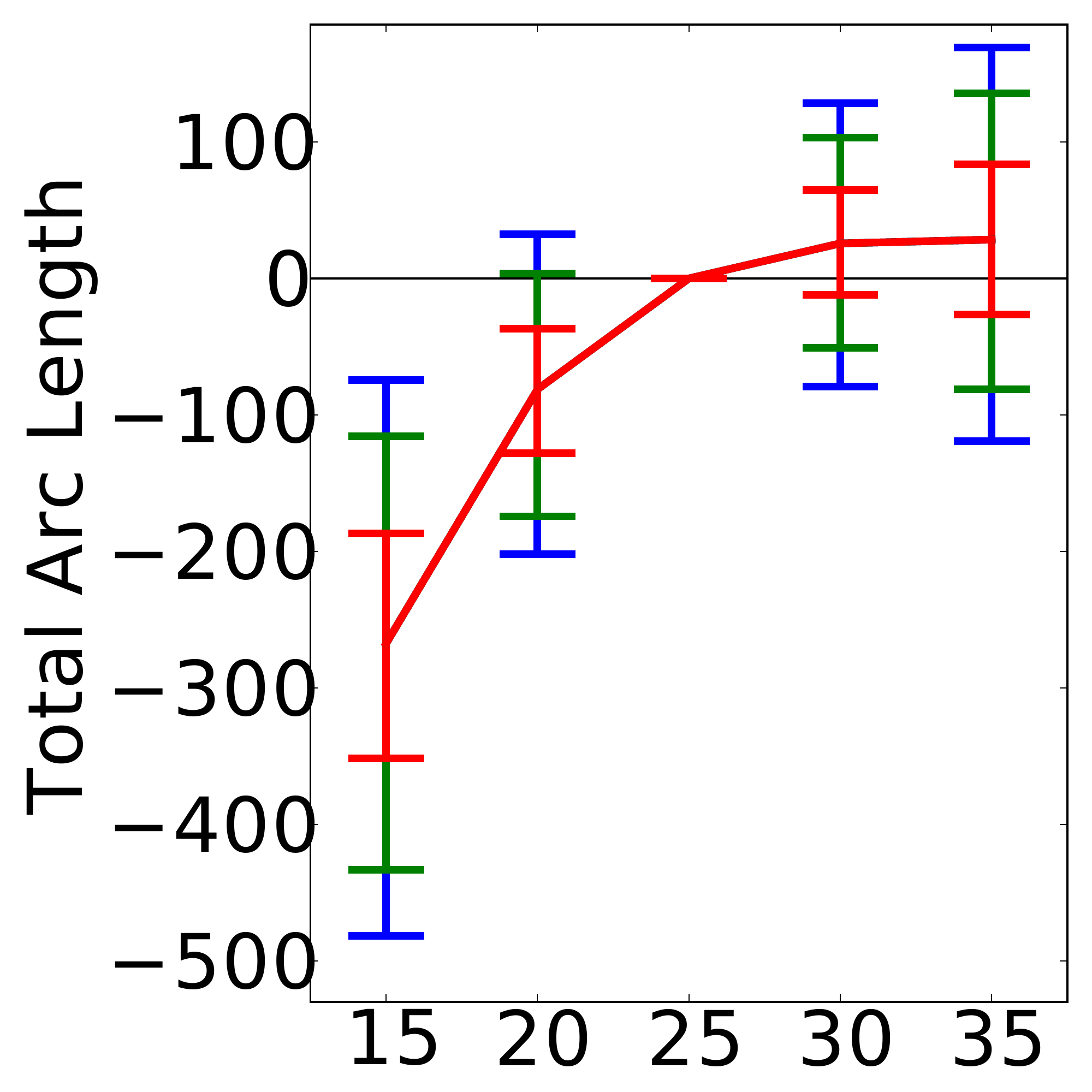}
	\end{subfigure}
	\begin{subfigure}[b]{0.32\linewidth}
		\centering
		\includegraphics[width=\linewidth]{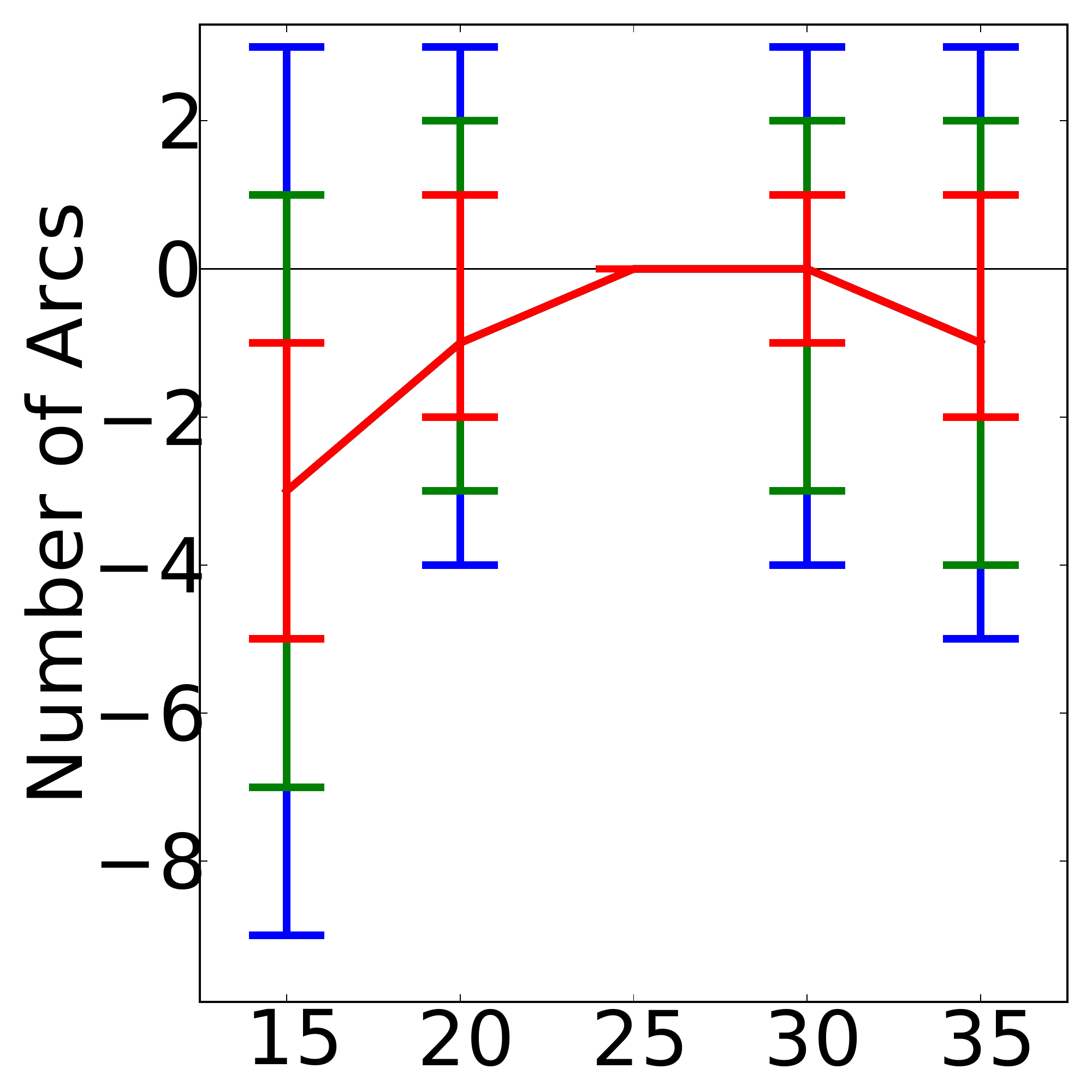}
	\end{subfigure}
	\\
	\begin{subfigure}[b]{0.32\linewidth}
		\centering
		\includegraphics[width=\linewidth]{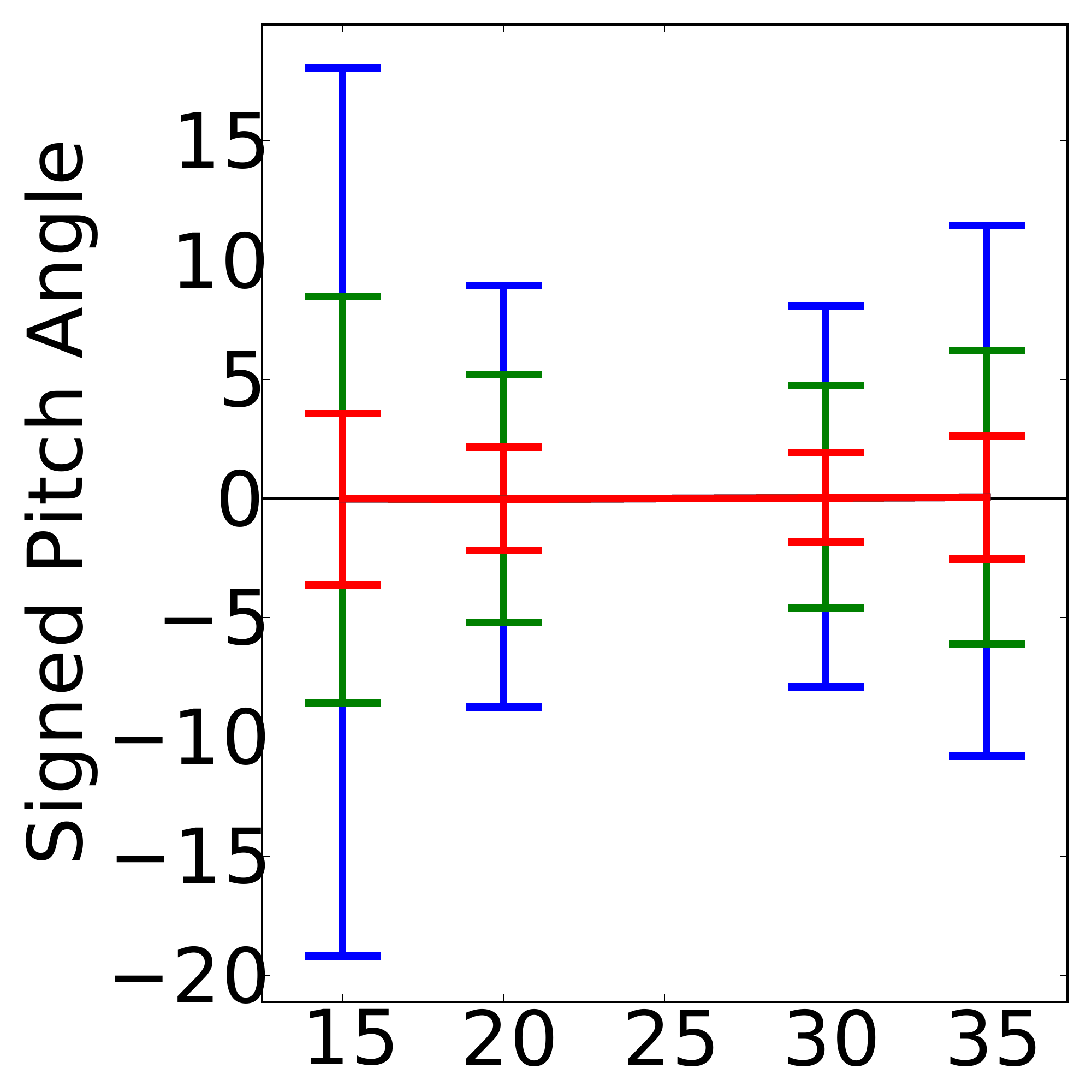}
	\end{subfigure}
	\begin{subfigure}[b]{0.32\linewidth}
		\centering
		\includegraphics[width=\linewidth]{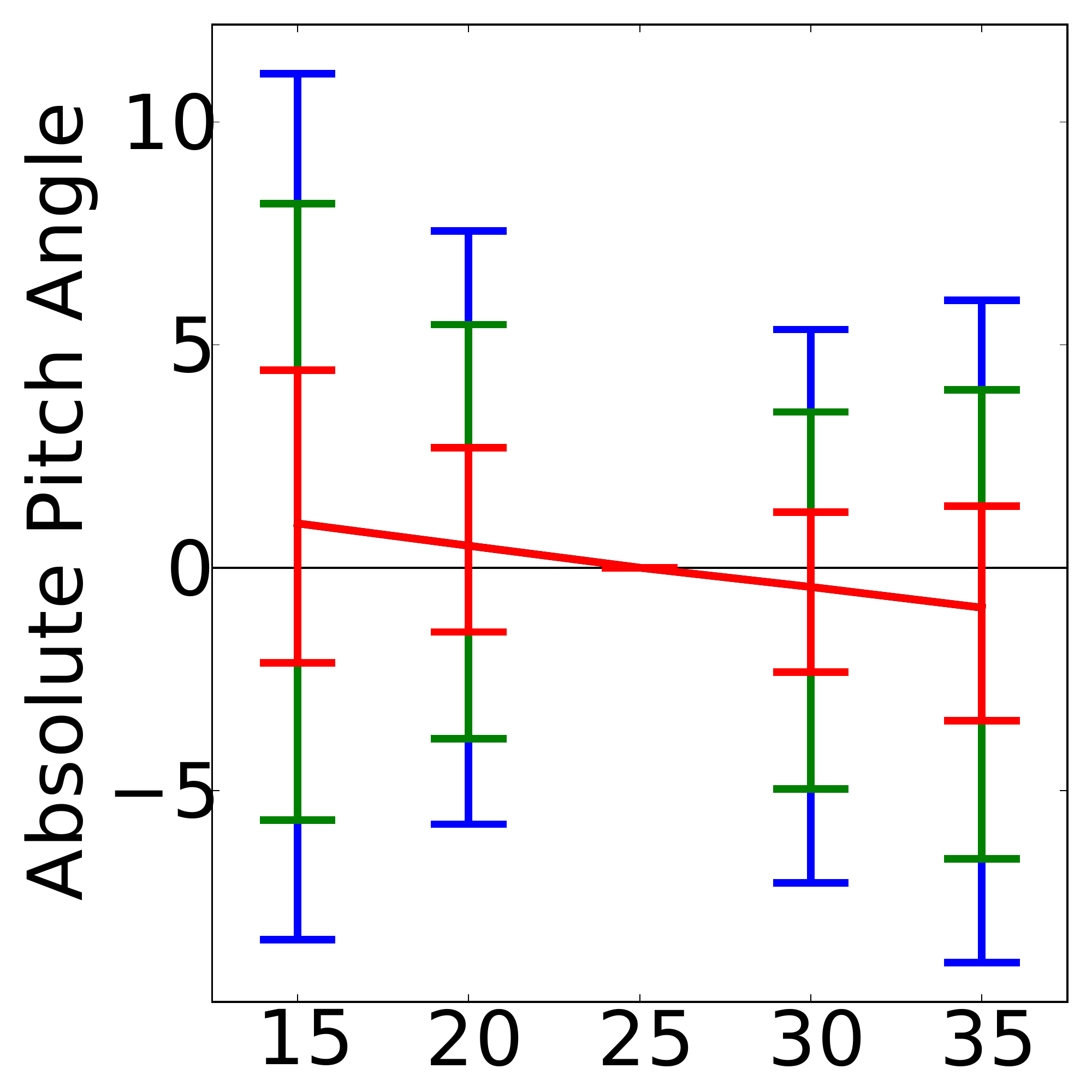}
	\end{subfigure}
	\begin{subfigure}[b]{0.32\linewidth}
		\centering
		\includegraphics[width=\linewidth]{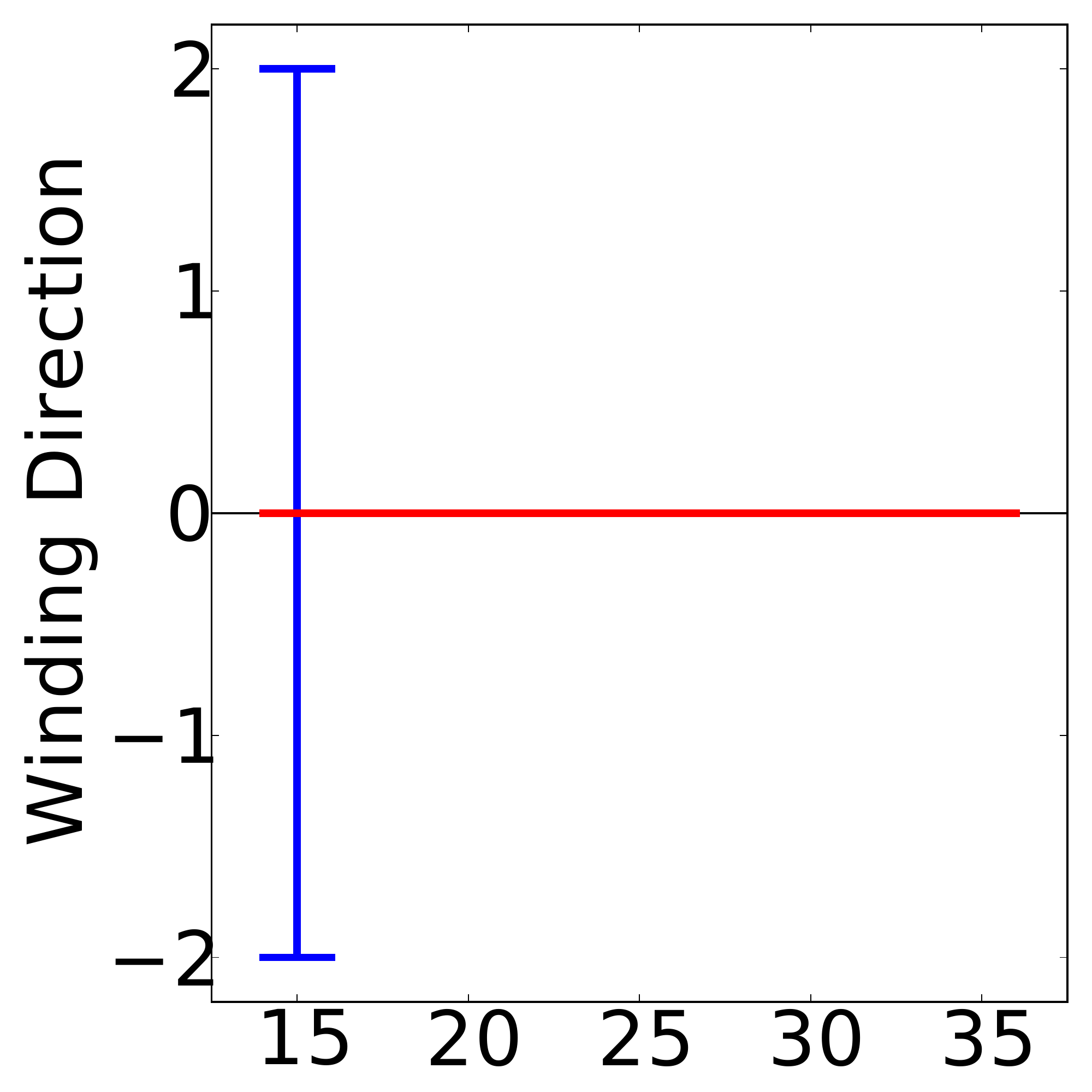}
	\end{subfigure}
	\sensitivitycaption{the size scale of the unsharp mask (i.e., the scale of the Gaussian blur subtracted from the image)}
	\label{fig:sensitivity_unsharpMaskSigma}
\end{figure}

Next, we assess the impact of the scale of the unsharp mask; the effects of changes to this parameter are shown in Figure \ref{fig:sensitivity_unsharpMaskSigma}.
The scale of the unsharp mask controls the width of the Gaussian blur subtracted from the image.
Consequently, smaller scales emphasize contrast in smaller image features and larger scales emphasize contrast in larger image features.
The effect of the unsharp mask scale on the total number of arcs is unique in that it is not monotonic; typical arc counts decrease with either an increase or decrease to the unsharp mask scale.
This confirms that the unsharp mask scale is within a range that emphasizes the typical size scales (widths) of spiral arms found in the standardized images.
Increasing the size scale of the unsharp mask increases the average arc length, perhaps because larger-scale features are emphasized.
The total arc length increases as a function of the unsharp mask scale, but less so at the largest unsharp mask scales, perhaps because the arc count and average arc length both decrease at unsharp mask scales smaller than the baseline value of 25, but move in opposite directions at unsharp mask scales larger than the baseline.

There is no substantial difference in the typical signed pitch angle, but the arm tightness increases slightly (the absolute pitch angle decreases slightly) with larger values for the unsharp mask scale.
The effect seen is very similar to the effect of increasing the unsharp mask amount, so the underlying mechanism (or mechanisms) may be similar.
We also note that decreasing the unsharp mask amount and decreasing the unsharp mask scale may both favor smaller portions of the spiral arms (due to the smaller portions' higher contrast for the unsharp mask amount and their smaller size for the unsharp mask scale).
Additionally, both algorithm parameters would be affected by the severity of brightness decreases between arm segments that could plausibly be merged together; as discussed previously, this severity could vary slightly by pitch angle, on average.
These observations further support the hypothesis that a common mechanism is present for the (slight) decrease in absolute pitch angle found when increasing the unsharp mask amount or unsharp mask scale.
The effect of the unsharp mask scale on typical winding direction is simpler: typical winding directions are unaffected, but some flips are noticeable at the smallest tested unsharp mask scale, likely because noise is more prevalent at smaller scales.

\begin{figure}
	\centering
	\begin{subfigure}[b]{0.32\linewidth}
		\centering
		\includegraphics[width=\linewidth]{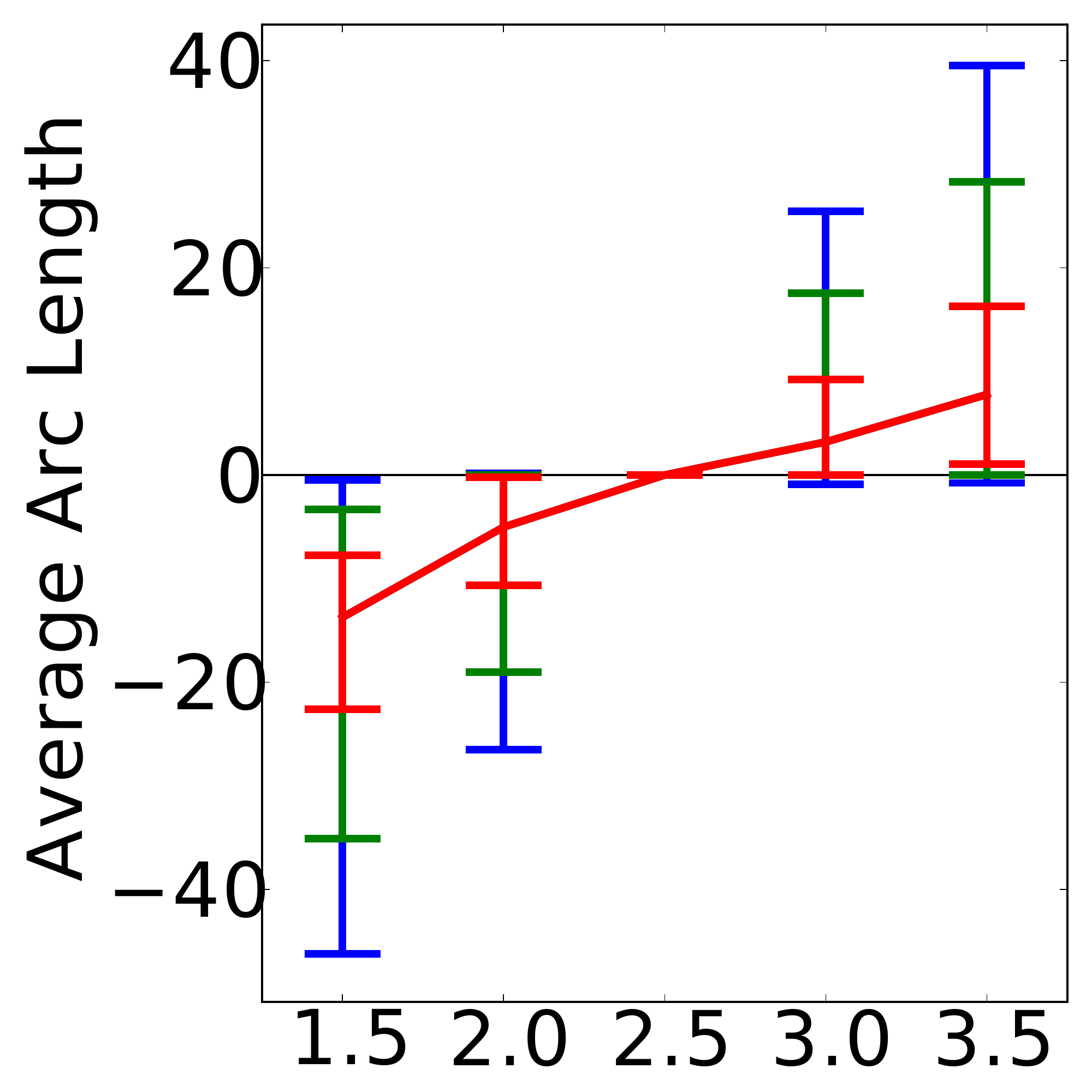}
	\end{subfigure}
	\begin{subfigure}[b]{0.32\linewidth}
		\centering
		\includegraphics[width=\linewidth]{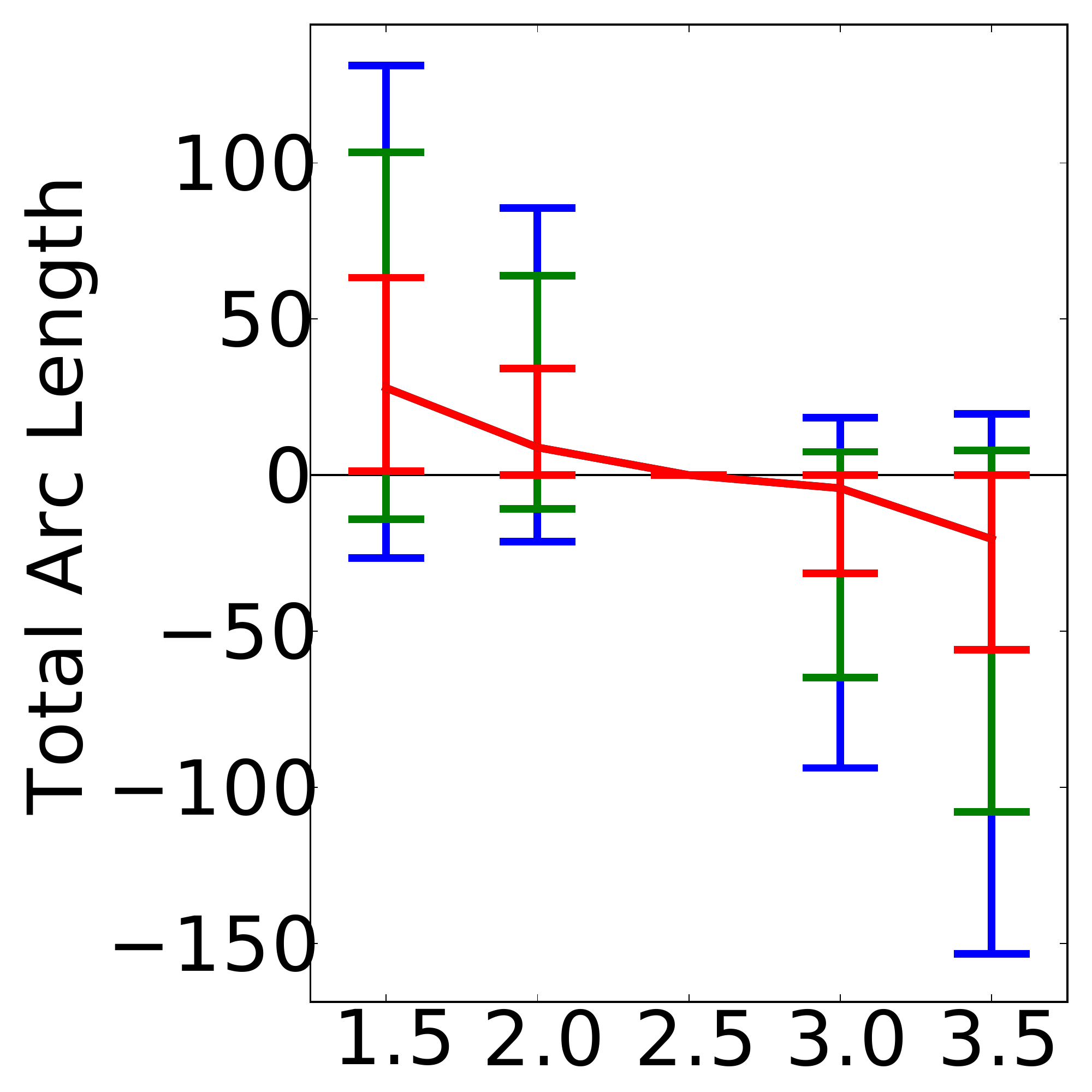}
	\end{subfigure}
	\begin{subfigure}[b]{0.32\linewidth}
		\centering
		\includegraphics[width=\linewidth]{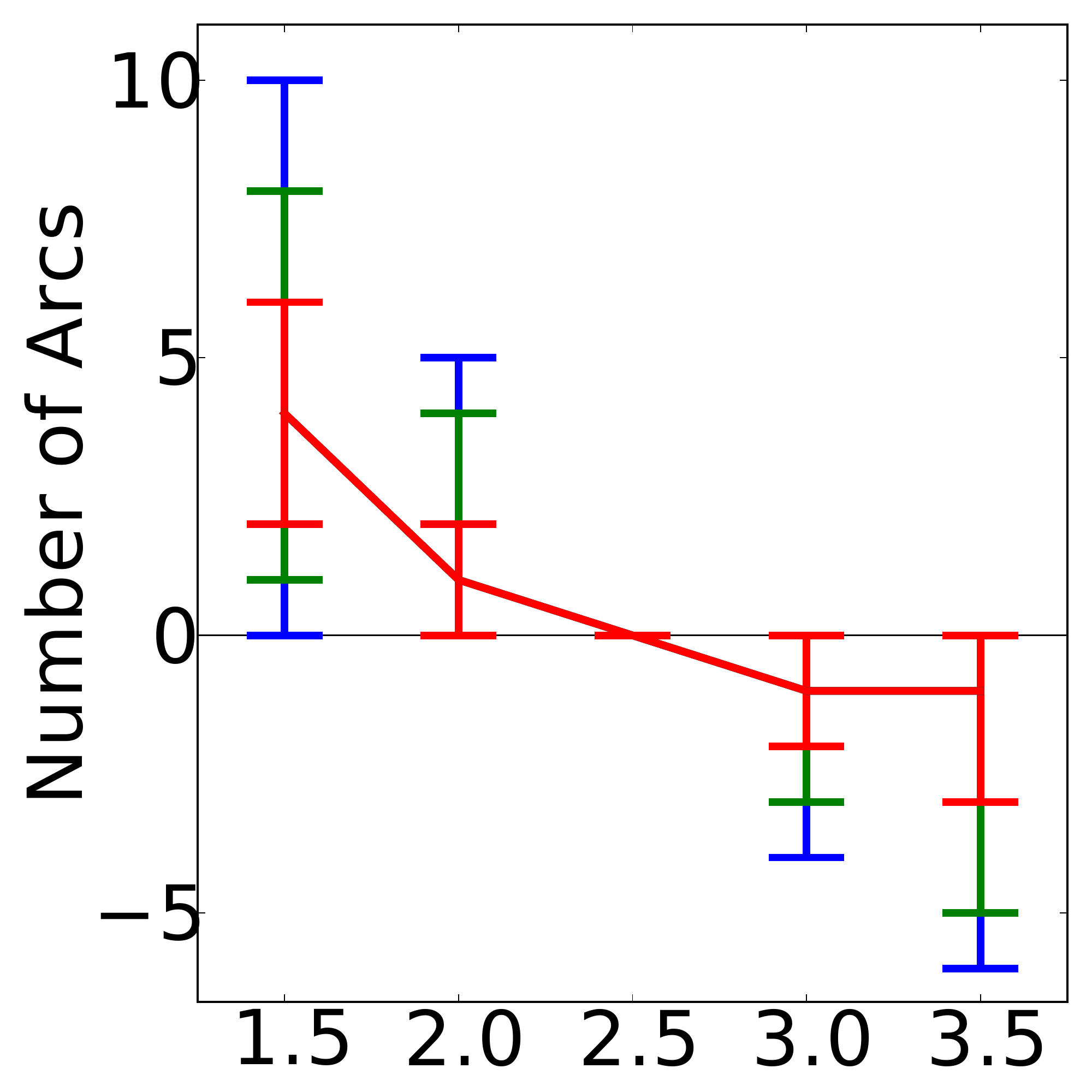}
	\end{subfigure}
	\\
	\begin{subfigure}[b]{0.32\linewidth}
		\centering
		\includegraphics[width=\linewidth]{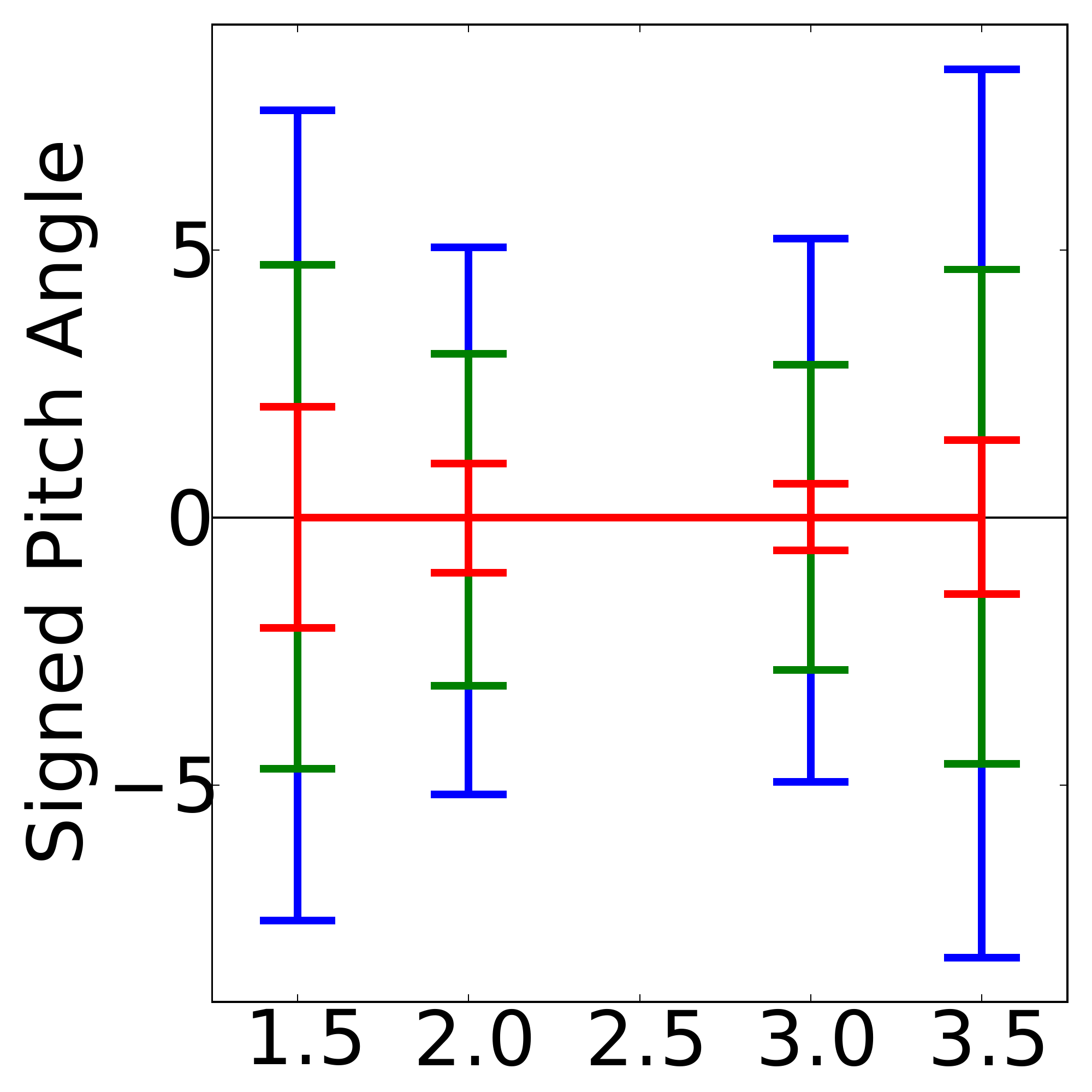}
	\end{subfigure}
	\begin{subfigure}[b]{0.32\linewidth}
		\centering
		\includegraphics[width=\linewidth]{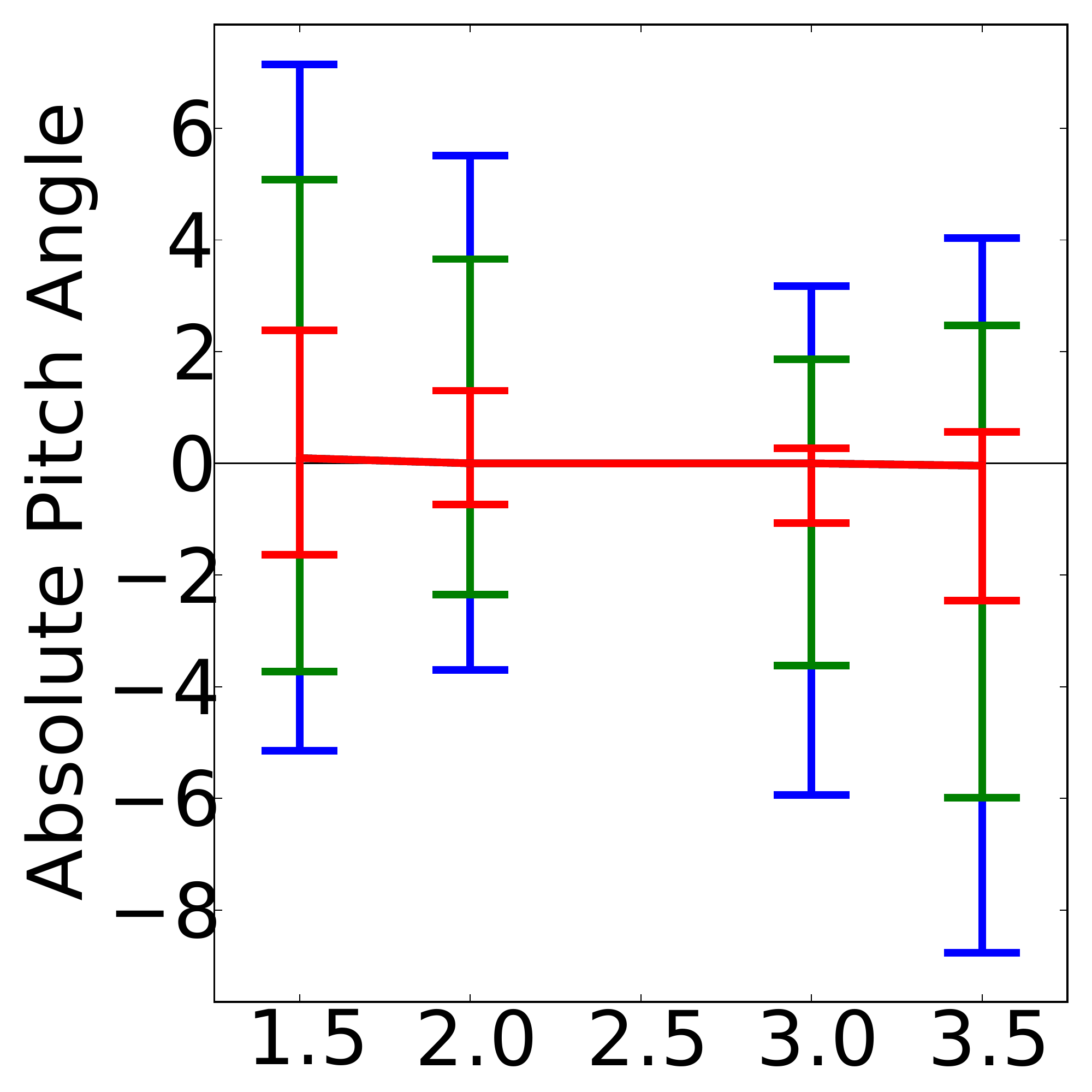}
	\end{subfigure}
	\begin{subfigure}[b]{0.32\linewidth}
		\centering
		\includegraphics[width=\linewidth]{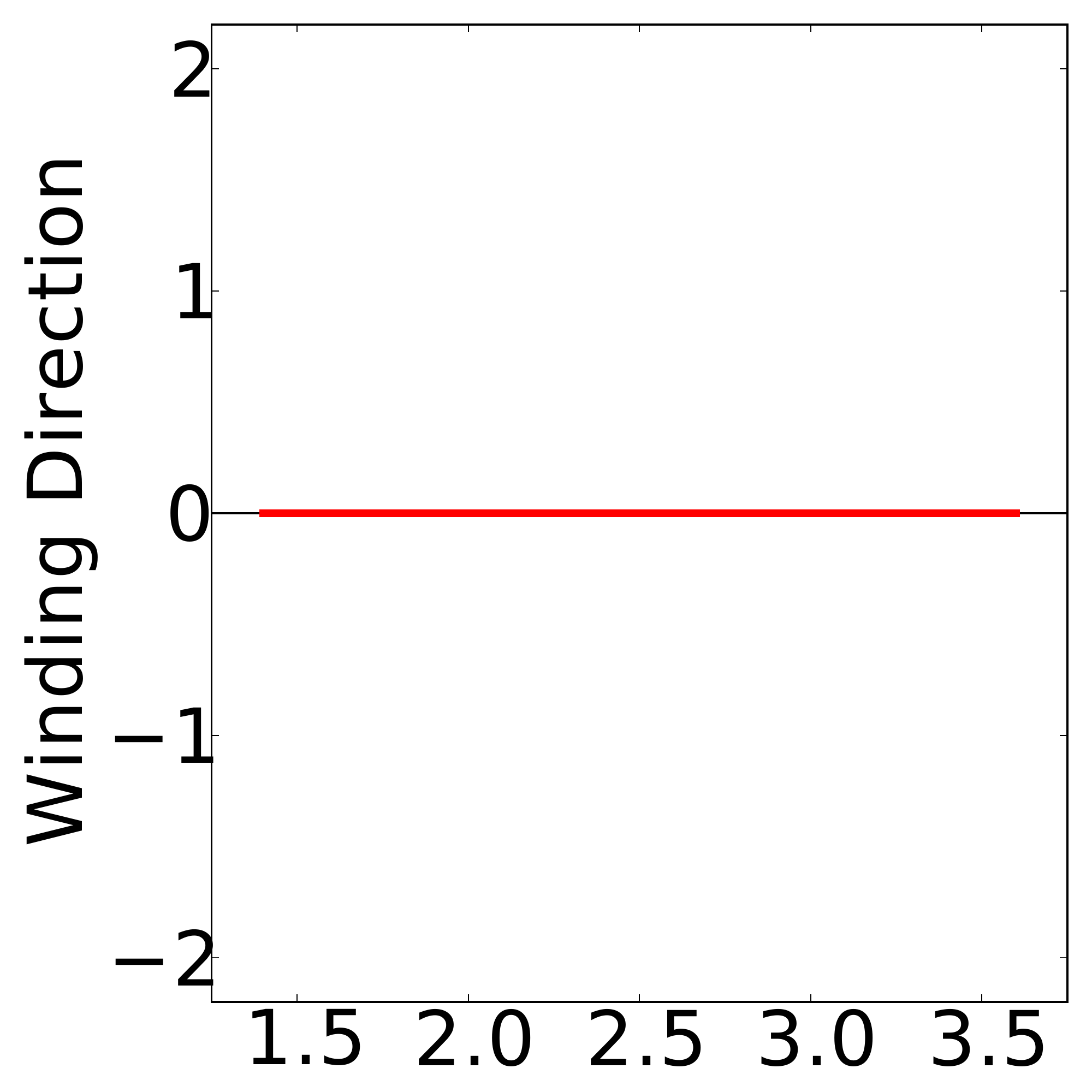}
	\end{subfigure}
	\sensitivitycaption{the maximum allowed combined-cluster to separate-cluster error ratio}
	\label{fig:sensitivity_errRatioThres}
\end{figure}

Figure \ref{fig:sensitivity_errRatioThres} displays the effect of varying the maximum allowed combined-cluster to separate-cluster error ratio, which controls the stringency applied when checking cluster merges (higher values are less stringent).
As the allowed ratio increases, the average arc length increases; this is expected because clusters get larger when more merges are allowed.
Reduced merge stringency also decreases the total number of arcs.
This parameter change could also conceivably increase the number of arcs because increased merge-check leniency could allow clusters to grow beyond the size-based output-inclusion threshold, but since fit-based merge checks probably have a stronger tendency to block merges of large clusters (there are more ways in which large clusters can deviate from the logarithmic spiral model), it makes sense that the net effect of less stringent checks is to reduce the number of arcs (since an arc merge decreases the count by one).
Less stringent checks also reduce the total arc length, although the impact of this parameter change on total arc length is less than the effect of most other algorithm parameters.
The reduction in total length is sensible because the merged arcs can sometimes overlap slightly in their angular range.
However, it is interesting that (for most galaxies) this effect is stronger than the extra arc length gained when allowing more merges of non-contiguous clusters during the secondary merging step.
The effects on the typical signed pitch angle and arm tightness (absolute pitch angle) are negligible to nonexistent, suggesting that difficult merge decisions do not, on average, change the typical pitch angle.
Similarly, effects on winding direction are rare or nonexistent.

\begin{figure}
	\centering
	\begin{subfigure}[b]{0.32\linewidth}
		\centering
		\includegraphics[width=\linewidth]{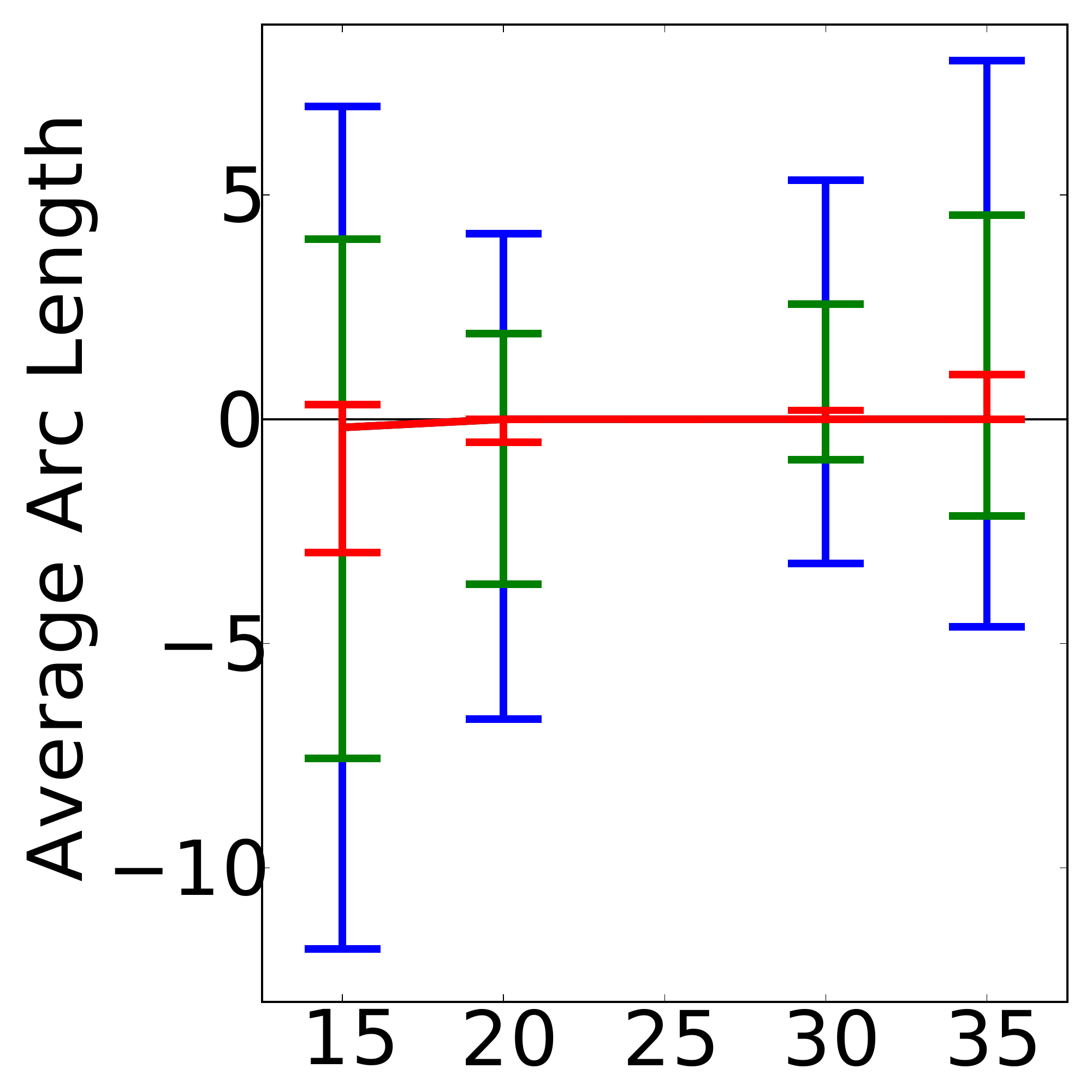}
	\end{subfigure}
	\begin{subfigure}[b]{0.32\linewidth}
		\centering
		\includegraphics[width=\linewidth]{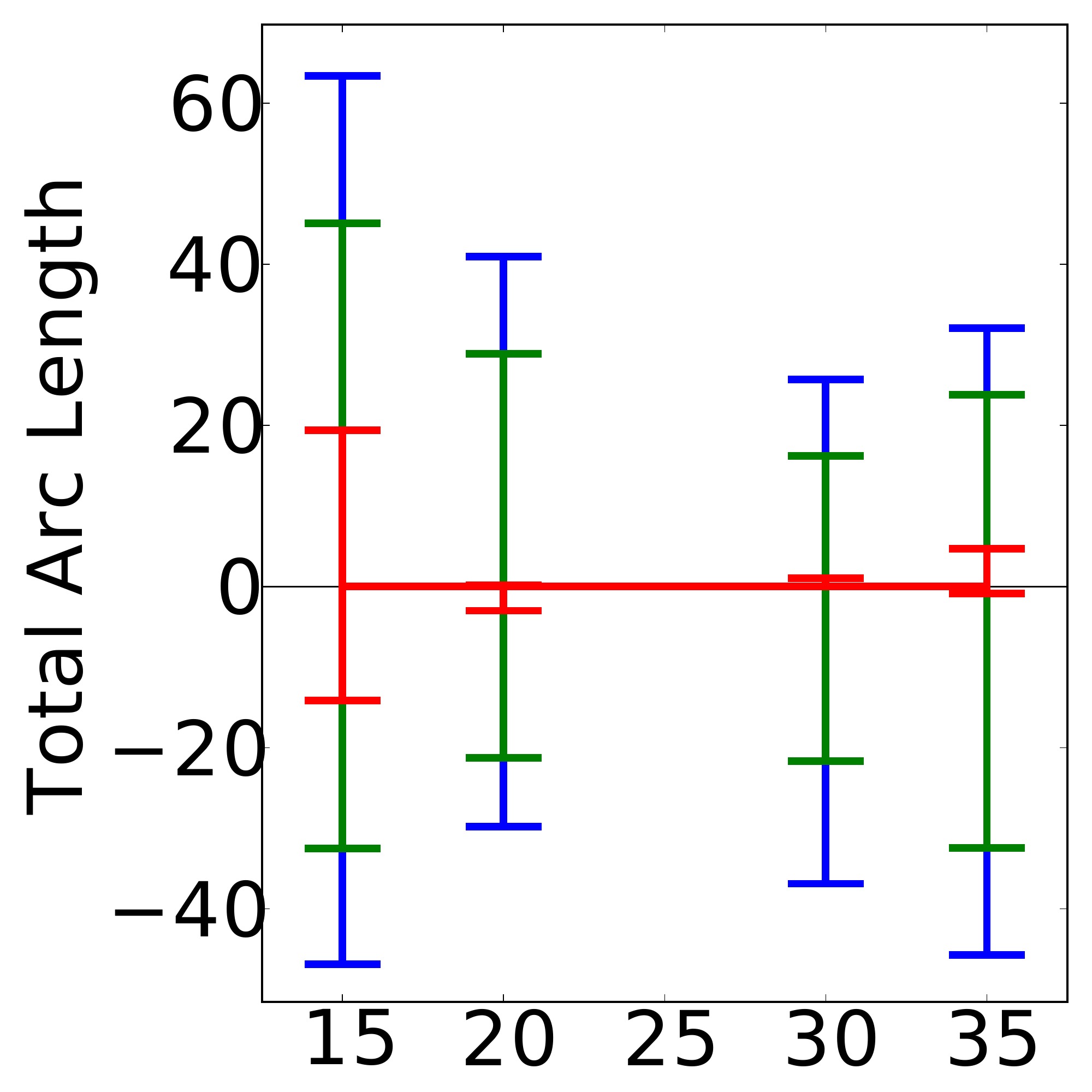}
	\end{subfigure}
	\begin{subfigure}[b]{0.32\linewidth}
		\centering
		\includegraphics[width=\linewidth]{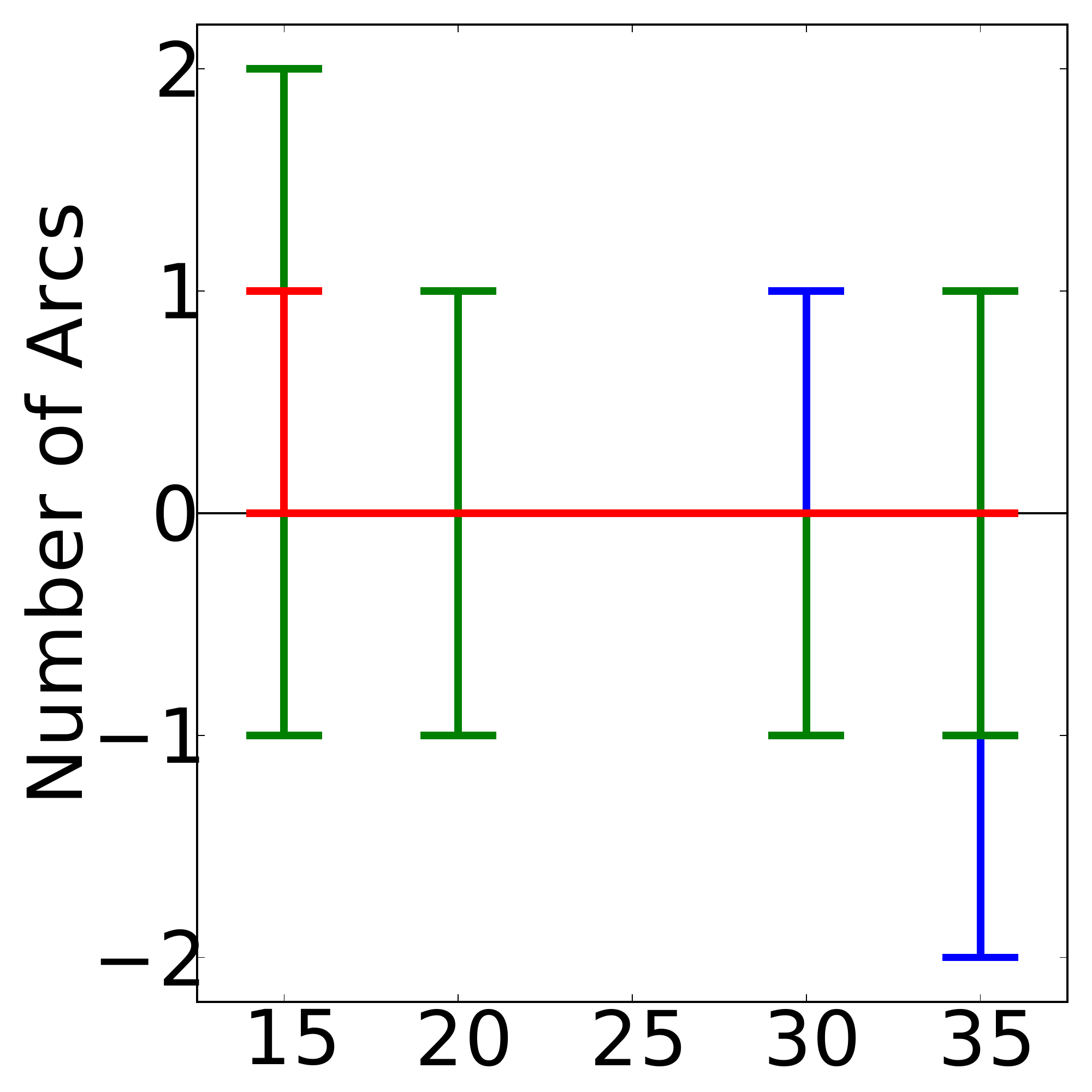}
	\end{subfigure}
	\\
	\begin{subfigure}[b]{0.32\linewidth}
		\centering
		\includegraphics[width=\linewidth]{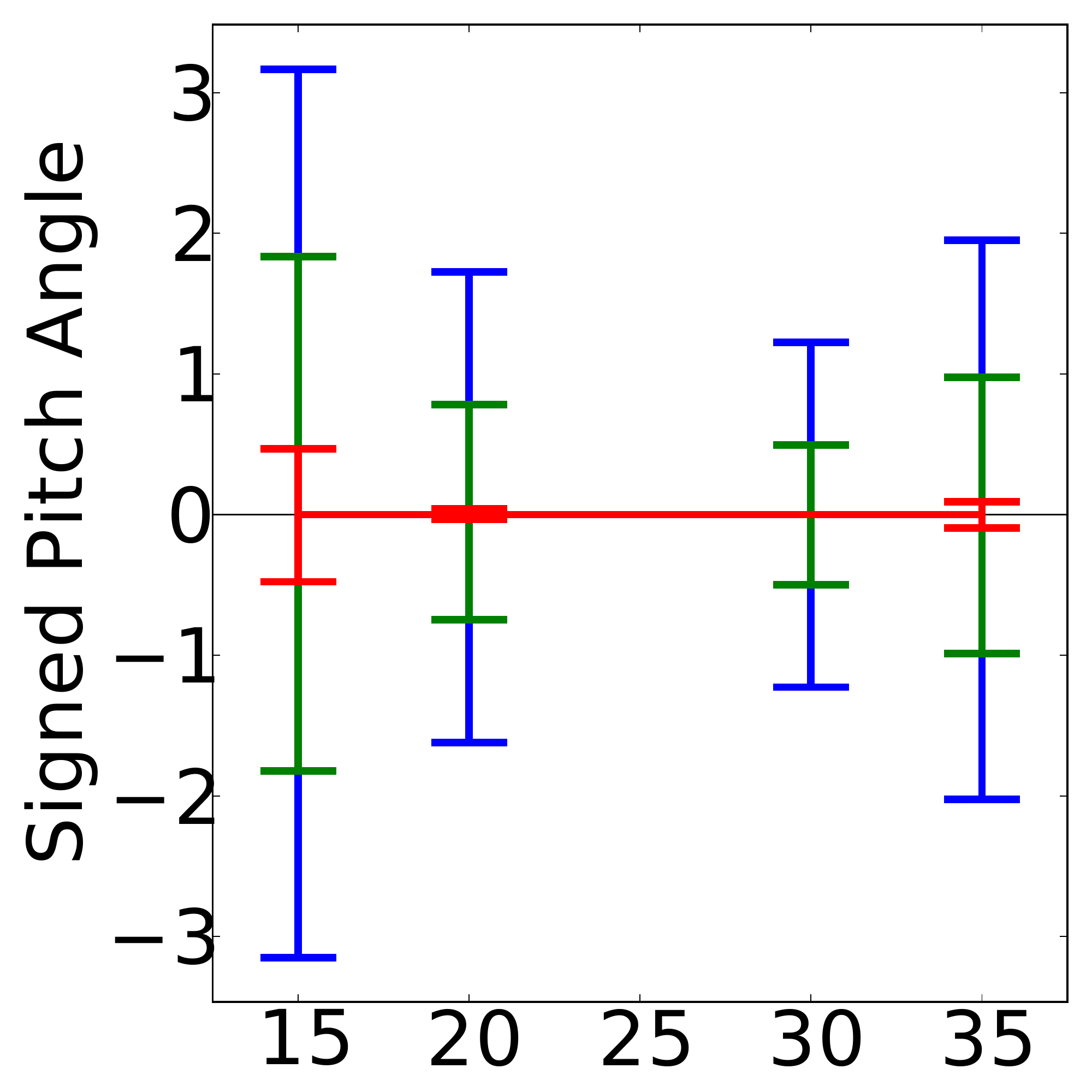}
	\end{subfigure}
	\begin{subfigure}[b]{0.32\linewidth}
		\centering
		\includegraphics[width=\linewidth]{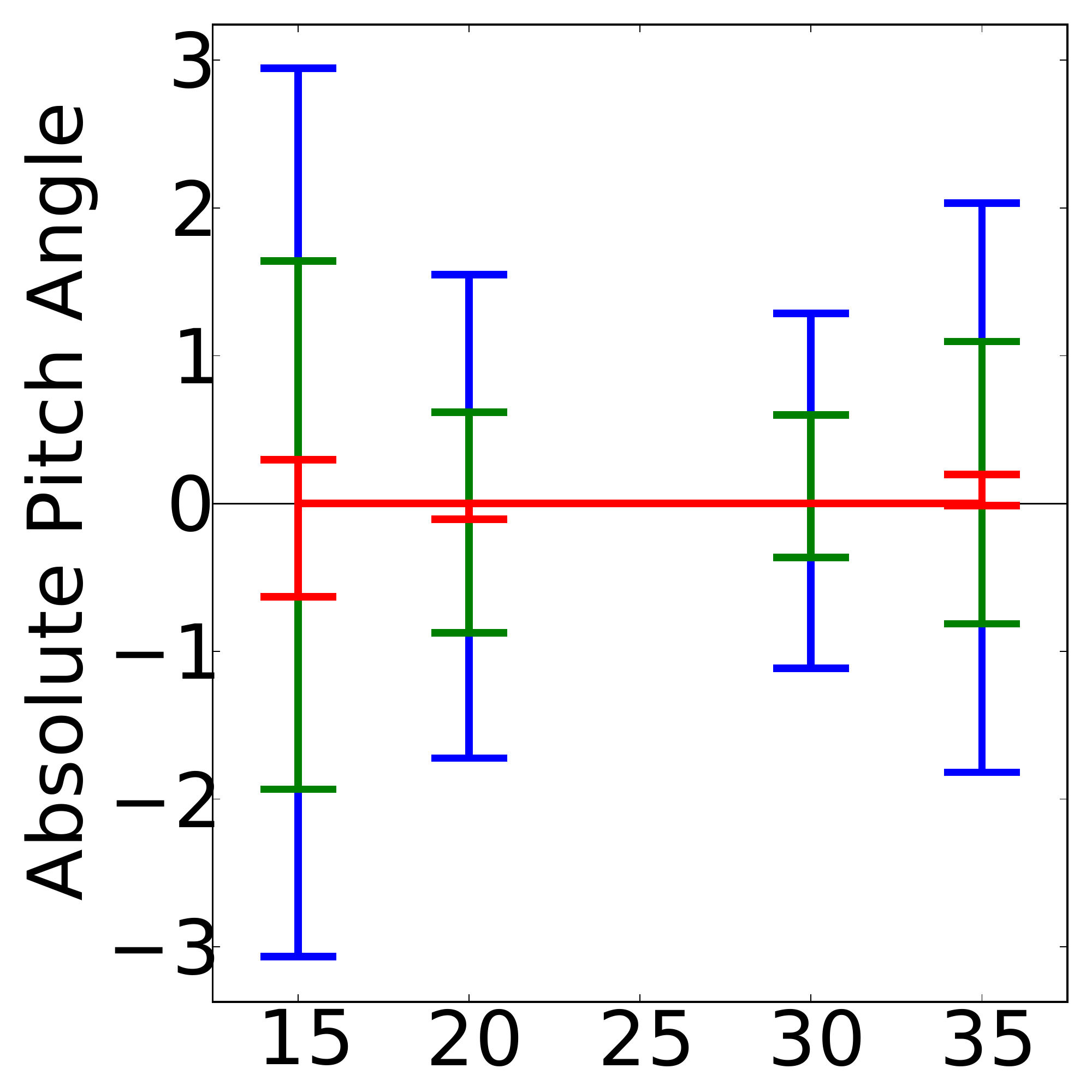}
	\end{subfigure}
	\begin{subfigure}[b]{0.32\linewidth}
		\centering
		\includegraphics[width=\linewidth]{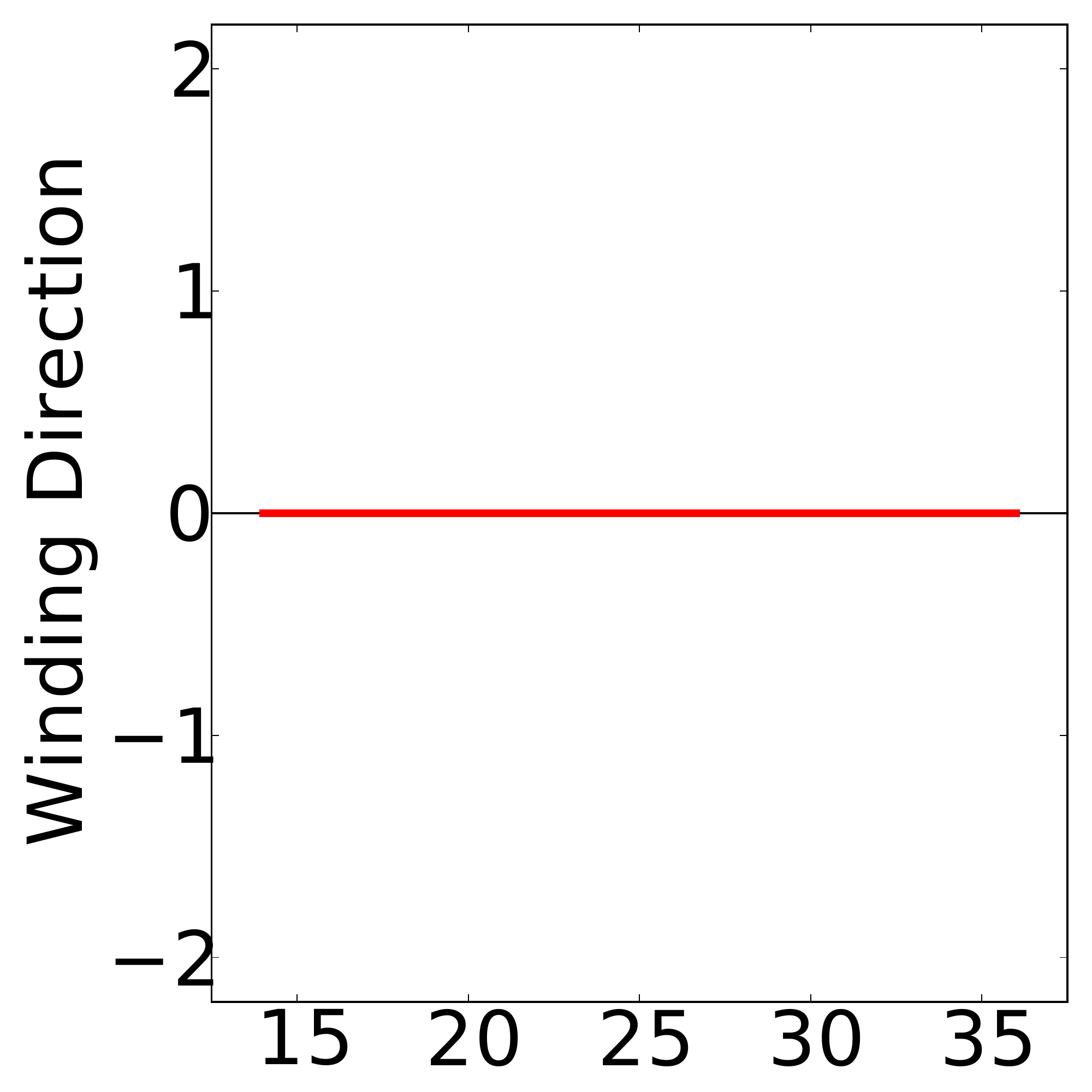}
	\end{subfigure}
	\sensitivitycaption{the minimum size each cluster must reach in order to trigger a fit-based merge check}
	\label{fig:sensitivity_mergeChkMinClusSz}
\end{figure}

This fit-based merge check is only used when both clusters reach a minimum size.
Figure \ref{fig:sensitivity_mergeChkMinClusSz} displays the effects of varying this minimum.
For all of the output measures considered, typical values do not change substantially; to the minimial extent that values do change (examining the heights and asymmetries of the bars), the effects of increasing this merge-check size threshold are similar to the effects of making the fit-based merge checks more lenient.
This is sensible because increasing the merge-check size threshold also makes the merge checks more lenient, but in a more limited way: checks are more important for large clusters, but this minimum is only relevant for very small clusters.
Since this minimum size was used in part to save computation time, it is good that changes to its value do not result in any substantial changes to typical values of any of the measurements.
The other aim of this minimum was to avoid performing a merge check before the cluster shape was known.
The smallest tested minimum value (which starts merge checks earlier) produces the most scatter, suggesting that this objective is being met.
We also note that scatter begins to increase again at the largest tested minimum value, which further supports the conclusion that the current parameter setting is a good one; the increased scatter may be due to allowing clusters to grow too far before starting merge checks.
In any case, however, there appears to be little or no effect on pitch angle, tightness (absolute pitch angle), or winding direction.

\begin{figure}
	\centering
	\begin{subfigure}[b]{0.32\linewidth}
		\centering
		\includegraphics[width=\linewidth]{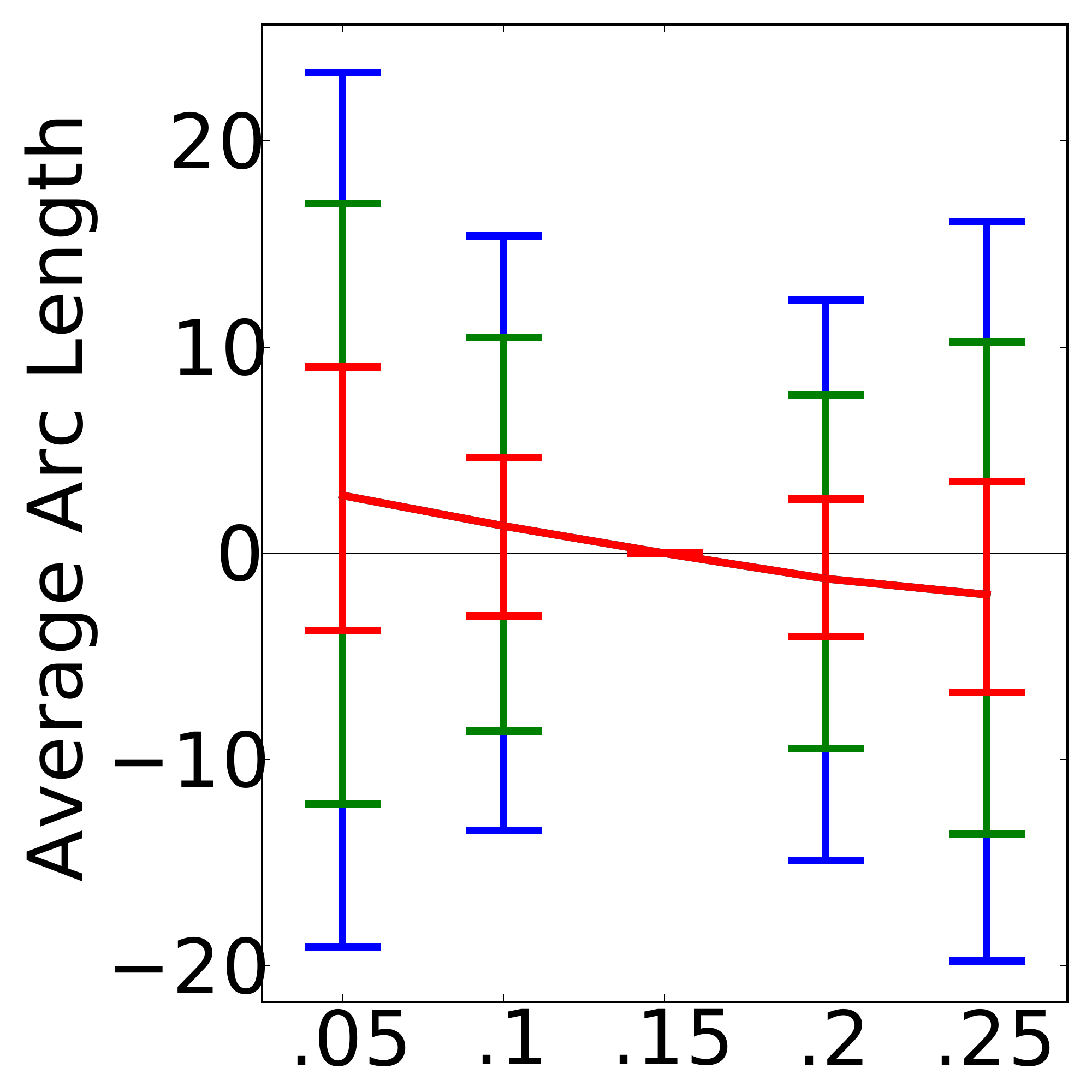}
	\end{subfigure}
	\begin{subfigure}[b]{0.32\linewidth}
		\centering
		\includegraphics[width=\linewidth]{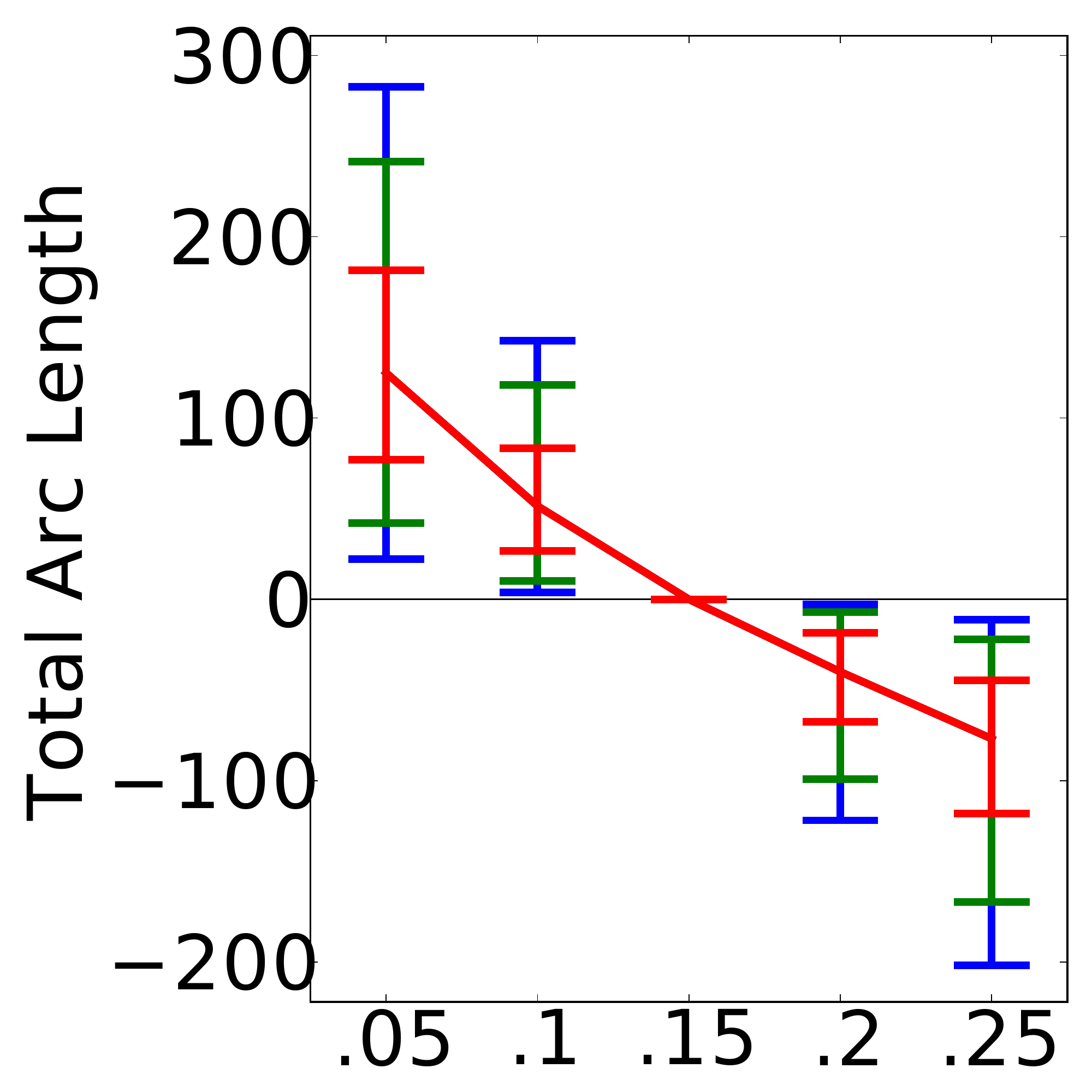}
	\end{subfigure}
	\begin{subfigure}[b]{0.32\linewidth}
		\centering
		\includegraphics[width=\linewidth]{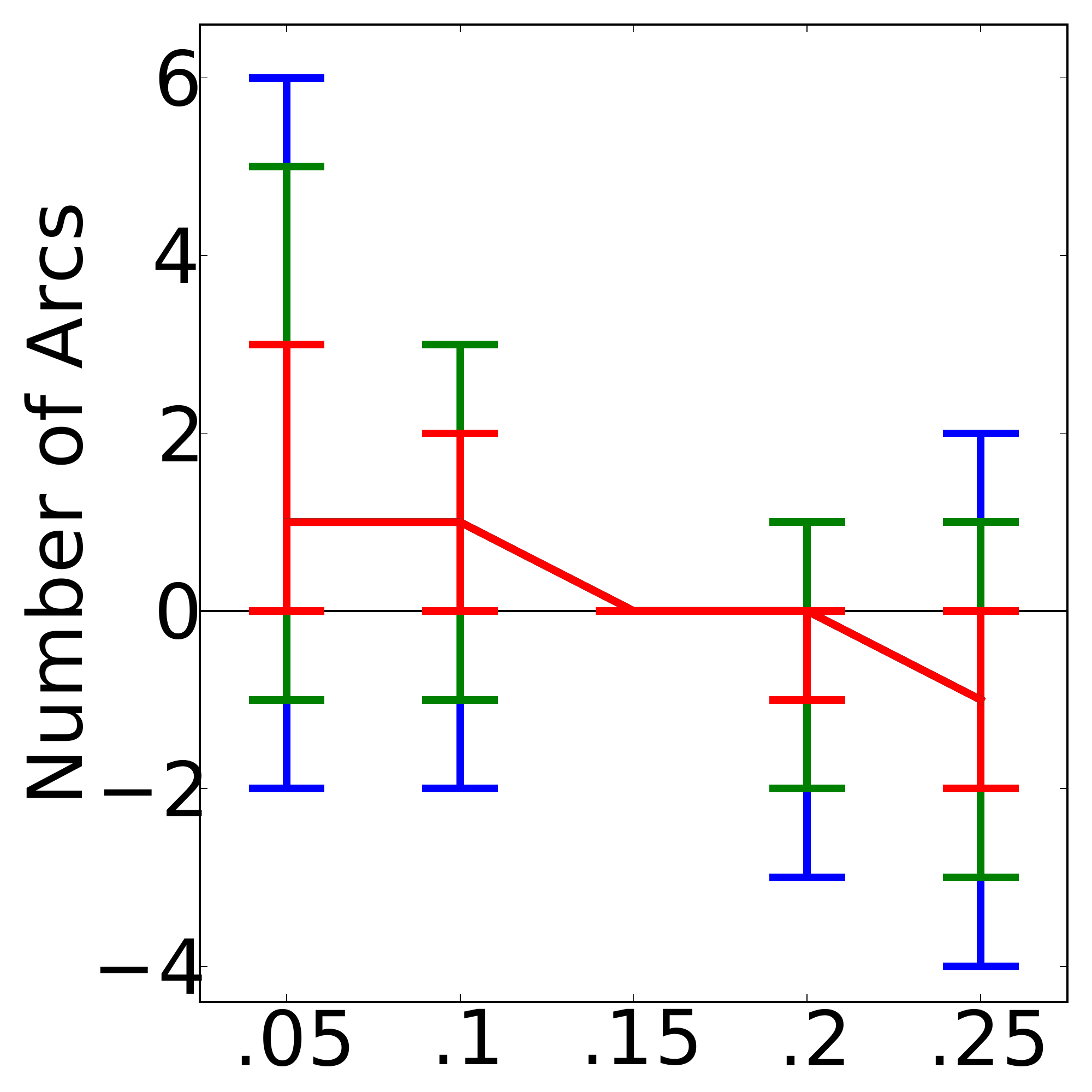}
	\end{subfigure}
	\\
	\begin{subfigure}[b]{0.32\linewidth}
		\centering
		\includegraphics[width=\linewidth]{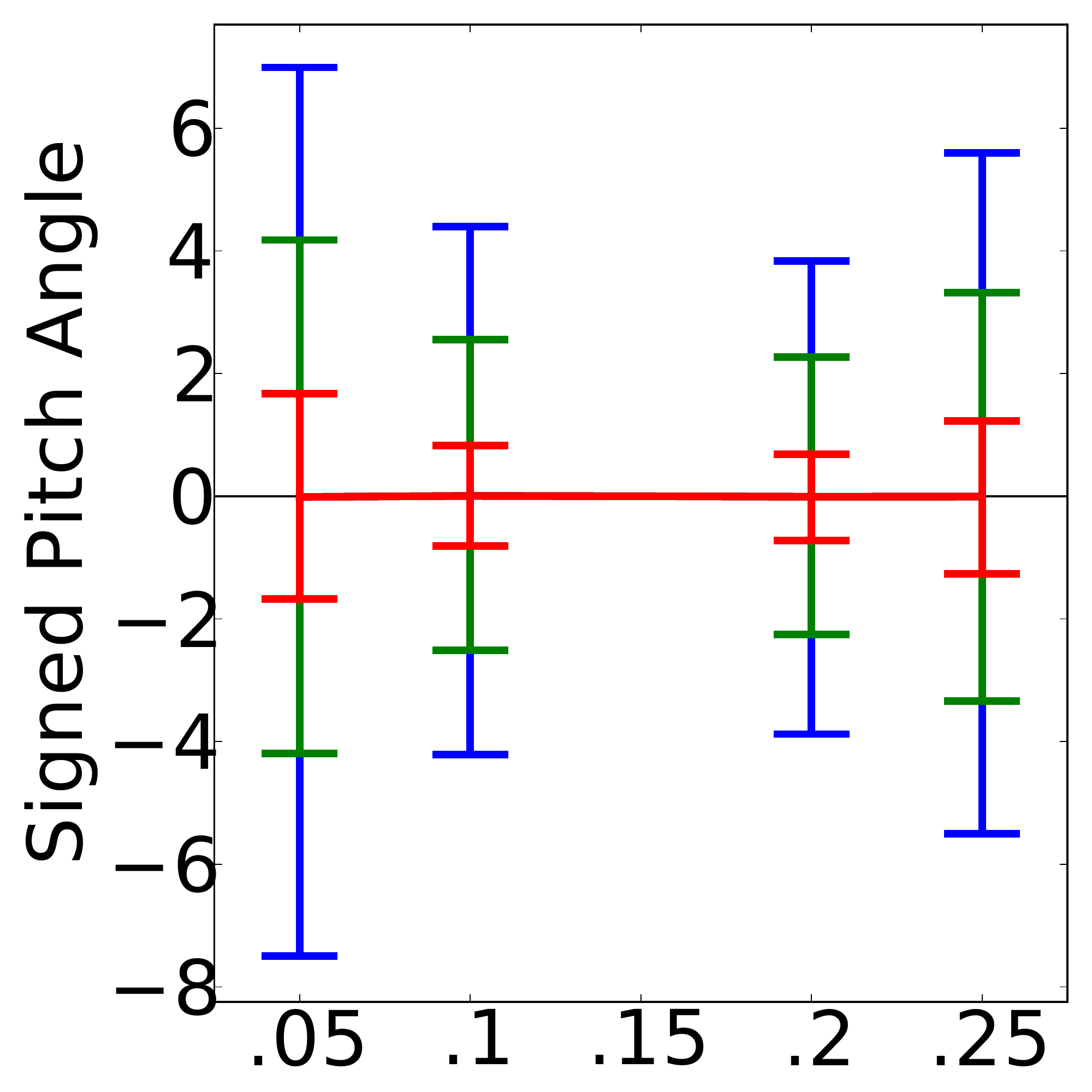}
	\end{subfigure}
	\begin{subfigure}[b]{0.32\linewidth}
		\centering
		\includegraphics[width=\linewidth]{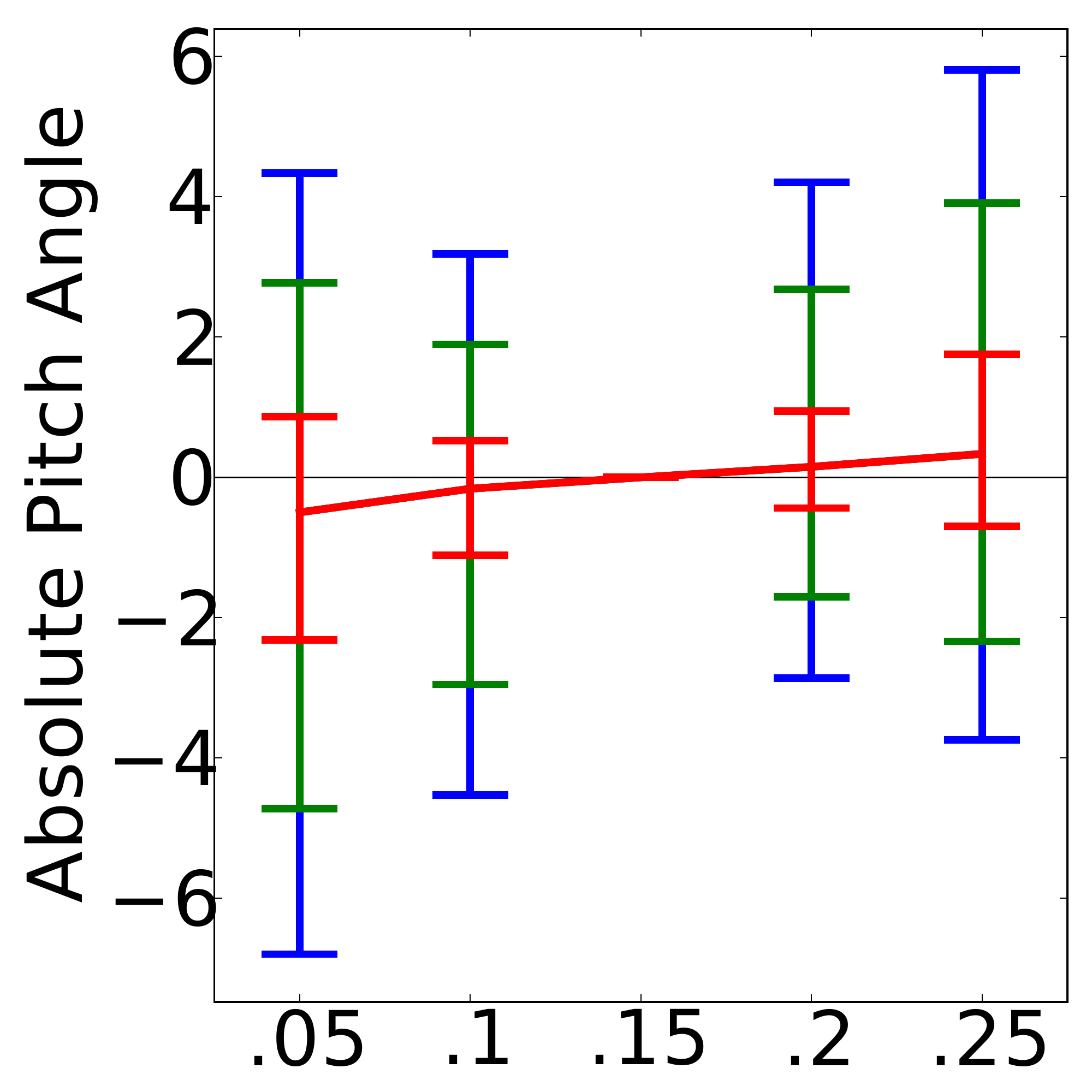}
	\end{subfigure}
	\begin{subfigure}[b]{0.32\linewidth}
		\centering
		\includegraphics[width=\linewidth]{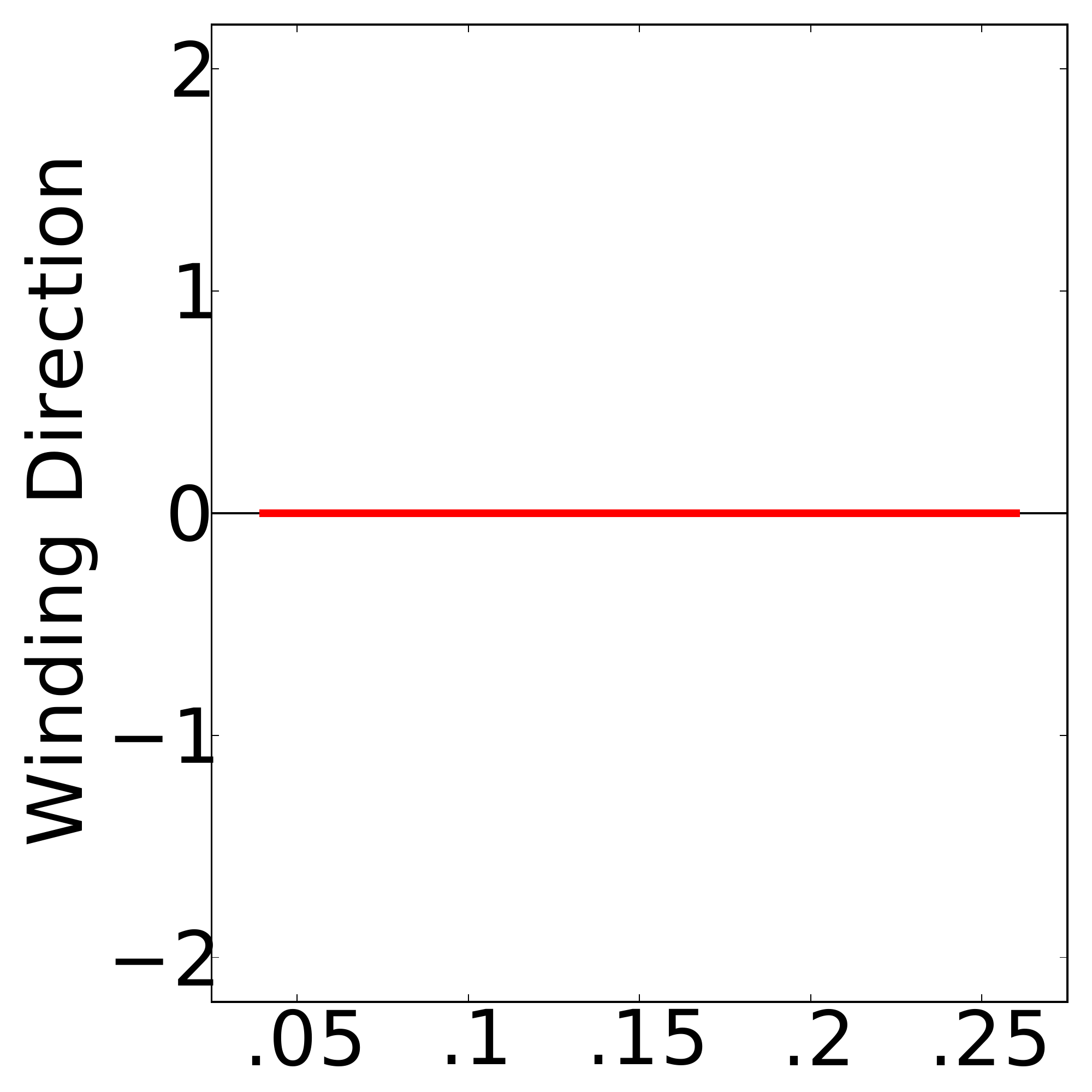}
	\end{subfigure}
	\sensitivitycaption{the minimum orientation similarity needed to continue clustering}
	\label{fig:sensitivity_stopThres}
\end{figure}

Finally, we investigate the effect of the stopping threshold applied to the pixel clustering (note that cluster similarities are determined by the maximum similarity of inter-cluster pixel pairs, that pixel similarities are determined by the dot product of the associated orientation vectors, and clustering continues in similarity order until the next cluster similarity falls below the threshold value considered here).
Comparing to Figure \ref{fig:sensitivity_unsharpMaskAmt}, which characterized the effect of the unsharp mask amount, we see that all trends are reversed, but less strong in terms of both the amount of scatter and the change in median values.
This correspondence in effect is sensible because both the clustering stopping threshold and the unsharp mask amount control the extent to which clusters can expand (with more expansion for a higher unsharp mask amount or a lower stopping threshold); higher unsharp mask intensities increase many ``good'' cluster-to-cluster similarities, while reducing the stopping threshold makes the pixel similarities larger relative to the stopping threshold.
Compared to changes in the unsharp mask amount, changes in the stopping threshold may have a weaker effect due to the range of stopping threshold values tested here, but it may also be because the stopping threshold uses more nuanced information (the next merge depends on cluster merge history, where the best merges were tried first) and is moderated by the fit-based merge checking.
Examining the effects of the stopping threshold on the individual measurements and keeping in mind that lower thresholds mean more cluster growth, it is sensible that a lower stopping threshold slightly increases average arc length (since existing clusters can grow by incorporating boundary pixels, and since merges reduce the cluster count, with these two factors overpowering the introduction of small clusters that newly exceed the size threshold), total arc length (as a straightforward consequence of cluster growth, along with a greater chance for smaller clusters to reach the minimum size threshold), and (to a slight extent) the total number of arcs (suggesting that, on average, slightly more clusters are introduced by exceeding a size threshold than are removed due to merging).
As was seen with the unsharp mask intensity, there is no noticeable effect on the typical signed pitch angle, but the arm tightness is increased somewhat (the absolute pitch angle is decreased somewhat) with more cluster growth.
Since this (weak) trend is similar to the observed effect of increasing the unsharp mask amount, and due to the previously-mentioned similarities of the stopping threshold and the unsharp mask amount, the underlying mechanism behind the arm-tightness change is likely similar.
Lastly, we find that typical winding direction is unaffected, as usual.

\section*{Acknowledgment}
We thank Aaron Barth for forcing us to think carefully about possible selection effects, which led directly to this paper; Aaron also helped us figure out exactly what parameters were important to think about when degrading the images. We thank Shawna Stahlheber for help with clear display of Figure \ref{fig:sp_result}. P. Silva was supported by CAPES (Coordination for the Improvement of Higher Education Personnel - Brazil) through the Science Without Borders fellowship for PhD Studies.

%%%%%%%%%%%%%%%%%%%%%%%%%%%%%%%%%%%%%%%%%%%%%%%%%%



\bibliographystyle{mnras}

\bibliography{library,wayne}
% % Alternatively you could enter them by hand, like this:
% \begin{thebibliography}
% \bibitem{Sunpy}
% Torrey, Paul, Snyder, Greg. 2016. Sunpy: Module for opening, manipulating, and plotting mock galaxy images produced with SUNRISE. GitHub repository. https://github.com/ptorrey/sunpy
% \end{thebibliography}

%%%%%%%%%%%%%%%%%%%%%%%%%%%%%%%%%%%%%%%%%%%%%%%%%%

%%%%%%%%%%%%%%%%% APPENDICES %%%%%%%%%%%%%%%%%%%%%

%%%%%%%%%%%%%%%%%%%%%%%%%%%%%%%%%%%%%%%%%%%%%%%%%%

% Don't change these lines
\bsp	% typesetting comment
\label{lastpage}
\end{document}